\documentclass[11pt,a4paper]{article}
\usepackage{jheppub}
\usepackage{graphicx}
\usepackage{amsmath,bm,braket}
\usepackage{slashed}
\usepackage{tikz}
\usetikzlibrary{patterns, arrows.meta}
\usepackage[compat=1.1.0]{tikz-feynman}
\usepackage{orcidlink}

\newcommand\nb{{\bar{n}}}
\newcommand\Nb{{\overline{N}}}
\def\lqcd{\Lambda_{\rm QCD}}
\def\nn{\nonumber \\ }
\def\rd{{\rm d}}
\def\msbar{\overline{\text{MS}}}
\def\lQ{\mathsf{L}_Q}

\title{Polarized Deep Inelastic Scattering as $x \to 1$ using Soft Collinear Effective Theory}

\author[a]{Jaipratap Singh Grewal \orcidlink{https://orcid.org/0000-0003-1256-2281},}
\author[a]{Aneesh V. Manohar \orcidlink{https://orcid.org/0009-0004-5497-8554},}
\author[b]{Jyotirmoy Roy \orcidlink{https://orcid.org/0000-0002-4574-5386}}
\affiliation[a]{Physics Department 0319, University of California San Diego, 9500 Gilman Drive, La Jolla, CA 92093-0319, USA}
\affiliation[b]{Department of Physics, Duke University, Durham, North Carolina 27708, USA}
\emailAdd{j1grewal@ucsd.edu}
\emailAdd{amanohar@ucsd.edu}
\emailAdd{jyotirmoy.roy@duke.edu}

\abstract{

We use Soft Collinear Effective Theory (SCET) to factorize the polarized Deep Inelastic Scattering (DIS) structure functions $g_1(x)$ and $g_2(x)$, and to sum  Sudakov double logarithms of $1-x$. The analysis is done both in terms of lightcone parton distributions and their moments. Computing $g_2$ requires subleading SCET operators which contain gluons. We calculate the one-loop matching coefficients from QCD onto these subleading SCET operators, and the one-loop matching from SCET onto the parton distribution function (PDF). The PDF in SCET is given by a bilocal operator, rather than the trilocal operator used in the QCD analysis of $g_2$ for generic $x$. We compute the one-loop anomalous dimension of the PDF operator for any $x$, and show that as $x \to 1$, it factors into a single-variable evolution.

We comment on the QCD anomalous dimensions of twist-three operators, their equation-of-motion relation, and connection to the SCET analysis. We briefly discuss the definition of axial operators in the BMHV scheme. As a side result, we derive the $1/N$ dependence of the QCD coefficient functions for $F_1$, $F_L$ and $g_1$ in the $N \to \infty$ limit, where $N$ is the moment, which is expected to hold to all orders in $\alpha_s$.
}

\begin{document}

\maketitle


\newpage

\section{Introduction} \label{sec:intro}

Deep-inelastic scattering, inclusive electron-proton scattering $e^- + p \to e^- + X$, has played an important role in precision tests of QCD.  The highest energy measurements of the unpolarized cross section were done at HERA \cite{H1:2015ubc} while the polarized cross section has been studied before by the COMPASS experiment \cite{COMPASS:2016jwv} and is currently being studied at Jefferson Lab \cite{JeffersonLabHallAg2p:2022qap}. Theory predictions for DIS will be tested to even higher precision at the future Electron-Ion Collider (EIC) \cite{AbdulKhalek:2021gbh}, which will measure both the unpolarized cross section and the polarization asymmetry. 

The DIS cross section is written in terms of the (Euclidean) momentum transfer $Q^2>0$, and the dimensionless Bjorken variable $x$. The invariant mass of the final hadronic state $X$ is
\begin{align}
M_X^2 &= M_p^2 + Q^2 \frac{1-x}{x}\,,
\label{1.1}
\end{align}
where $M_p$ is the proton mass. 

In the region $x < 1$, $M_X^2 \sim Q^2 \gg \lqcd$, and the perturbation series for the DIS cross section has logarithms $\left[ \alpha_s \ln Q^2/\mu^2 \right]^n$ which need to be summed so that perturbation theory remains valid. In the usual QCD analysis, one matches the product of two electromagnetic currents to a PDF operator at the scale $Q$ using an operator product expansion (OPE). The PDF operator is evolved from $Q$ to a low scale $\mu_0$, where its proton matrix element gives the non-perturbative PDF $f(x,\mu_0^2$). The renormalization group evolution of the PDF operator from $Q$ to $\mu$ sums the logarithms $\alpha_s \ln Q^2/\mu^2$. Higher dimension operators in the OPE give $\lqcd/Q$ corrections, which are referred to as higher-twist or power corrections. 

In the kinematic region $x \to 1$, the final hadronic state becomes jet-like, with an invariant mass $M_X^2 \ll Q^2$, and in this regime, we denote $X$ by $J$.  DIS as $x \to 1$ refers to the regime $\lqcd^2 \ll M_J^2 \ll Q^2$ where there are many hadronic states in the final state sum, and QCD perturbation theory is valid. As $x \to 1$, the perturbation series has Sudakov double-logarithms $\left[ \alpha_s \ln^2 (1-x) \right]^n$ which need to be summed to maintain the validity of perturbation theory. This summation can be done using SCET~\cite{Bauer:2000ew, Bauer:2001yt,Bauer:2001ct,Bauer:2002nz,Beneke:2002ph} --- (a) The QCD current is matched onto the SCET current at the scale $Q$. (b) The current is evolved to the jet scale $M_J$. (c) An OPE of the time-ordered product of two SCET currents is performed at the jet scale to match onto the PDF operator, which is (d) then evolved to a low scale $\mu$. SCET sums the Sudakov double-logarithms $[\alpha_s \ln^2(1-x)]^n $ as well as the single-logs $[\alpha_s \ln Q^2/\mu^2]^n$, but only gives the $x$ dependence as $x \to 1$, since it expands in $1-x$. SCET power corrections give $(1-x)$ corrections, as well as the usual $\lqcd/Q$ power corrections. 

For a spin-1/2 target such as the proton, there are four DIS structure functions --- $F_1$ and $F_2$ for unpolarized scattering, and $g_1$ and $g_2$ for the polarization asymmetry.  The $x \to 1$ summation for the unpolarized case has been studied before~\cite{Manohar:2003vb}.  In this paper, we study the $x\to 1$ resummation for the polarized case using a SCET analysis to order $\lambda$, where $\lambda$ is the SCET expansion parameter\footnote{$\lambda$ is defined explicitly in Section~\ref{sec:review}.}. $g_1$ is leading twist, and like $F_1$, requires matching the QCD electromagnetic current onto the leading-order SCET current, which was computed in~\cite{Manohar:2003vb} to one-loop. The results for $g_1$ are almost identical to the previously obtained results for $F_1$~\cite{Manohar:2003vb}. $g_2$ is twist-three, and requires matching the QCD current onto order $\lambda$ operators which contain gluons. The tree-level matching was given in~\cite{Inglis-Whalen:2021bea,Luke:2022ops}. Here we compute the one-loop matching onto the subleading order $\lambda$ operators. The anomalous dimension of the subleading operators was computed in~\cite{Goerke:2017lei}. We compute the one-loop OPE matching onto twist-three PDF operators and their one-loop anomalous dimensions. This gives the full order $\alpha_s$ result for $g_2$ including leading-log resummation as $x \to 1$.  As a side result, we obtain the gluon contribution to $F_1$ and $g_1$ as $x \to 1$, which is order $\lambda^2 \sim 1/N$, as well as the $1/N$ dependence of the various OPE coefficient functions.

The structure functions $F_1$, $F_2$, and $g_1$ are determined in terms of parton distributions in a single variable, and the DGLAP evolution kernel is also a function of a single variable. $g_2(x)$, while still a structure function of a single variable, is determined in terms of parton distributions and evolution kernels which depend on two variables. This greatly complicates the QCD analysis of $g_2$. It is known that the $g_2$ evolution kernel simplifies to a single-variable DGLAP-like kernel as $x \to 1$~\cite{Ali:1991em,Geyer:1996isa}. We will see that the SCET analysis automatically gives expressions for $g_2$ and its evolution kernel as a function of two variables, $x$, which is a parton momentum fraction, and $u$, which is a SCET momentum fraction label. The evolution of the PDF factors into separate single variable evolutions in $x$ and $u$. The SCET analysis of $g_2$ as $x \to 1$ gives results in a much simpler form than the full QCD result.

The $g_1$ structure function is given by matrix elements of axial operators involving $\gamma_5$. These axial operators have to be properly defined --- in the BMHV scheme, they require a finite correction which was studied in detail to three-loop order in~\cite{Larin:1991tj,Larin:1993tq} for the case of the axial current. We compute the finite corrections at one-loop order for the entire tower of twist-two axial operators that are needed for $g_1$.  The results are given in Appendix~\ref{app:BMHV}.

Throughout our analysis, we refer to the hadronic target as the proton. However, the QCD and SCET analysis is completely general and does not depend on the hadronic target. The matching at $Q$, the SCET evolution from $Q$ to $M_J$, the OPE at $M_J$ onto PDF operators, and the PDF running from $M_J$ to $\mu_0$ all make no reference to the hadronic target.
The only place where the target enters is when taking the non-perturbative target matrix element of the PDF operator at the local scale $\mu_0$, which is given in terms of quark and gluon PDFs that depend on the target. Thus, our analysis extends to arbitrary spin targets~\cite{Jaffe:1988up,Hoodbhoy:1988am}.  We also refer to the lepton as the electron, but clearly, the results hold for muon scattering as well.

Section~\ref{sec:review} summarizes DIS kinematics and the SCET power counting.  The results of~\cite{Manohar:2003vb}
for unpolarized DIS as $x \to 1$ are reviewed in Section~\ref{sec:unpolarized}. Section~\ref{sec:g1} summarizes the QCD analysis of the polarized structure function $g_1$, and the SCET analysis as $x \to 1$.  The subleading operators in the matching of the current onto SCET are reviewed in Section~\ref{sec:g2}. The one-loop matching onto the subleading operators is given in
Section~\ref{sec:matchingQ}, their one-loop anomalous dimension is given in Sec.~\ref{sec:IW}, and the one-loop OPE at the jet scale in Section~\ref{sec:matchingJ}. The anomalous dimension of the twist-three PDF operator is computed in Section~\ref{sec:anomdim}. There are some matching consistency conditions which are verified in Section~\ref{sec:consistency}. The SCET analysis determines the order in $1/N$ of the QCD matching coefficients, which is given in Section~\ref{sec:coeff}. The results are expected to hold to all orders in $\alpha_s$.  In Section~\ref{sec:qcd_vs_scet}, we compare with QCD treatments of $g_2$. In Appendix~\ref{app:ops}, we discuss some properties of quark-gluon twist-three operators. The equation-of-motion which eliminates twist-three quark operators is given in Appendix~\ref{app:eom}. Appendix~\ref{app:BMHV} discusses the definition of axial operators in the BMHV scheme. Appendix~\ref{app:PDF} gives the general form of PDF operators in SCET (and QCD). The endpoint behavior of the PDF $h_q(x,u)$ as $u \to 0$ and $u \to 1$ is studied in Appendix~\ref{app:F}.

The entire computation will be performed, as in~\cite{Manohar:2003vb}, for a single quark flavor with unit charge. At one-loop order, the general result is obtained by summing the answers over quark flavors weighted by the square of the quark charges.

\section{Notation and Review}\label{sec:review}

\subsection{SCET Power Counting and Notation}

SCET \cite{Bauer:2000ew,Bauer:2001yt,Bauer:2002nz,Beneke:2002ph} is an effective field theory for strongly interacting particles with high energy but small invariant mass, and can be used for DIS in the $x \to 1$ limit \cite{Becher:2006mr,Hoang:2015iva} since the final state is jet-like.
We use the formalism of SCET established in \cite{Goerke:2017ioi,Inglis-Whalen:2021bea,Luke:2022ops} where each SCET sector is full QCD, and operators coupling different sectors are expanded.

We work in the Breit frame, where $q$ has no time component, and the target is back-scattered. The null vector $\nb^\mu  = (1,0,0,-1)$ is the incoming target direction, and $n^\mu = (1,0,0,1)$ is the direction of the incoming lepton and outgoing final state jet.  A generic momentum $k$ has lightcone components $k^+ \equiv n \cdot k$, $k^- \equiv \overline n \cdot k$ and $k_\perp$ in the $x-y$ plane. We will use $\parallel$  and $\perp$ to denote directions in the $t-z$ plane, and $x-y$ planes, respectively.

In the Breit frame, partons in the target are collinear to the $\nb$ direction, with SCET power counting $k^+ \sim Q$, $k^- \sim Q \lambda^2$, $k_\perp \sim Q \lambda$ with $\lambda \sim \lqcd/Q$, since their invariant mass is of the order of a hadronic scale and much smaller than their energy.  Partons in the final state jet when $x \to 1$ are collinear to the $n$ direction, with SCET power counting $k^+ \sim Q \lambda^2$, $k^- \sim Q $, $k_\perp \sim Q \lambda$ with $\lambda \sim M_J/Q \sim
\sqrt{1-x}$.  QCD is matched onto SCET at the scale $Q$, and operators of subleading power are suppressed by $\lambda$. The value of $\lambda$ is determined at a lower scale depending on whether the SCET operator matrix element is in the initial hadron or final jet, so the order $\lambda$ operators include both $\lqcd/Q$ and $\sqrt{1-x}$ corrections. 

The momentum transfer $q$ in the Breit frame has components $q^+ = -Q$, $q^- = Q$ and $q_\perp=0$,
\begin{align}
q^\mu &= \frac{Q}{2} \left(n^\mu - \overline n^\mu \right),  & P^\mu = \frac12 (n \cdot P) \overline n^\mu +  \frac12 \frac{M_p^2}{n \cdot P} n^\mu \approx  \frac12 (n \cdot P) \overline n^\mu \,,
\label{1.3}
\end{align}
where $P^\mu$ is the proton momentum and $M_p$ is the proton mass.
We will neglect $\lqcd^2/Q^2$ power corrections, so we can use $P^-=0$, meaning $P$ only has a $\nb^\mu$ component. It is convenient to introduce
\begin{align}
g_\perp^{\mu \nu} &= g^{\mu \nu} - \frac12 (n^\mu \nb^\nu + n^\nu \nb^\mu), &
\epsilon_\perp^{\mu \nu} &= \frac12 \epsilon^{\mu \alpha  \nu  \beta} n_\alpha \nb_\beta, & \epsilon_\perp^{12} = -\epsilon_\perp^{21} = 1 \,,
\label{1.4}
\end{align}
which are the metric tensor and the antisymmetric tensor restricted to the $\perp$ directions, respectively.

The $N{}^{\text{th}}$ moment of a function $f(x)$ is defined as
\begin{align}
M_{N} [f] &\equiv \int_0^1 \rd x\ x^{N-1}\ f(x)\,.
\label{1.2}
\end{align}
In moment space $1-x \sim 1/N$ so that $\lambda^2 \sim 1/N$ and the jet mass is $M_J^2 \sim Q^2/N$. We also use the notation
\begin{align}
 \lQ &= \log \frac{Q^2}{\mu^2} \,,
\label{1.7}
\end{align}
and the harmonic number
\begin{align}
H_k &= \sum_{j=1}^k \frac{1}{j} \,.
\label{1.8}
\end{align}
Our sign conventions are $D_\mu = \partial_\mu + i g A_\mu$, and $\epsilon_{0123}=+1$.  We use the $\msbar$ scheme. Many quantities depend on the renormalization scale $\mu$, which is often left implicit.

\subsection{Kinematics}

We briefly summarize the results for polarized DIS that we need --- a more detailed review can be found in \cite{Manohar:1992tz}. The DIS process is $e^-(k) + p (P) \to e^-(k^\prime) + J$ where $k$ is the incoming electron momentum,  $k^\prime$ is the outgoing electron momentum and $P$ is the proton momentum, as show in Fig.~\ref{fig:1}. The Lorentz-invariant kinematic variables are
\begin{align}
Q^2 &= - q^2 > 0, & x &= \frac{Q^2}{2 P \cdot q} = \frac{1}{\omega}, & y &= \frac{P \cdot q}{P \cdot k}\,.
\label{2.1}
\end{align}
$y$ is the lepton energy fraction lost in the lab frame. In the Breit frame
\begin{align}
    x = \frac{Q}{P^+}.
\end{align}
The hadronic tensor is defined as
\begin{figure}
    \centering
        \begin{tikzpicture}
        
        \draw[decoration={pre length=6pt, markings,mark=at position 0.5 with {\arrow{Stealth}}},postaction={decorate}] 	(-3,0.5)  -- (-1,0.5) ;
        \draw[decoration={pre length=6pt, markings,mark=at position 0.5 with {\arrow{Stealth}}},postaction={decorate}] 	(-1,0.5)  -- (0.5,1.3) ;
        
        \draw (-3.5,0.5) node [align=center] {$e^-(k)$};
        \draw (0.8,1.6) node [align=center] {$e^-(k^\prime)$};
        \draw (-0.2,0.2) node [align=center] {$q$};
        \draw (-2.25,-2.25) node [align=center] {$p(P)$};
        \draw (2.9,-1) node [align=center] {$J$};

        \filldraw[]  (-1,0.5)  circle (0.04);

        \draw [decorate,decoration={snake}] (-1,0.5) -- (0.5,-1);

        \draw [line width=0.6mm, black,decoration={pre length=6pt, markings,mark=at position 0.5 with {\arrow{Stealth}}},postaction={decorate} ] (-2,-2) -- (0.5,-1) ;
        
        \draw [black] (0.4,-1) -- (2.5,-0.6) ;
        \draw [black] (0.4,-1) -- (2.5,-0.8) ;
        \draw [black] (0.4,-1) -- (2.5,-1) ;
        \draw [black] (0.4,-1) -- (2.5,-1.2) ;
        \draw [black] (0.4,-1) -- (2.5,-1.4) ;

        \filldraw[fill=white] (0.5,-1) circle (0.7);
        \filldraw[fill=white, opaque, draw=black, thick, pattern=north west lines] (0.5,-1) circle (0.7);

        \end{tikzpicture}
    \caption{    \label{fig:1} Deep Inelastic Scattering. As $x \to 1$, the final hadronic state $J$ becomes jet-like, with invariant mass much smaller than $Q$ but large compared to $\lqcd$.}
\end{figure}
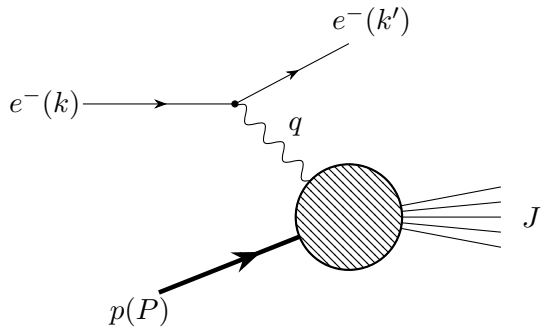
\begin{align}
    W_{\mu\nu}(P,S,q) &= \frac{1}{4\pi} \int d^4z \, e^{iq \cdot z} \braket{ P,S|[j_\mu(z),j_\nu(0)]|P,S } \nn
    &= -F_1 \left(g_{\mu\nu} - \frac{q_\mu q_\nu}{q^2} \right) + \frac{F_2}{P \cdot q}\left(P_\mu - \frac{(P \cdot q)\, q_\mu}{q^2} \right)\left(P_\nu - \frac{(P \cdot q)\, q_\nu}{q^2} \right) \nn
    & + \frac{ig_1}{P \cdot q}\epsilon_{\mu\nu\lambda\sigma} q^\lambda S^\sigma  + \frac{ig_2}{(P \cdot q)^2} \epsilon_{\mu\nu\lambda\sigma} q^\lambda (P \cdot q \, S^\sigma - S \cdot q \, P^\sigma) \, ,
    \label{2.2}
\end{align}
where $j$ is the electromagnetic current, and the structure functions $F_1$, $F_2$, $g_1$ and $g_2$ are dimensionless functions of $x$ and $Q^2$. The decomposition Eq.~\eqref{2.2} is valid for spin-1/2 targets, and assumes $P$ and $T$ symmetry.  We will use uppercase letters $S^\mu,\, P^\mu$ for hadronic (proton) quantities, and lower case letters $s^\mu,\, p^\mu$ for partonic quantities. $S^\mu$ is the spin of the proton, and is normalized to $S^2 = -M_p^2$. The $Q^2$ dependence of structure functions for large $Q^2$, which is logarithmic, is computable in perturbative QCD.  Using lightcone coordinates in the Breit frame, we can write
\begin{align}
W^{\mu \nu} &=  - F_1 g^\perp_{\mu \nu} +\frac{F_L}{8x} 
\left(  \nb + n  \right)^\mu\left(  \nb + n \right)^\nu - i g_1 \epsilon_\perp^{\mu \nu} \frac{n \cdot S}{n \cdot P}   \nn 
&  + i \frac{g_1+g_2}{Q}x \left[ - (n + \nb)^\mu \epsilon_\perp^{\nu \sigma} + (n + \nb)^\nu \epsilon_\perp^{\mu \sigma} \right] S_\sigma  \,.
\label{2.20}
\end{align}
The DIS literature usually uses $F_2$ and $F_L=F_2-2xF_1$ as the two independent unpolarized structure functions. In SCET, Eq.~\eqref{2.20} shows that it is better to use $F_1$ and $F_L$, since they separate the dependence on $\parallel$ and $\perp$ indices. $F_1$ and $g_1$ contribute to $W^{\mu \nu}$ with $\mu,\nu$ both $\perp$ indices, $F_L$ with both $\parallel$ indices, and $g_1+g_2$ with one $\perp$ and one $\parallel$ index. We will see the benefit of using $F_1$ instead of $F_2$ in Section~\ref{sec:coeff}.

The spin-averaged DIS cross section is
\begin{align}
    \frac{d^2\sigma}{\rd x\, \rd y}     &=  \frac{4 \pi \alpha^2} {x y Q^2}\left[xy^2F_1 + \left(1-y-\frac{x^2 y^2 M_p^2}{Q^2} \right) F_2 \right]\,.
    \label{2.3}
\end{align}
The cross section asymmetry for a target polarized parallel to the beam direction is
\begin{align}
    \frac{d^2 \Delta \sigma}{\rd x\, \rd y}
    &=  \frac{8 \pi \alpha^2} {x y Q^2} \left[ x y (y-2)  g_1 +  \frac{ 2 M_p^2 x^3 y^2}{Q^2} (y g_1 + 2 g_2)  \right]\,,
    \label{2.4}
\end{align}
and for a target polarized transverse to the beam direction is
\begin{align}
    \frac{d^2\Delta \sigma}{\rd x\, \rd y\, \rd \phi}
    &=  \frac{8 \pi \alpha^2} {x y Q^2} \left[ -\frac{\sqrt{2} M_p x^2 y \sqrt{1-y}  }{Q} (y g_1 + 2 g_2)  \right] \cos \phi \,,
    \label{2.5}
\end{align}
where $\phi$ is the azimuthal angle between the scattered electron and the transverse polarization direction. Polarized cross section measurements for the proton require both a polarized target and polarized beam, and are done by flipping either the lepton or target polarization. $g_1$ can be measured using a longitudinally polarized target, and the cross section asymmetry is leading order in $M_p^2/Q^2$. The contribution of $g_2$ to the cross section for longitudinal polarization is suppressed by $M_p^2/Q^2$. The transverse asymmetry is order $M_p/Q$, and both $g_1$ and $g_2$ contribute at this order. In QCD, structure functions have a twist expansion in powers of $M_p^2/Q^2$. $F_1$, $F_2$, and $g_1$ get contributions at leading twist, with higher twist corrections suppressed by powers of $M_p^2/Q^2$. The QCD contribution to $g_2$ starts at twist-three, so we need subleading operators of order $\lambda$ in the SCET analysis. $g_2$ is a dimensionless function of $x$ and is order unity. Its contribution to the cross section is order $M_p/Q$.

\subsection{OPE} \label{sec:ope}

The QCD analysis of DIS uses an OPE at the scale $Q$ to match the time-ordered product of two electromagnetic currents
\begin{align}
    t^{\mu\nu}  &= i \int d^4z \, e^{iq \cdot z} \, T \left( j^\mu(z) j^\nu(0) \right) \,,
    \label{2.7}
\end{align}
onto local operators. The hadronic matrix element of the time-ordered product is
\begin{align}
T_{\mu\nu}(P,S,q) &=i \int d^4z \, e^{iq \cdot z} \braket{ P,S| T \left( j^\mu(z) j^\nu(0) \right) |P,S } \nn
    &= -\widetilde F_1 \left(g_{\mu\nu} - \frac{q_\mu q_\nu}{q^2} \right) + \frac{\widetilde F_2}{P \cdot q}\left(P_\mu - \frac{(P \cdot q)\, q_\mu}{q^2} \right)\left(P_\nu - \frac{(P \cdot q)\, q_\nu}{q^2} \right) \nn
    & + \frac{i \widetilde g_1}{P \cdot q}\epsilon_{\mu\nu\lambda\sigma} q^\lambda S^\sigma  + \frac{i \widetilde g_2}{(P \cdot q)^2} \epsilon_{\mu\nu\lambda\sigma} q^\lambda (P \cdot q \, S^\sigma - S \cdot q \, P^\sigma) \, .
    \label{2.60}
\end{align}
with a tensor decomposition analogous to $W_{\mu \nu}$. The structure functions $\widetilde{\mathcal{F}}$ with $\mathcal{F}=F_1,\ F_2,\ g_1,\ g_2$ have branch cuts in the physical region $-1 \le x \le 1$ with discontinuity $\text{Disc} \, \widetilde{\mathcal{F}} = 4 \pi i\, \mathcal{F}$, which relates the structure functions in $T_{\mu \nu}$ to those in $W_{\mu \nu}$.

The OPE gives an expansion of $t_{\mu \nu}$ in powers of $\omega=1/x$ around $\omega=0$. The coefficients of powers of $\omega$ can be related to moments of the structure functions, which are discontinuities across the cut, using Cauchy's theorem.
The OPE for the polarized structure function $g_1$ at leading twist (twist-two) gives the moment sum rule\footnote{The OPE quark coefficients for the moments of $F_1$, $F_2$, $F_L$ and $g_1$ will be denoted by $C_{1q,N}$, $C_{2q,N}$, $C_{Lq,N}=C_{2q,N}-C_{1q,N}$ and $C_{\Delta q,N}$, and similarly for the gluon coefficients.}
\begin{align}
2 M_{N} [g_1] &= \sum_{i=q,g} C_{\Delta  i,N}(Q,\mu) A_{i,N}(\mu), \qquad (N\ \text{odd})
\label{2.8}
\end{align}
where the sum on $i$ is over quark and gluon contributions.  The quark coefficient $ C_{\Delta  q,N}=1$ at tree-level and the gluon coefficient $ C_{\Delta  g,N}$ starts at $O(\alpha_s)$. The axial matrix elements $A_{q,N}(\mu)$ are defined via the proton matrix element of twist-two quark operators renormalized at scale $\mu$,
\begin{align}
O_{A,q,N} &= \frac12 \overline \psi \slashed{n} \gamma_5 \left( i n \cdot D \right)^{N-1}  \psi\,, \nn
\braket{P,S | O_{A,q,N} | P,S} &= A_{q,N}(\mu)\, (n \cdot S)\, (n \cdot P)^{N-1} \,.
\label{2.10}
\end{align}
The operator in Eq.~\eqref{2.10} is normalized so that the tree-level matrix element in a free-quark state has $A_{q,N}=1$.

The renormalization group (RG) improved expressions for $g_1$ are given by evaluating $ C_{\Delta i,N}(Q,\mu)$ at $\mu \sim Q$, where $\log Q^2/\mu^2$ is not large, and using the RG to evolve it to a fixed scale $\mu_0$ of order a hadronic scale, where the matrix elements are evaluated.

The tower of twist-two operators can be inverted in moment space to obtain bilocal PDF operators~\cite{Manohar:1990jx,Manohar:1990kr}
\begin{align}
O_{\Delta q}(r^+) &= \frac{1}{4\pi} \int \rd \xi \ e^{-i \xi r^+} \ \overline \psi(n \xi) \slashed{n} \gamma_5 \psi(0)\,, \nn
O_{ \Delta \overline q}(r^+) &= \frac{1}{4\pi} \int \rd \xi \ e^{-i \xi r^+} \ \overline \psi(0) \slashed{n} \gamma_5 \psi(n \xi) \,,
\label{2.14}
\end{align}
whose proton matrix elements define the polarized quark and antiquark PDFs $f_{\Delta q}(x)$ and  $f_{\Delta \overline q}(x)$,
\begin{align}
\braket{P,S | O_{\Delta q}(x P^+) | P,S} &= f_{\Delta q}(x)\,, \nn
\braket{P,S | O_{\Delta \overline q}(x P^+) | P,S} &= f_{\Delta \overline q}(x) \,,
\label{2.15}
\end{align}
which satisfy $f_{\Delta q}(-x) = f_{\Delta \overline q}(x)$. The general form of PDF operators is discussed in Appendix~\ref{app:PDF}.

The axial matrix elements are moments of the polarized PDF
\begin{align}
    M_N[ f_{\Delta q}(x) + f_{\Delta\overline{q}}(x)] &= A_{q,N}(\mu), \qquad (N \text{ odd})
    \label{2.16}
\end{align}
The OPE Eq.~\eqref{2.7} involves the product of two electromagnetic currents, so the PDF combinations which enter the DIS cross section are charge-conjugation even, and are the sum of quark and antiquark contributions.

The tree-level OPE for the antisymmetric part of $t^{\mu \nu}$ is~(see~\cite{Manohar:1992tz} for a review)
\begin{align}
t^{[\mu\nu]}=\sum_{n=1,3,5}^\infty 2\  \frac{2^{n} q_{\mu_2}
\ldots q_{\mu_n} }{ (-q^2)^{n}}
\ i\epsilon^{\mu\nu \alpha\mu_{1}}\,q_\alpha \  \frac12 \overline \psi \gamma^{\mu_1} \gamma_5 \left( i D^{\mu_2} \right)  \cdots \left( i D^{\mu_n} \right)  \psi \,.
\label{2.17}
\end{align}
The hadronic matrix element of the quark operator in Eq.~\eqref{2.17} only has $+$ components, so the operator that contributes to $t^{[\mu \nu]}$ is
\begin{align}
O_q^\sigma &=  \frac12 \overline \psi \gamma^{\sigma} \gamma_5 \left( i n \cdot D \right) ^{N-1}  \psi = O_{2,q}^\sigma + \frac{N-1}{N} O_{3,q}^\sigma \, ,
\label{2.70}
\end{align}
which has twist-two and twist-three pieces. The twist-two piece is
\begin{align}
O_{2,q}^\sigma &=  \frac1{2N} \overline \psi \gamma^{\sigma} \gamma_5 \left( i n \cdot D \right) ^{N-1}  \psi 
+ \sum_{r+s=N-2}  \frac1{2N} \overline \psi  \slashed{n} \gamma_5 \left( i n \cdot D \right)^r (iD^\sigma) (i n \cdot D)^s \psi 
\label{2.71}
\end{align}
and is completely symmetric in the Lorentz index $\sigma$ and the $N-1$ indices contracted with $n^\mu$. Its hadronic matrix element is
\begin{align}
\braket{P,S | O_{2,q}^\sigma  | P, S} & = A_{q,N} \left[ \frac{1}{N} S^\sigma (n \cdot P)^{N-1} + \frac{N-1}{N} (n \cdot S) P^\sigma (n \cdot P)^{N-2} \right] \,,
\label{2.73}
\end{align}
and is completely symmetric in $S$ and $P$.
The matrix element $A_{q,N}$ is the same as for the twist-two operator for $g_1$ defined in Eq.~\eqref{2.10}, since $n_\sigma O_{2,q}^\sigma =  O_{A,q,N}$.
The twist-three piece is
\begin{align}
O_{3,q,N}^\sigma 
&=\frac12 \overline \psi \gamma^{\sigma } \gamma_5 \left( i n \cdot D \right)^{N-1}  \psi 
- \frac{1}{N-1} \sum_{r+s=N-2} \frac12 \overline \psi \slashed{n} \gamma_5 \left( i n \cdot D \right)^r  (i D^\sigma)  \left( i n \cdot D \right)^s \psi \,.
\label{2.39}
\end{align}

The contribution of the twist-two operator $O_{A,q,N}$ to $T_{\mu \nu}$ (including the matching coefficient  $ C_{\Delta q,N}$ which is unity at tree-level) is
\begin{align}
T_{\mu \nu} &= \braket{P,S | t_{\mu \nu} | P,S} \nn
& =
\sum_{n=1,3,5}^\infty  2\ C_{\Delta q,N} A_{q,N}  \frac{2^{n}}{(-q^2)^{n}} i\epsilon^{\mu\nu \alpha \sigma}\,q_\alpha \left[ \frac{1}{N}S^{\sigma} (q \cdot P)^{N-1} +  \frac{N-1}{N} P^\sigma (q \cdot S) (q \cdot P)^{N-2} \right] \nn
&= 
\sum_{n=1,3,5}^\infty  2\  C_{\Delta q,N} A_{q,N}\ \omega^n i\epsilon^{\mu\nu \alpha \sigma}\,q_\alpha \left[ \frac{S^\sigma}{(q \cdot P)} - \frac{N-1}{N}  \frac{  \big(S^\sigma (q \cdot P) - P^\sigma (q \cdot S) \big) }{(q \cdot P)^{2}} \right] \,.
\label{2.40}
\end{align}
Comparing with Eq.~\eqref{2.2}, the first term leads to the quark term for $g_1$ discussed earlier in Eq.~\eqref{2.8}, and the second term is the Wilczek-Wandzura contribution to $g_2$
\begin{align}
2 M_{N} [g_2^\text{WW} ] &= - \frac{N-1}{N} C_{  \Delta i,N}(Q,\mu) A_{i,N}(\mu) = - \frac{N-1}{N}\  2 M_{N} [g_1]\,. \qquad (N\ \text{odd})
\label{2.37}
\end{align}
In $x$-space,
\begin{align}
g^{\text{WW}}_2(x) &= -g_1(x) + \int_x^1 \frac{\rd y}{y} g_1(y) \,.
\label{2.25}
\end{align}
$g^{\text{WW}}_2(x)$ is purely kinematic in origin. It arises because the spin-dependent twist-two operators have matrix elements proportional to $(n \cdot S)(n \cdot P)^{N-1}$, which must be rewritten in terms of the $g_1$ and $g_2$ tensors in the decomposition of $W^{\mu\nu}$. Thus Eq.~\eqref{2.25} holds with the complete expression for $g_1$, including the gluon contributions and higher order QCD corrections.

The total expression for $g_2$ is
\begin{align}
g_2(x) &= g^{\text{WW}}_2(x) + \overline g_2 (x) \,,
\label{2.26}
\end{align}
where $\overline g_2$ is the contribution from the twist-three operators $O_{3,q}$ in Eq.~\eqref{2.70}. Both $g^{\text{WW}}_2(x) $ and $ \overline g_2 (x)$ are dimensionless functions of the same order in power counting, so $ \overline g_2 (x)$  is not formally suppressed relative to $g^{\text{WW}}_2(x) $.

The proton matrix element of the twist-three operators $O_{3,q,N}^\sigma$ defines the matrix elements $B_{q,N}$,
\begin{align}
\braket{P,S| O_{3,q,N}^\sigma |P,S} &= B_{q,N}(\mu) \left[  S^\sigma (n \cdot P)^{N-1} - P^\sigma (n \cdot S) (n \cdot P)^{N-2}  \right],
\label{2.21}
\end{align}
and contributes to $\overline g_2$ but not $g_1$. Its contributions to the moments of $\overline g_2$ are given by
\begin{align}
2M_{N} [\overline g_2] &=  B_{q,N} (\mu) \,, \qquad N \text{ odd} \,,
\label{2.23}
\end{align}
since the matching coefficient is unity at tree-level.
Note that there are no twist-three operators for $N=1$, and the first moment of the Wilczek-Wandzura contribution Eq.~\eqref{2.8} vanishes,
which leads to the Burkhardt-Cottingham sum rule~\cite{Burkhardt:1970ti} $M_1[g_2]=0$, as long as the first moment integral is convergent. The matrix elements $B_{q,N}$ and $A_{q,N}$ are of order unity, so the twist-three contribution $\overline g_2$ is the same order as the twist-two contribution $g^{\text{WW}}_2$ and cannot be neglected.

It was pointed out by Shuryak and Vainshtein~\cite{Shuryak:1981pi} that the above twist-three analysis~\cite{Kodaira:1978sh,Kodaira:1979ib,Kodaira:1979pa} is incorrect, because the twist-three quark operator $O_{3,q,N}$ can be eliminated using the equations of motion in favor of three-parton quark-gluon operators Eq.~\eqref{2.30} which start at order $g$. The moments of $\overline{g}_2$ cannot be written as in Eq.~\eqref{2.23} with matrix elements $B_{q,N}$ of quark-only twist-three operators. 

The correct analysis requires the operators~\cite{Kodaira:1994ge}
\begin{align}
V_{r,s}^\sigma &= i g  \overline \psi   (i n \cdot D)^{r-1} G^{\sigma \alpha} n_\alpha (i n \cdot D)^{s-1} \slashed{n} \gamma_5   \psi \,, \nn
U_{r,s}^\sigma &= g  \overline \psi   (i n \cdot D)^{r-1}  \widetilde G^{\sigma \alpha} n_\alpha (i n \cdot D)^{s-1}    \slashed{n}  \psi \,,
\label{2.30}
\end{align}
where for the $N^\text{th}$ moment,  $N=r+s+1$ and $1 \le r,s \le N-2$, $\widetilde G_{\alpha \beta} = \frac12 \epsilon_{\alpha \beta \lambda \sigma}G^{\lambda \sigma}$.\footnote{
We have contracted the symmetrized $\mu$-indices in the definition~\cite{Kodaira:1994ge} with $n$, and used $g \to -g$ because our sign convention for the covariant derivative is the opposite of~\cite{Kodaira:1994ge}.}
Some properties of the quark-gluon operators, and the equation-of-motion identity are given in Appendix~\ref{app:ops} and Appendix~\ref{app:eom}. The equation-of-motion gives the linear combination Eq.~\eqref{a2.34} for tree-level matching. However, at one-loop, the coefficients of the individual $V_{r,s}^\sigma $ and $U_{r,s}^\sigma $ can be different, so the matching is not given by a single overall coefficient $B_{q,N}$. We compute the one-loop matching in Section~\ref{sec:matchingJ}.

The charge-conjugation even combinations of $V_{r,s}^\sigma$ and $U_{r,s}^\sigma$ that enter DIS are
\begin{align}
& V_{r,s}^\sigma - V_{s,r}^\sigma\,, &  & \text{and} & U_{r,s}^\sigma + U_{s,r}^\sigma\,.
\label{2.33}
\end{align}
There are $N-2$ charge-conjugation even twist-three operators, and the number of operators grows with the moment. The anomalous dimension is a $(N-2) \times (N-2)$ matrix. The anomalous dimension of the non-singlet twist-three operators was computed in moment space in Refs.~\cite{Bukhvostov:1983eob,Ratcliffe:1985mp,Ji:1990br}, and in $x$ space in Refs.~\cite{Ratcliffe:1985mp,Balitsky:1987bk,Ali:1991em,Geyer:1996isa}.

Similar to the twist-two case, one can combine the twist-three operators into trilocal PDF operators~\cite{Balitsky:1987bk,Geyer:1996isa}
\begin{align}
F_\pm(r_1^+,r_2^+) &=  \int \frac{{\rd} \xi_1}{2\pi} \int \frac{{\rd} \xi_2}{2\pi}\, e^{-i \xi_2 r_2^+}\, e^{i \xi_1 r_1^+} \overline \psi (n \xi_2) \left[ \pm i g G_{n \sigma} (0) \slashed{n}\gamma_5+ g \widetilde G_{n \sigma}(0) 
\slashed{n} \right] \psi(n \xi_1)
\label{2.35}
\end{align}
whose double-moments give $U^\sigma_{r,s}\pm V^\sigma_{r,s}$, and whose proton matrix elements give a PDF
\begin{align}
\braket{P,S | F(x_1 P^+ ,x_2 P^+) | P,S} &= f(x_1,x_2) \,,
\label{2.36}
\end{align}
which is a function of two variables.

The analysis of $g_2$ is much more complicated than the other structure functions because of the trilocal PDF operator. The anomalous dimension kernel of this operator also depends on two variables~\cite{Balitsky:1987bk,Ali:1991em,Geyer:1996isa}, whereas the structure function $g_2(x)$ itself depends only on a single variable. Equivalently, the $N^{\text{th}}$ moment of $g_2$ is a single number, whereas it gets contributions from $N-2$ operators. Thus, a measurement of $g_2(x,Q_1^2)$ at a single scale $Q_1$ is not sufficient to determine $g_2(x,Q_2^2)$ at a different scale $Q_2$.

In the $x \to 1$ limit, the QCD evolution of the twist-three PDF distribution simplifies to an evolution with single variable kernels~\cite{Ali:1991em,Geyer:1996isa}. We will see that SCET explains why this occurs, and also greatly simplifies the QCD analysis.

\section{ \texorpdfstring{Unpolarized DIS as $ x \to 1$: $F_1$ and $F_L$}{Unpolarized DIS as x goes to 1: F1 and F2}}\label{sec:unpolarized}

The SCET analysis of unpolarized DIS as $x \to 1$ at leading order in $\lambda$ is given in Ref.~\cite{Manohar:2003vb}. We briefly summarize the results here. 
The SCET operators are constructed out of gauge-invariant objects\footnote{Our sign convention for the covariant derivative is the opposite of Ref.~\cite{Luke:2022ops}.}
\begin{align}
    \label{building_blocks}
    \chi_{n} (x) &= W^\dagger_{n}(x) \psi_{n} (x) \,,  \nn
   \mathcal{B}_n^{\mu_1 \cdots \mu_N}(x) &=  {W}_n^\dagger(x)\, i D_n^{\mu_1}\cdots i D_n^{\mu_N} {W}_n(x) \,,
\end{align}
where $\psi_n$ satisfies
\begin{align}
    \label{3.1a}
P_n \psi_n &= \psi_n \,,
\end{align}
and
\begin{align}
    \label{3.1b}
P_n &= \frac{ \slashed{n} \slashed{\nb}}{4} \,, &P_\nb &= \frac{ \slashed{\nb} \slashed{n}}{4} \,, &
P_n + P_\nb &= 1 \,,
\end{align}
are projection operators.
The collinear Wilson lines, needed to preserve gauge invariance, are defined as
\begin{align} 
\label{Wline_defns}
{W}_n(x) &= \mathcal{P} \exp \left( -ig \int_{-\infty}^{0} \!\! \rd s \ \bar n\cdot A^a_{n}(x+\nb s) T^a  \right) \,.
\end{align}
Note that the subscript on a Wilson line corresponds to the sector with which it interacts rather than its direction along the lightcone. $W_n$ contains $n$-collinear fields,  but is in the $\nb$ direction. $\chi_n$ is the gauge invariant quark field satisfying $\slashed{n} \chi_n=0$, and $\mathcal{B}_n^{\mu_1 \cdots \mu_N}(x)$ is a gauge invariant version of the gauge field and its derivatives. We will need $\mathcal{B}_n^{\mu}(x) \sim -g A^\mu_n(x) + \ldots$ with a single Lorentz index. The definitions of $\nb$-collinear fields are given by $n \leftrightarrow \nb$.

With our choice of kinematics, the incoming partons are in the $\nb$ direction and the outgoing partons are in the $n$ direction.
At the scale $Q$, the electromagnetic current $j_\text{QCD}^\nu=\overline\psi \gamma^\nu \psi $ in QCD is matched onto the SCET current at leading power, %
\begin{align}
  j_\text{QCD}^\nu &=  C_J(\mu)\ j^\nu_{\text{SCET}}(x) + O(\lambda)\,, & j^\nu_{\text{SCET}}(x) & = \overline \chi_n(x) \gamma^\nu_\perp \chi_{\nb}(x) \,.
\label{3.18}
\end{align}
The Lorentz index for the leading order SCET current is a $\perp$ index. In SCET, 
\begin{align}
    W^{\mu\nu}(P,S,q) &= \frac{1}{4\pi} \int d^4z \, e^{iq \cdot z} 
    \braket{P,S| j^{\dagger \mu}_{\text{SCET}}(z)  j^\nu_{\text{SCET}}(0) |P,S}
     \label{3.41}
\end{align}
gives the quark contribution to $W^{\mu \nu}$, and the antiquark contribution is given by adding the charge-conjugate contribution. The second term in the commutator in Eq.~\eqref{2.2} does not contribute. The antiquark contribution must be added explicitly, since there is no crossing symmetry in SCET. The current Eq.~\eqref{3.18} describes an incoming quark in the $\nb$ direction and an outgoing quark in the $n$ direction. The charge-conjugate operator has an incoming antiquark in the $\nb$ direction and an outgoing antiquark in the $n$ direction, and lives in a different SCET sector.

The one-loop matching coefficient is~\cite{Manohar:2003vb}
\begin{align}
C_J(\mu)&= 1 + \frac{\alpha_s}{4 \pi} C_F \left[ -\lQ^2 + 3\lQ  + \frac{\pi^2}{6} - 8 \right] \,.
\label{3.3}
\end{align}
The operator SCET current is evolved to $M_J$ using the SCET anomalous dimension~\cite{Manohar:2003vb}
\begin{align}
\mu \frac{\rd}{\rd \mu} j^\nu_{\text{SCET}} &= \gamma_J \, j^\nu_{\text{SCET}}\,,  &  \gamma_J =\frac{\alpha_s}{4\pi} C_F \left(-4 \lQ + 6 \right)\,.
\label{3.2}
\end{align}
The anomalous dimension contains a logarithm $\lQ$, and the RGE sums the Sudakov double logarithms.

\begin{figure}
\centering
\includegraphics[width=10cm]{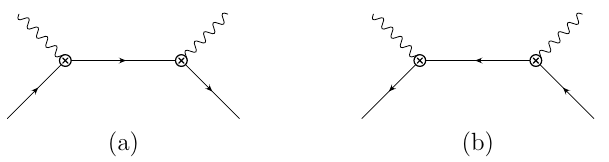}
\caption{\label{fig:ope} Time ordered product of two currents at tree-level matching onto the (a) quark PDF and (b) antiquark PDF operators.}
\end{figure}

At the scale $M_J$, the time-ordered product of two SCET currents is matched using an OPE (see Fig.~\ref{fig:ope}) onto the quark PDF operator
\begin{align}
O_q(r^+) &= \frac{1}{4\pi} \int \rd \xi \,  e^{- i \xi r^+} \overline \chi_{\nb} (n \xi) \, \slashed{n} \, \chi_{\nb}(0) \,,
\label{3.20}
\end{align}
and the antiquark PDF operator $O_{\overline q}(r^+)= - O_q(-r^+)$ given by charge conjugation. The antiquark contributions in the OPE are related to the quark contributions by reversing the direction of the fermion arrow in the diagram, as shown in Fig.~\ref{fig:ope}.
 In Eq.~\eqref{3.20} the Wilson lines in $\chi_{\nb}(0)$ and $\overline \chi_{\nb}(n\xi)$ can be combined into a single Wilson line from $0$ to $n\xi$,
\begin{align}
O_q(r^+) &= \frac{1}{4\pi} \int \rd \xi \,  e^{- i \xi r^+} \overline \psi_{\overline n} (n \xi) W(n\xi,0) \ \slashed{n} \ \psi_{\overline n}(0) \,.
\label{3.4}
\end{align}
 The matrix elements of the PDF operators give the quark and antiquark PDFs,
\begin{align}
\braket{P | O_q(x P^+) | P} &=  f_q(x,\mu) &
\braket{P | O_{\overline q} (x P^+) | P} &=  f_{\overline q}(x,\mu) \,,
\label{3.13}
\end{align}
with $f_q(-x) = - f_{\overline q}(x)$.\footnote{The physical region is $0 \le x \le 1$. The moment sum rules integrate over all $x$, and the negative $x$ part of the $x$ integral is the antiquark contribution, e.g.\ in Eq.~\eqref{2.16}. The $g_1$ moment sum rules for odd $N$ give the matrix elements of the sum of quark and antiquark distributions, which is charge-conjugation even. There are no $g_1$ moment sum rules for even $N$, as those would have generated charge-conjugation odd matrix elements. See~\cite{Jaffe:1983hp} for a detailed discussion on the allowed $x$ range for PDFs.}
\begin{figure}
    \centering
    \includegraphics[width=\linewidth]{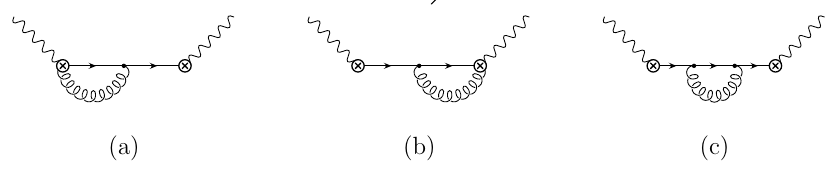}
    \caption{SCET loop graphs that generate the 1-loop matching onto the quark PDF. The gluon from the current vertex arises from the $W_n$ Wilson line. The graphs that match onto the antiquark PDF are the same, but with arrows reversed. The quark and gluon are in the $n$-sector (jet sector), which is integrated out.  External particles (quarks and/or gluons) couple to the vertices and are in the $\nb$-sector, and not shown. The overlap/zero-bin prescription~\cite{Manohar:2006nz, Goerke:2017ioi} ensures that only the $n$-sector contributes to the matching.}
    \label{fig:g1_soft_matching}
\end{figure}
The factorization formula for the matching at the jet scale is
\begin{align} 
W^{\mu\nu} &= \frac{1}{4\pi} \int \rd x \ e^{i q \cdot x} \ j_\text{SCET}^{\dagger \mu(x)}\ j_\text{SCET}^\nu(0)  \nn
&=  \frac12 (-g_\perp^{\mu \nu}) \int \frac{\rd r^+}{r^+} \mathcal{M}\left({-q^+}/{r^+} \right) \left[O_q(r^+) + O_{\overline q}(r^+) \right] \,,
\label{3.5}
\end{align}
and the matching coefficient to one-loop on integrating out the $n$-collinear jet, generated by the graphs in Fig.~\ref{fig:g1_soft_matching}, is~\cite{Manohar:2003vb} 
\begin{align}
\mathcal{M}(z) = \delta(1-z) +  \frac{\alpha_s}{4 \pi}C_F \Bigg \{ 4  {\left[ \frac{\ln(1-z)}{(1-z)} \right]_+ }  &+ \left[ 4  \lQ
 -3 \right] { \frac{1}{(1-z)_+} }  \nn
 &+ 
 \left[  2 \lQ^2
-3 \lQ  -\pi^2  + 7  \right] \delta(1-z) \Bigg\} \,.
\label{3.6}
\end{align}
Comparing with Eq.~\eqref{2.20}, we see that Eq.~\eqref{3.5} implies $F_L$ vanishes at leading power so that $C_{Lq,N} =\mathcal{O}(1/N)$,
and Eq.~\eqref{3.6} is the one-loop matching contribution to the quark coefficient $C_{1q}$ for $F_1$.

Below $M_J$, there is only one SCET sector, so the theory is the same as QCD, since there is nothing to expand in. The PDF operator Eq.~\eqref{3.4} is the same as the QCD PDF operator, and the Wilson line $W(n\xi,0)$ is the QCD Wilson line along the $n$ direction from $0$ to $n\xi$. The SCET calculation is identical to the QCD calculation and gives the entire kernel, including the complete $x$-dependence. The anomalous dimension below $M_J$ of the PDF operator is the usual DGLAP anomalous dimension,
\begin{align}
\mu \frac{\rd}{\rd \mu} O_q(r^+) &=  \int \frac{\rd s^+}{s^+} P_{qq}\left(  {r^+}/{s^+}\right) O_q(s^+) \,, \nn
P_{qq}(z) &= \frac{\alpha_s}{4\pi} \,  4C_F \left[  \frac{2}{(1-z)}_{\!+} -z-1  + \frac32\, \delta(1-z)  \right] \,.
\label{3.8}
\end{align}

The analysis in Ref.~\cite{Manohar:2003vb} was done in both $x$-space and moment space. The scale separation between $Q$ and $M_J$ is cleaner in moment space, where the $N{}^{\rm th}$ moment picks out the scale $Q^2/\overline{N}$, $\overline{N} = N e^{\gamma_E}$. This can be seen from the expression for the one-loop matching $\mathcal{M}$ in moment space
\begin{align}
M_N[ \mathcal{M}] &=1 +  \frac{\alpha_s}{4 \pi }C_F \left\{ 4 \sum_{j=1}^{N-1} \frac {H_j }{ j} + \left(3 - 4  \lQ \right) H_{N-1}  +  2 \lQ^2  - 3  \lQ  - \pi^2 + 7 \right\} \,, \nn
&\to 1+  \frac {\alpha_s }{4 \pi }C_F \left[2  \ln^2 \frac{Q^2  }{ \overline{N} \mu^2}  - 3   \ln \frac {Q^2  }{ \overline{N} \mu^2}  - \frac {2 \pi^2 }{ 3} + 7 \right] \,,\qquad \text{as}\ N \to \infty \,.
\label{3.10}
\end{align}
Neglecting $1-x$ terms is more precise in moment space --- the expansion neglects terms which vanish as $N \to \infty$, since $1-x \sim 1/N$. The moments of $\delta(1-x)$ are $M_N[\delta(1-x)]=1$, whereas $M_N[1]=1/N$, so constant terms in $x$ space are subleading as $x \to 1$. The terms which survive as $N \to \infty$ are singular as $x \to 1$. The matching condition has no large logarithms if $\mu \sim Q/\sqrt{\overline{N}}$, which is the jet scale in moment space.

The moments of the DGLAP kernel Eq.~\eqref{3.8} are
\begin{align}
M_N \left[ P_{qq} \right] &=\frac {\alpha_s(\mu)  }{ 4 \pi}\, 2 C_F \left[ -4H_N + \frac {2 }{ N(N+1)} + 3 \right]  \nn
& \to \frac{\alpha_s(\mu)  }{ 4 \pi}\, 2 C_F \left[ -4 \ln \overline{N} + 3  \right]   \qquad \text{as}\ N \to \infty \,.
\label{3.11}
\end{align}

\section{ \texorpdfstring{Polarized DIS as $ x \to 1$: $g_1$}{Polarized DIS as x goes to 1: g1} }\label{sec:g1}

The analysis for the polarized structure function $g_1$ is almost identical to the unpolarized case. The matching from QCD to SCET at $Q$ and the SCET RGE from $Q$ to $M_J$ are unchanged. The new feature is that we must include spin-dependent terms in the OPE Eq.~\eqref{3.5} at $M_J$. These terms contribute to the antisymmetric part of $W_{\mu \nu}$. The OPE produces the structure
\begin{align}
-g_\perp^{\mu \nu} O_q(r^+) - i \epsilon_\perp^{\mu \nu} O_{\Delta q}(r^+) \,,
\label{3.40}
\end{align}
using the identity 
\begin{align}
    \gamma^\mu \gamma^\alpha \gamma^\nu = g^{\mu\alpha}\gamma^\nu + g^{\nu\alpha}\gamma^\mu - g^{\mu\nu}\gamma^\alpha -i\epsilon^{\mu\alpha\nu\lambda}\gamma_\lambda\gamma_5 \,,
\label{3.21}
\end{align}
where
\begin{align}
O_{\Delta q}(r^+) &= \frac{1}{4\pi} \int \rd \xi e^{- i \xi r^+} \overline \psi (n \xi) W(n\xi,0) \ \slashed{n} \, \gamma_5 \ \psi(0) \,,
\label{3.15}
\end{align}
is the polarized PDF operator~\cite{Manohar:1990kr,Manohar:1990jx} whose matrix elements give the polarized PDF $f_{\Delta q}(x,\mu)$
\begin{align}
\braket{P | O_{\Delta q}(x P^+) | P} &=  f_{\Delta q}(x,\mu) \,.
\label{3.25}
\end{align}
The polarized antiquark operator $O_{\Delta \overline q}(r^+) = O_{\Delta q}(-r^+) $ is given by charge conjugation, which also gives $f_{\Delta q}(-x) = f_{\Delta\overline q}(x)$.

The structure function contributions can be read off from Eq.~\eqref{2.20}.
The unpolarized contribution $-g_\perp^{\mu \nu}$ in Eq.~\eqref{3.40} gives the structure function $F_1$. The matching coefficient for $g_1$ is the same as Eq.~\eqref{3.6} for $F_1$, since they occur in the combination Eq.~\eqref{3.40}.

The operator $O_{\Delta q}$ involves $\gamma_5$, so there are subtleties related to the definition of $\gamma_5$. These are discussed in Appendix~\ref{app:BMHV} as they don't play a role as $x \to 1$. The one-loop anomalous dimension of $O_{\Delta q}$ is the same as for the unpolarized PDF $O_{q}$, Eq.~\eqref{3.8}. The entire analysis of $g_1$ at leading power (up to terms which vanish as $N \to \infty$) is identical to $F_1$, for which the results are given in~\cite{Manohar:2003vb}.
Using the solutions to the RG equations 
\begin{align} \label{3.22}
    \Gamma_J(\mu_h,\mu_l) &= \int_{\mu_l}^{\mu_h} \frac{{\rm{d}}\mu}{\mu} \, \gamma_J(\mu) \,, \nn
    \Gamma_{qq,N}(\mu_h,\mu_l) &=  \int_{\mu_l}^{\mu_h} \frac{{\rm{d}}\mu}{\mu} \, M_N[P_{qq}(\mu)] \,,
\end{align}
where $M_N[P_{qq}(\mu)]$ is given in Eq.~\eqref{3.11},
the complete resummed factorization formula for $g_1$ in moment space is
\begin{align} \label{3.23}
   2 M_N(g_1) = \left( C_J(Q) \right)^2 \, e^{2\,\Gamma_J(Q,\mu_J)} \, M_N\left[\mathcal{M}\left(\mu_J\right)\right] \,  e^{\Gamma_{qq,N}(\mu_J,\mu_0)} \, A_{q,N}(\mu_0) \,,
\end{align}
where the jet scale is $\mu_J = Q/\sqrt{\Nb}$ and moments of the PDF are evaluated at a reference momentum $\mu_0$ below the jet scale. This formula includes all  $\alpha_s$ terms, and sums  the double-log series $[\alpha_s \ln^2 (1-x) ]^n$ as well as the leading-log series $[\alpha_s \ln M_J/\lqcd]^n$.

\section{ \texorpdfstring{Polarized DIS  as $ x \to 1$: $g_2$ Operators}{Polarized DIS as x goes to 1: g2 Operators} }\label{sec:g2}

The SCET analysis of $g_2$ is much simpler than the QCD analysis. It naturally gives the twist-3 quark-gluon operators at order $\lambda$, and the twist-three PDF as a bilocal operator.

SCET cleanly extracts the twist-3 part $\overline{g}_2$. From Eq.~\eqref{2.20}, $g_2$ can be obtained from the $W_{\mu \nu}$ tensor structure
\begin{align}
W^{\mu \nu} &=    i \frac{g_1+g_2}{Q}x \left[ - (n + \overline n)^\mu \epsilon_\perp^{\nu \sigma} + (n + \overline n)^\nu \epsilon_\perp^{\mu \sigma} \right]S_\sigma \,,
\label{4.38}
\end{align}
and
\begin{align}
g_1+g_2 &= g_1 + g^{\text{WW}}_2 + \overline g_2 =  \int_x^1 \frac{\rd y}{y} g_1(y) + \overline g_2 \to \overline g_2 \,, \qquad
\text{as} \  x \to 1
\label{4.39}
\end{align}
using Eq.~\eqref{2.25}. The twist three part $\overline g_2$ is given directly from the tensor structure in Eq.~\eqref{4.38}, which has one $\parallel$ ($n$ or $\nb$)  index and one $\perp$ index.

\begin{figure}
    \centering
    \includegraphics[width=0.95\linewidth]{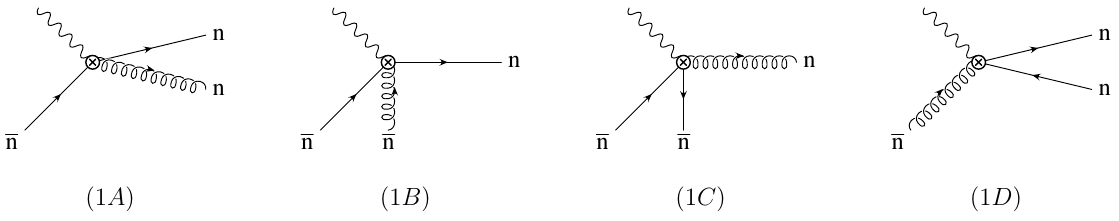}
    \caption{\label{fig:1B} The $O^{(1A)}$, $O^{(1B)}$, $O^{(1C)}$ and $O^{(1D)}$ operators in the expansion of the current at order $\lambda$.}
\end{figure}

The SCET analysis of $g_2$ requires both the leading order and order $\lambda$ terms in the matching of $j^\nu_{\text{QCD}}$. The order $\lambda$ contribution to $W^{\mu \nu}$ is given by the interference term between the leading order and order $\lambda$ contributions. The expansion of the QCD current to order $\lambda$ was computed in~\cite{Freedman:2011kj,Goerke:2017ioi,Inglis-Whalen:2021bea,Luke:2022ops} and matches onto $q \overline q g$ SCET operators. SCET directly gives operators containing gluons without using the equations of motion. There are four kinematic sectors which are shown in Fig.~\ref{fig:1B}: \\
\textbf{topology A:} incoming $\nb$-collinear quark and outgoing $n$-collinear quark and gluon,\\
\textbf{topology B:} incoming $\nb$-collinear quark and gluon and an outgoing $n$-collinear quark,\\
\textbf{topology C:}  incoming $\nb$-collinear quark and antiquark and outgoing $\nb$-collinear gluon,\\
\textbf{topology D:}  incoming $\nb$-collinear gluon and outgoing $n$-collinear quark and antiquark. There are also the hermitian conjugate amplitudes, which exchange incoming and outgoing particles.

\begin{figure}
\begin{center}
\includegraphics[width=8cm]{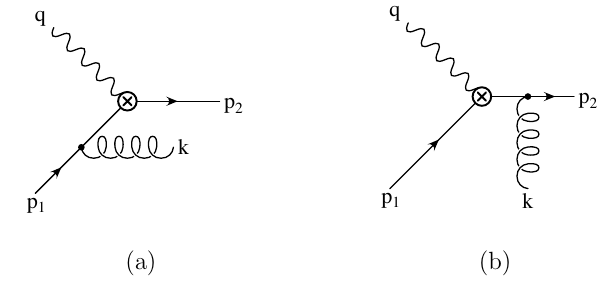}
\end{center}
\caption{\label{fig:subleading} Tree graphs for the order $\lambda$ matching of the QCD current. The quarks and gluons can be in the $n$ or $\nb$ sector, and incoming or outgoing, depending on topology $A$ -- $D$. The expansion has to be done separately for each topology.}
\end{figure}

Using the notation employed in \cite{Luke:2022ops}, the QCD current is given by
\begin{align} \label{4.3}
    j^\nu_{\text{QCD}}(x) &= C_J(\mu)\ j^\nu_{\text{SCET}}(x) + \frac{1}{Q} \sum_{i\in\{A,B,C,D\}}\int_0^1 {\rm{d}} u \, C^{(1i)}(u,\mu)O^{(1i)}(x,u,\mu) + \text{h.c.} +  O\left( \lambda^2  \right),
\end{align} 
where the superscript factor of $1$ denotes order $\lambda^1$ in the power counting. The tree-level matching is generated by the QCD graphs shown in Fig.~\ref{fig:subleading}. 

%
%

%
%
The subleading operators are\footnote{We have simplified the $\gamma$-matrix structure in~\cite{Luke:2022ops}.}
\begin{align} \label{4.14}
    O^{(1A)\mu}(x,u) &= - (n^\mu + \overline{n}^\mu) \, \left[  \mathcal{B}_n^{\alpha}\left(x\right) \right]_{u Q} \, \left[\overline{\chi}_n(x) \right]_{(1-u)Q} \, 
    \gamma^\perp_\alpha \,  \chi_\nb(x) \,, \nn
    O^{(1B)\mu}(x,u) &= - (n^\mu + \overline{n}^\mu) \, \overline{\chi}_n(x) 
    \gamma^\perp_\alpha \,  \left[ \chi_\nb(x) \right]_{(1-u)Q} \left[  \mathcal{B}_\nb^{\alpha}\left(x\right) \right]_{u Q} \,,  \nn
    O^{(1C)\mu}(x,u) &= \mathcal{B}_n^\alpha (x) \Bigg ( \frac{1}{u} \, \left[\overline{\chi}_\nb(x)\right]_{-uQ} \gamma^\perp_\alpha \frac{\slashed{n}}{2}\gamma_\perp^\mu \left[\chi_\nb(x)\right]_{(1-u)Q} \nn
    &\hspace{3cm} - \frac{1}{u} \left[\overline{\chi}_\nb(x)\right]_{-(1-u)Q} \gamma_\perp^\mu \frac{\slashed{n}}{2} \gamma^\perp_\alpha \left[\chi_\nb(x)\right]_{uQ} \Bigg) \,, \nn
    O^{(1D)\mu}(x,u) &=  \mathcal{B}_\nb^{\alpha}(x) \Bigg( \frac{1}{u}\left[\overline{\chi}_n(x)\right]_{(1-u)Q} \gamma_\perp^\mu  \frac{\slashed{\nb}}{2} \gamma^\perp_\alpha \left[\chi_n(x)\right]_{-uQ} \nn
    &\hspace{3cm} - \frac{1}{u} \left[ \overline{\chi}_n(x) \right]_{uQ} \gamma^\perp_\alpha  \frac{\slashed{\nb}}{2} \gamma_\perp^\mu  \left[\chi_n(x)\right]_{-(1-u)Q} \Bigg) \,,
\end{align}   
and start at order $g$. The minus sign on the label $[\chi_n(x)]_{-uQ}$ means that we pick out the outgoing antifermion with momentum $\overline n \cdot p =+ u Q$. The two terms in $O^{(1C)\mu}(x,u)$ and in $O^{(1D)\mu}(x,u) $ are negative charge-conjugates of each other.
The operators in the $A$ and $B$ topologies are related by taking the hermitian conjugate and switching $n \leftrightarrow \nb$, as are the operators in the $C$ and $D$ topologies. The index $\mu$ for the $1A$ and $1B$ operators is a $\parallel$ index, and for the $1C$ and $1D$ operators is a $\perp$ index.
There is also the operator
 \begin{align} \label{4.17}
        O^{(1\perp)\mu}(x) &= \nb^\mu \big[ i\partial^\alpha \bar{\chi}_n(x) \big] \gamma_\alpha^\perp \chi_\nb(x)  - n^\mu \bar{\chi}_n(x) \gamma_\alpha^\perp \big[i\partial^\alpha \chi_\nb(x) \big]
\end{align}
which vanishes in the Breit-frame as it describes partons carrying non-zero $\perp$-momentum. It is related to the leading operator $j^\mu_{\text{SCET}}$ by reparameterization invariance~\cite{Manohar:2002fd,Goerke:2017ioi} and thus has the same matching coefficient $C_J$ as $j^\mu_{\text{SCET}}$ to all orders.

The tree-level matching coefficients are $C^{(i)}(u) = 1$~\cite{Goerke:2017ioi}. The variable $u$ denotes the fraction of the total sector momentum carried by a particle.  For example, the matrix element of $O^{(1B)\mu}(x,u)$ is
\begin{align}
\langle p_2|O^{(1B)\mu}(x,u)|p_1, \, k \rangle &= \delta\left(u - \frac{k^+}{Q} \right)(n^\mu + \nb^\mu) {gT^a} \varepsilon^a_\alpha \bar{u}_2(p_2) \Delta^{\alpha \beta}(k)\gamma_\beta^\perp P_\nb \, u(p_1) \,,
\end{align}
where $\Delta^{\alpha\beta} = g_\perp^{\alpha\beta} - \nb^\alpha k^\beta/k^-$, $P_\nb = \slashed{\nb}{\slashed{n}}/4$ and $\varepsilon^a_\alpha$ is the gluon polarization. Integrating over $u$ in Eq.~\eqref{4.3} sets $u = k^+/Q$, the momentum fraction carried by the gluon i.e.\ by the gauge invariant operator $\mathcal{B}^\alpha_{\overline n}$, while momentum conservation sets the momentum fraction carried by the quark $1 - u = p_1^+/Q$. The domain of $u$ is set by the on-shellness of the external states.  The SCET expansion assumes that collinear particles have momentum of order $Q$ and is only valid for $u \in (0,1)$, away from the endpoints, so that $u$ and $1-u$ are order unity, and not parametrically small.

The interference of $O^{(1B)\mu}$ with $j^\mu_{\text{SCET}}$ in the time-ordered product at the jet scale $M_J$ generates $g_2$ as we show in Section~\ref{sec:matchingJ}. Operators in the $D$ topology are needed for the gluon contribution to the DIS structure functions. $D$ topology operators are order $\lambda$, so the gluon contribution to the DIS structure functions is order $\lambda^2 \sim 1/N$, as can be verified from the full QCD result. Operators in the $A$ and $C$ topologies generate order $\lambda$ corrections to the $F_1, F_L$ and $g_1$ structure functions. The order $\lambda$ corrections to the $F_1, F_L$ structure functions were computed in \cite{Luke:2022ops}.

\section{\texorpdfstring{One-Loop Matching onto the Subleading SCET Current at $Q$}{One-Loop Matching at Q}} \label{sec:matchingQ}

A complete order $\alpha_s$ analysis of $g_2$  requires the matching coefficient $C^{(1B)}$ to one loop. $C^{(1B)}$ has been calculated to tree-level before~\cite{Goerke:2017ioi}. We derive the one-loop matching from QCD onto all the order $\lambda$ SCET operators in this section.
\begin{figure}
    \centering
    \includegraphics[width=0.95\linewidth]{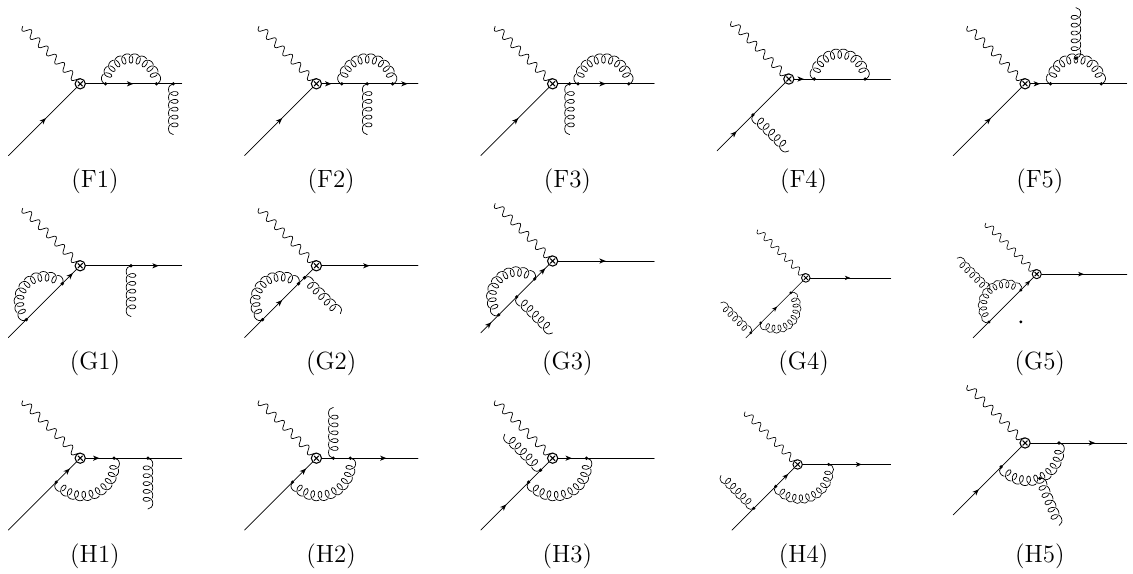}
    \caption{QCD loop graphs needed for one-loop matching onto SCET. The graphs have to be computed separately for each topology. For the $B$ topology, graphs (F3), (F4), (G1)--(G5) are scaleless and vanish.  For the $D$ topology, graphs (F3), (F4), (G1), and (G2) are scaleless and vanish. One-particle reducible graphs contribute when using on-shell matching.}
    \label{fig:3}
\end{figure}
The matching calculation is done using on-shell matching in dimensional regularization in the $\msbar$ scheme with all scales apart from the matching scale $Q$ set to zero. This makes all  EFT loop graphs scaleless and thus zero. The loop graphs to be calculated in QCD are shown in Fig.~\ref{fig:3}. As these involve an external gluon, they are computed using the background field method to obtain a gauge-invariant matching. Once computed, these can be expanded in the SCET limit using the methods outlined in \cite{Goerke:2017ioi} to obtain the matching coefficients in Eq.~\eqref{4.18}. For the $B$ topology, the graphs have an incoming $\nb$-collinear quark and gluon, and an outgoing $n$-collinear quark. For the $D$ topology, they have an incoming $\nb$-collinear gluon and outgoing $n$-collinear quark and antiquark. The $A$ and $C$ topologies have the same matching as the $B$ and $D$ topologies, respectively.

The one-loop matching coefficient $C_J$ for the leading power operator $j^\mu_{\text{SCET}}$ has been calculated before in~\cite{Manohar:2003vb}, and was reproduced here as a check on our calculation. In~\cite{Manohar:2003vb}, this matching was calculated using the $q \to q$ matrix element, while we obtained the same result using the $q \to q g$ matrix element. The two results have to agree, because matching from a UV theory onto the EFT is an operator relation and independent of matrix elements. $C_J$ does not depend on the momentum fraction $u$, which is another non-trivial check on our calculation. The results for the one-loop coefficients are
\begin{align} \label{4.18}
    C_J(\mu) &=  C^{(1\perp)}(\mu)= 1 + \frac{\alpha_s(\mu)}{4\pi}C_F \left[ - \lQ^2 + 3\lQ - 8 + \frac{\pi^2}{6} \right] \,, \nn
    C^{(1A)}(\mu,u) &=C^{(1B)}(\mu,u) = 1 + \frac{\alpha_s(\mu)}{4\pi}C_F \Bigg[ - \lQ^2 + \left( 1 -  \frac{2 \ln \overline u }{ u} \right)\lQ \nn
    &\hspace{6.1cm} - \frac{ \ln^2 \overline u}{ u } + \frac{4  \ln \overline u }{ u } + \frac{\pi^2}{6} - 3 \Bigg] \nn
    &\hspace{3cm} + \frac{\alpha_s(\mu)}{4\pi} C_A \left[ \frac{ \ln \overline u }{u}\ \lQ + \frac{ \ln^2 \overline u}{2  u }-\frac{2  \ln \overline u }{ u } - \frac{ \ln u }{\overline u} \right] \,, \nn
    C^{(1C)}(\mu,u) &=  C^{(1D)}(\mu,u) =  1 + \frac{\alpha_s(\mu)}{4\pi} C_F \left[ \big( 2 \ln u + 3 \big)\lQ + \ln^2 u + \frac{2 u}{\overline u}\ln u + 5 \ln \overline u  - 9 \right] \nn
    & \hspace{3cm}+ \frac{\alpha_s(\mu)}{4\pi}C_A \Bigg[ - \lQ^2 - \left( \ln \overline u + 2 \ln u \right)\lQ \nn
    &\hspace{4.5cm} - \frac{1}{2} \ln^2 \overline u - \ln^2 u -  \ln \overline u - \frac{ u}{\overline u} \ln u + \frac{\pi^2}{6} + 1  \Bigg] \,,
\end{align}
where $\overline u = 1-u$. The one-loop matching for $g_2$ was computed previously in~\cite{Ji:2000ny}. The $x \to 1$ limit of the results in~\cite{Ji:2000ny} differs from Eq.~\eqref{4.18}, though many terms are similar.\footnote{The result in~\cite[Eqs.~(22), (23)] {Ji:2000ny} gives the one-loop matching coefficients onto trilocal operators. The coefficients are functions of the ratios $x_B/x$ and $x_B/y$, as required by boost invariance. However, the actual expressions contain terms which depend on all three variables, not just on the two boost invariant ratios.}

The one-loop matching coefficients are for the QCD current and make no reference to DIS. They are general and can be applied to order $\alpha_s$ SCET analyses of any hard-scattering process such as Drell-Yan or dijet production. For time-like processes such as Drell-Yan, the coefficients are given by analytic continuation, $Q^2 \to -q^2 - i0^+$.

The expressions for the one-loop matching coefficients diverge logarithmically in the limit $u \to 0$ and $u \to 1$, which is when one of two particles in the same sector is soft. The coefficients are thus only valid for $u \in (0, 1)$ and not in the endpoint region where $u$ or $1-u$ are parametrically small, as mentioned before. The factorization formula for $g_2$ derived in Section~\ref{sec:matchingJ} gives $g_2$ as the integral over $u$ of $C(u)$ with a non-perturbative PDF $h_q(x,u)$. Since the divergence of $C(u)$ at the endpoints is logarithmic, the integral converges as long as $h_q(x,u)$ does not diverge as fast as $1/u$ at the endpoints, i.e.\ as long as the divergence is $1/u^a$ with $a<1$.  The behavior of $h_q(x,u)$ near the endpoint is studied in Appendix~\ref{app:F}, following the method of~\cite{vanBijleveld:2025ekz}. We find that $h_q(x,u)$ behaves as $u$ as $u \to 0$ or as $1$ as $u \to 1$, so that the $u$ integral is convergent.

$1/u$ endpoint singularities were studied in~\cite{Beneke:2020ibj} in the context of off-diagonal resummation in DIS as $x \to 1$. Coefficient functions with $1/u$ behavior at the endpoints are generated by power corrections to the structure functions. They do not occur in the leading contribution to $g_2$, which is twist-three, but do occur in higher order power corrections.


\section{Anomalous Dimension of the order \texorpdfstring{$\lambda$}{lambda} SCET Operators}\label{sec:IW}

The operator $O^{(1B)}$ is evolved from $Q$ to $M_J$ using the SCET anomalous dimension
\begin{align} \label{4.20}
    \mu\frac{d}{d\mu}O^{(1B)\nu}(u) = \int_0^1 {\rm{d}}v \, \gamma^{(1B)}(u,v) \, O^{(1B)\nu}(v) \,.
\end{align}
The anomalous dimension was computed in~\cite{Freedman:2014uta,Goerke:2017lei}.\footnote{\label{foot:10} Our convention for the anomalous dimension Eq.~\eqref{4.20} differs from~\cite{Goerke:2017lei} by a sign, so Eq.~\eqref{4.21} is the negative of the anomalous dimension
in~\cite{Goerke:2017lei}.}  For future convenience, we write
\begin{align} \label{4.21a}
    \gamma^{(1B)}(u,v) &= \widetilde \gamma^{(1B)}(u,v) + \gamma_J \delta(u-v)\,,
\end{align}
where $\gamma_J$ is given in Eq.~\eqref{3.2} and
\begin{align} \label{4.21}
   \widetilde \gamma^{(1B)}(u,v) & = \frac{\alpha_s}{4\pi} \Bigg\{ \delta(u-v) \left[-4 C_F \ln \overline{v} - 2 C_A \left( 1 + \ln \frac{v}{\overline{v}} \right) \right] \nn
    &-  2 \left( 2C_F - C_A \right) \overline u \left[ \frac{uv}{\overline u \, \overline{v}}\theta(1-u-v) + \frac{uv+u+v-1}{u v} \theta(u+v-1) \right] \nn
    &- 2C_A \overline u \Bigg[ \frac{\overline{v}-uv}{u \overline{v}}  \theta(u-v) + \frac{\overline{u} - uv}{\overline u v} \theta(v-u) \nn
    &\hspace{3.5cm} - \frac{1}{\overline u \, \overline{v}} \left( \overline{u} \frac{\theta(u-v)}{u-v} + \overline{v} \frac{\theta(v-u)}{v-u} \right)_{\!\!+} \Bigg] \Bigg\}\,.
\end{align}
The $+$ function is defined in \cite[(3.14)]{Goerke:2017lei} and $\overline u = 1-u$, $\overline v = 1-v$. $  \widetilde \gamma^{(1B)}(u,v) $ depends only on $u$ and $v$, and the entire $Q/\mu$ dependence of the anomalous dimension is in $\gamma_J$.
The coefficient function has the RG equation
\begin{align} \label{4.60}
    \mu\frac{d}{d\mu}C^{(1B)\nu}(u) = -\int_0^1 {\rm{d}}v \, \gamma^{(1B)}(v,u) \, C^{(1B)\nu}(v) \,.
\end{align}

The anomalous dimension of the $1C$ operator is also given in~\cite{Goerke:2017lei},\footnote{ \label{foot:11} The anomalous dimension for our definition of $O^{(1C)}(x,u)$ is $-v/u$ times the anomalous dimension given in \cite[(3.13)]{Goerke:2017lei} as our operator is defined to have a tree-level matching $C^{(1C)}(u) = 1$ and is $u$ times the operator in~\cite{Goerke:2017lei}. The minus sign is explained in footnote \ref{foot:10}.} but we do not need it here. The anomalous dimensions of the $1A$ and $1D$ operators are equal to those of the $1B$ and $1C$ operators, respectively.

\section{Matching onto the PDF Operator at the Jet Scale \texorpdfstring{$M_J$}{}} \label{sec:matchingJ}

At the scale $M_J$, the order $\lambda$ piece of $W^{\mu\nu}$,
\begin{align} \label{4.22}
W^{(\lambda)\, \mu\nu} &=  C_J \int_0^1 du \, \frac{1}{Q}C^{(1B)}(u)\, \left[ W_{JO} ^{(\lambda)\, \mu\nu} (u)  + W_{OJ} ^{(\lambda)\, \mu\nu} (u) \right]  \,, \nn
W_{JO}^{(\lambda)\,\mu\nu} (u)  &=  \frac{1}{4\pi}  \int \rd x \ e^{i q \cdot x}\, j_\text{SCET}^{\dagger\mu}(x) \, O^{(1B)\nu}(0,u)\,, \nn
W_{OJ}^{(\lambda)\,\mu\nu} (u)  &=  \frac{1}{4\pi}  \int \rd x \ e^{i q \cdot x}\, O^{\dagger(1B)\mu}(x,u) \, j_\text{SCET}^{\nu}(0) \,, 
\end{align}
is matched onto a twist-three $q \overline q g$ PDF operator by integrating out the $n$-collinear jet. Eventually, one takes the forward matrix element of the PDF operator between proton states to generate the non-perturbative piece.
The final expression produces the $g_2$ tensor structure, because $j^\mu_{\text{SCET}}$ has a $\perp$ index, and $O^{(1B)\nu}$ has a $\parallel$ index.

\begin{figure}
    \centering
    \includegraphics[width=10cm]{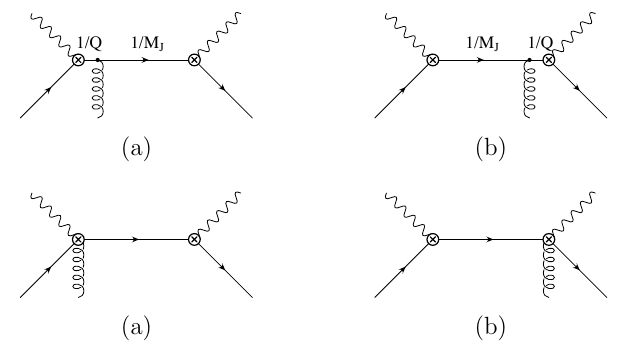}
    \caption{Comparison of the QCD (top) and SCET (bottom) OPEs for $g_2$. The QCD OPE, where $M_J$ is order $Q$, generates trilocal quark-gluon operators. As $x \to 1$, $Q \gg M_J$ and these turn into bilocal SCET operators.}
    \label{fig:qcd_vs_scet}
\end{figure}

The tree-level partonic matrix element in SCET is 
\begin{align} \label{4.10}
    \langle p_2|W^{(\lambda)\, \mu\nu}_{JO} (u) |p_1,k\rangle &=  gT^a  ( n^\nu + \nb^\nu )\,  \varepsilon_\perp^{a \alpha} \,  \delta\left(1- \frac{-q^+}{p_2^+}\right) \, \delta\left( u - \frac{k^+}{Q} \right)\frac{1}{2p_2^+}\overline{u}_2 \gamma_\perp^\mu \frac{\slashed{n}}{2} \gamma^\perp_\alpha u_1 \,,
    \end{align} 
where $p_1$ is the quark momentum, $k$ is the gluon momentum, $p_2=p_1+k$, and $\varepsilon^a_\alpha$ is the gluon polarization. 
The parameter $u$ in $O^{(1B)}(x,u)$ labels a momentum $uQ$, meaning the operator is quasi-local --- it has a non-locality of order $1/Q$ at the scale $Q$, but is local when considered at the parametrically lower scale $M_J$.  The PDF parameter $z = -q^+/p_2^+$ arising from the OPE describes a non-locality of order $1/M_J$ along the lightcone in the theory below $M_J$. Thus, the QCD trilocal operators look like bilocal SCET operators, with one separation of order $1/Q$ treated as local. The comparison is shown in Fig.~\ref{fig:qcd_vs_scet}.

The matrix element in Eq.~\eqref{4.10} is reproduced by the (non-hermitian) quark-gluon PDF operator
\begin{align} \label{4.50}
    H_q^\mu(r^+,u) &= -\frac{1}{4\pi} \int d\xi \, e^{-i\xi r^+}  \overline{\chi}_{\nb} (n\xi) \gamma_\perp^\mu \frac{\slashed{n}}{2} \gamma^\perp_\alpha [ \mathcal{B}_{\nb}^{\alpha}(0)]_{uQ} [\chi_\nb (0)]_{(1-u)Q},
\end{align}
whose tree-level partonic matrix element is
\begin{align} \label{4.51}
    \langle{p_2 | H_q^\mu(x p_2^+,u) | p_1,k\rangle} &= \frac{gT^a}{Q} \, \varepsilon_\perp^{a \alpha}  \, \delta\left( u - \frac{k^+}{Q} \right)\delta(1-x)\frac{1}{2p_2^+}\overline{u}_2 \gamma_\perp^\mu \frac{\slashed{n}}{2} \gamma^\perp_\alpha u_1 \,.
\end{align}
The antiquark-gluon PDF operator is given by charge conjugation,
\begin{align} \label{4.50a}
    H_{\overline q}^\mu(r^+,u) &=- \frac{1}{4\pi} \int d\xi \, e^{-i\xi r^+}  [\overline{\chi}_{\nb}(0)]_{(1-u)Q}[ \mathcal{B}_{\nb}^{\alpha}(0)]_{uQ} \gamma_\perp^\alpha \frac{\slashed{n}}{2} \gamma^\perp_\mu \, \chi_\nb (n \xi) \,, \nn
&= H^{\mu \dagger}_q(-r^+,u) \,.
\end{align}
The OPE Eq.~\eqref{4.22} is
\begin{align} \label{4.52}
W_{JO}^{(\lambda)\, \mu\nu} (u) &=  ( n^\nu + \nb^\nu )\, \int \frac{\rd r^+}{r^+} \ \mathcal{M}\left( -q^+/r^+\right) \left[ H_q^\mu(r^+,u) 
+ H_{\overline q}^\mu(r^+,u)  \right] \,,
 \end{align} 
with matching coefficient $\mathcal{M}(z)=\delta(1-z)$ at tree-level.

The matrix element of the quark-gluon PDF operator $H_q^\mu(r^+,u)$ in a proton is
\begin{align} \label{4.53}
\braket{P,S | H_q^\mu(x P^+,u) | P, S} &= i\, \epsilon_{\perp}^{\mu \sigma}S_{\sigma}\   h_q(x,u) \,,
\end{align}
where the quark-gluon PDF $h_q(x,u)$ defined by Eq.~\eqref{4.53} is real. $h_q(x,u)$ is an interference PDF and does not have to be positive. Similarly, the antiquark-gluon PDF $h_{\overline q}(x,u)$ is the matrix element
\begin{align} \label{4.53a}
\braket{P,S | H_{\overline q}^\mu(x P^+,u) | P, S} &= i\, \epsilon_{\perp}^{\mu \sigma}S_{\sigma}\   h_{\overline q}(x,u) \,.
\end{align}
The form of Eq.~\eqref{4.53} and Eq.~\eqref{4.53a}  is required by parity and time-reversal invariance.  By dimensional analysis, $h_q(x,u)$, $h_{\overline q}(x,u)$  are dimensionless functions, since we have normalized the spin $S_\sigma$ to have dimensions of mass.

Repeating the above analysis for $W_{JO}^{\mu \nu}(u) $ gives
\begin{align} \label{4.65}
W_{OJ}^{(\lambda)\, \mu\nu} (u) &=  ( n^\mu + \nb^\mu )\, \int \frac{\rd r^+}{r^+} \ \mathcal{M}\left( -q^+/r^+\right) \left[ \overline H_q^\nu(r^+,u) 
+ \overline H_{\overline q}^\nu(r^+,u)  \right] \,,
 \end{align}
where
\begin{align} \label{4.66}
\overline    H_q^\nu(r^+,u) &= -\frac{1}{4\pi} \int d\xi \, e^{-i\xi r^+}   [\overline{\chi}_{\nb}(n\xi)]_{(1-u)Q}[ \mathcal{B}_{\nb}^{\alpha}(n \xi)]_{uQ} \gamma_\perp^\alpha \frac{\slashed{n}}{2} \gamma^\perp_\mu \, \chi_\nb (0) \,,
\end{align}
and $\overline    H_{\overline q} ^\nu(r^+,u) = \overline    H_q^{\nu \dagger }(-r^+,u)$. 
We can use translation invariance of the forward matrix element and flip the sign of $\xi$ to write
\begin{align} \label{4.67}
\overline    H_q^\nu(r^+,u) &= -\frac{1}{4\pi} \int d\xi \, e^{i\xi r^+}   [\overline{\chi}_{\nb}(0)]_{(1-u)Q}[ \mathcal{B}_{\nb}^{\alpha}(0)]_{uQ} \gamma_\perp^\alpha \frac{\slashed{n}}{2} \gamma^\perp_\mu \, \chi_\nb (n \xi) \,, \nn
&=  H_q^{\dagger \nu}(r^+,u)  \,.
\end{align}
The matrix element of the hermitian conjugate is
\begin{align} \label{4.54}
\braket{P,S | H_q^{\nu \dagger} (x P^+,u) | P, S} &= - i\, \epsilon_{\perp}^{\nu \sigma}S_{\sigma}\  h_q(x,u)\,.
\end{align} 
The minus sign relative to Eq.~\eqref{4.53a}  is needed to generate the $g_2$ tensor structure in Eq.~\eqref{2.20}.  
Using Eq.~\eqref{4.50a} and Eq.~\eqref{4.54} gives $h_{\overline q}(x,u) =  -h_q(-x,u)$.

Combining the above equations and comparing with Eq.~\eqref{2.20} gives the factorization formula
\begin{align} \label{4.55}
x \left[ g_1(x) +g_2(x) \right] &= C_J \int_0^1 \rd u\ C^{(1B)}(u)  \int_x^1 \frac{\rd y}{y} \ \mathcal{M}\left({x}/{y}\right) \ \left[h_q(y,u) + h_{\overline q}(y,u)\right]\,.
\end{align} 
The matching coefficient $\mathcal{M}$ is the \emph{same} as for $F_1$ and $g_1$, as is explained at the end of this section.
The l.h.s.\ can be identified with $x\,\overline g_2(x)$ (see Eq.~\eqref{4.39}) since we are working to leading order in $\lambda$ as $x \to 1$. Note that the $g_2$ tensor structure in Eq.~\eqref{2.20} is only generated \emph{after} taking the hadronic matrix element of the PDF operator. The PDF operator $H_q^\mu$ carries a Lorentz index associated with $W^{\mu\nu}$, unlike the PDFs needed for $F_1,F_L$ or $g_1$. $g_2$ enters $W^{\mu\nu}$ as part of the coefficient of the tensor structure $\epsilon_\perp^{\mu \sigma}S_\sigma$. As $S_\sigma$ is a hadronic quantity,  $\epsilon_\perp^{\mu \sigma}S_\sigma$ has to be generated after taking the hadronic matrix element of  $H^\mu_q$. 

We have chosen to define the PDF operator in Eq.~\eqref{4.50} in terms of a product of three $\gamma$-matrices, rather than use the identity
\begin{align} \label{4.11}
    \gamma_\perp^\sigma \slashed{n} \gamma_\perp^\tau &= -g_\perp^{\sigma\tau}\slashed{n} - i\epsilon_\perp^{\sigma\tau}\slashed{n}\gamma_5 \, ,
\end{align}
to identify each term in the matrix element with the $G$ and $\widetilde{G}$ terms in the trilocal PDF defined in Eq.~\eqref{2.35}. We find it easier to use the definition in Eq.~\eqref{4.50} as it does not involve an explicit $\gamma_5$, meaning the subtleties mentioned in Appendix~\ref{app:BMHV} do not enter. It also simplifies the computation of the anomalous dimension of the PDF operator given in Section~\ref{sec:anomdim}. The relation between the SCET PDF and the QCD PDF given in Eq.~\eqref{2.36} will be discussed in Section~\ref{sec:qcd_vs_scet}.

The one-loop matching coefficient $\mathcal{M}$ onto $H^\mu_q$ is given by the same graphs as in Fig.~\ref{fig:g1_soft_matching}. SCET decouples the $n$- and $\nb$-sectors if the overlap/zero-bin prescription is implemented correctly. As the matching is generated by integrating out the $n$-sector, it does not see the $\nb$-sector structure of the SCET operators. This leads to the result that the one-loop matching coefficient $\mathcal{M}$ is the same as for $F_1$ given in Eq.~\eqref{3.6}, Eq.~\eqref{3.10}, and does not depend on SCET momentum fraction labels. A check of this result is given in Section~\ref{sec:consistency}. 

\begin{figure}
\begin{center}
\includegraphics[width=9cm]{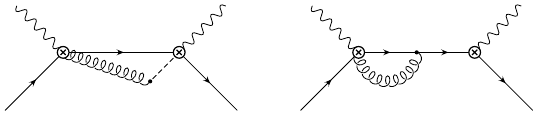}
\end{center}
\caption{\label{fig:1A} Graphs contributing to the product of  $j^{\dagger\mu_\perp}_\text{SCET}$ and $O^{(1A)\nu_\parallel} $, which vanish. The collinear Wilson line has been shown as the dashed line. The gluon from the left vertex is from the $\mathcal{B}_\alpha$ field.}
\end{figure}

In principle, at order $\alpha_s$, there are also contributions from the interference of $j^\mu_\text{SCET}$ and $O^{(1A)\nu}$ as shown in Fig.~\ref{fig:1A}.
%
%
These graphs vanish because the $O^{(1A)}$ vertex involves $\mathcal{B}^\alpha_n \gamma_\alpha^\perp$, and there is no $\perp$ direction in the loop \cite{Luke:2022ops}.

\section{Anomalous Dimension of the PDF Operator} \label{sec:anomdim}

The running of the PDF operator $H^\mu_q$ requires calculating the anomalous dimension of the quark-gluon PDF in SCET. Unlike the PDF operators $O_q$ and $O_{\Delta q}$ for the unpolarized and polarized quark PDFs, which had the same anomalous dimensions in QCD and SCET, $H^\mu_q$ has no analog in QCD as the momentum fraction $u$ is not a label that exists in QCD.
\begin{figure}
\begin{center}
    \includegraphics[width=1\linewidth]{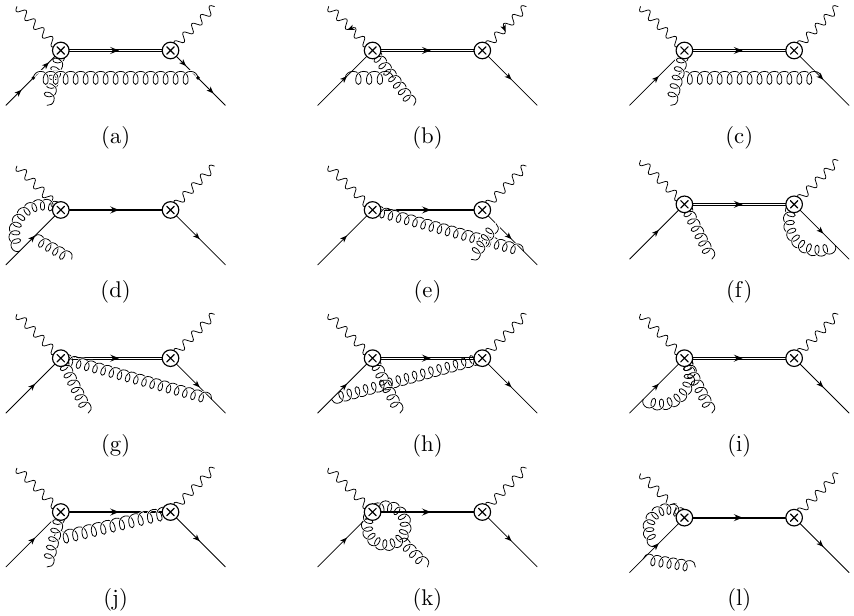}
  \end{center}
    \caption{Loop graphs that contribute to the PDF running. Gluons from the vertex arise from either $\mathcal{B}_\alpha$ or a Wilson line.
    Graph (l) is one-particle reducible, but contributes for on-shell matching. Graphs where one end of a gluon line is from a Wilson line vertex, and the other end from a Wilson line or $\mathcal{B}_\alpha$ vertex vanish, since $n^2=0$ and $n^\alpha_\perp=0$, and are not shown. The double line denotes the separation $n \xi$ in the lightcone Fourier transform, and is \emph{not} a Wilson line. }
    \label{fig:5}
\end{figure}
%
%
The PDF anomalous dimension for $H^\mu_q$ is computed in a theory with only the $\nb$ sector, and so can be computed for all $x$, not just $x \to 1$, since $1-x$ is no longer an expansion variable. This is similar to the DGLAP evolution kernel, whose full $x$ dependence can be computed in SCET.

The graphs that contribute to the anomalous dimension are shown in Fig.~\ref{fig:5}. The one-gluon and two-gluon Feynman rules needed for the calculation are given in Appendix G of \cite{Goerke:2018pns}. Extra care needs to be taken as individual graphs are rapidity divergent. We do the loop $l^-$ integrals by contour integration, and  $l_\perp$ integrals using dimensional regularization. The remaining individual $l^+$ integrals are rapidity-divergent, but the rapidity divergences cancel in the sum, so the $l^+$ integral is done after combining the individual diagrams.

The resulting anomalous dimension is
\begin{align} \label{4.26}
    \mu\frac{d}{d\mu}H^\nu(x,u) = \int_0^1 {\rm{d}}v \, \int_x^1 \frac{{\rm{d}}z}{z} \, \gamma(z ; u,v) \, H^\nu\left(x/z,v\right) \,,
\end{align}
where
\begin{align} \label{4.23}
    \gamma(z ; u,v) = &P_{qq}(z) \, \delta(u-v) +\widetilde \gamma^{(1B)}(u,v)\, \delta(1-z) \nn
    & + \frac{\alpha_s}{4\pi} \bigg\{ 4C_F \, z \, \delta(u-v) + 2C_A \delta(u-v) \nn
    &  + 2(2C_F - C_A) \left[ -\frac{z}{\overline{v}} + \left( \frac{z}{v\overline{v}} - \frac{1}{1-z} \right) \theta(0 \leq z \leq v) \right] \delta(u-v) \nn
    &+ 2C_A \left[ \left( \frac{\overline{v}}{u} - \frac{1}{v} \right) - \frac{\theta(0 \leq z \leq \overline{v})}{1-z} \right] \delta(1-z+u-v) \nn
    &   + 4C_A \frac{\delta(1-z+u-v) - \delta(u-v)}{1-z} \bigg\} \,.
\end{align}
$P_{qq}(z)$ is the Altarelli-Parisi kernel Eq.~\eqref{3.8} and $\widetilde\gamma^{(1B)}(u,v)$ is given in Eq.~\eqref{4.21}. It is useful to extract from $\gamma(z;u,v)$ the first two terms in the first line, as will be seen in Section~\ref{sec:consistency}.
The anomalous dimension Eq.~\eqref{4.23} can be used to evolve the twist-three PDF from $M_J$ to a reference scale $\mu_0$. The combination
$ \gamma(z;u,v)$ is only a function of $z,u,v$ with no dependence on $Q$ or $\mu$. $  \gamma(z ; u,v)$  is an anomalous dimension computed in a single sector theory, and can have no dependence on $Q$.

As $z \to 1$, all the terms except $P_{qq}(z)$ and $\widetilde \gamma^{(1B)}(u,v)\delta(1-z)$ are subleading in $\lambda$, and have vanishing $z$-moments as $N \to \infty$. For example, the last term becomes 
\begin{align} \label{4.24}
    \frac{\delta(1-z+u-v) - \delta(u-v)}{1-z} &= \delta^\prime(u-v), \ \ \ \text{ as } z \to 1
\end{align}
whose $z$-moments $\delta^\prime(u-v)/N$ vanish as $N \to \infty$. This gives
\begin{align} \label{4.25}
    \gamma(z; u,v) = &P_{qq}(z)\, \delta(u-v) +\widetilde \gamma^{(1B)}(u,v)\, \delta(1-z) + \mathcal{O}(\lambda)\,,
\end{align}
so that as $x \to 1$, the $H^\mu_q$ anomalous dimension is the sum of $P_{qq}$ and $\widetilde \gamma^{(1B)}$. The anomalous dimension simplifies as $x \to 1$ into a sum of two anomalous dimensions, each of which gives the evolution in a single variable, so that $x$ and $u$ evolve independently. The $N^\text{th}$ moment w.r.t.\ $z$ dropping the $ \mathcal{O}(\lambda)$ terms is
\begin{align} \label{4.25a}
   M_N[ \gamma(z; u,v) ] = &M_N[P_{qq}(z)]\,  \delta(u-v) +\widetilde \gamma^{(1B)}(u,v) \,.
\end{align}

The RG improved result for the moments of $g_1(x)+g_2(x)$ from Eq.~\eqref{4.55} is
\begin{align}
 M_N[ x\, (g_1 +g_2) ] &= \int_0^1 du \int_0^1 \rd v \int_0^1 \rd w\, C_J(Q) \, C^{(1B)}(u,Q) \,  e^{2\Gamma_J(Q,\mu_J)}\, \widetilde \Gamma^{(1B)} (u,v | Q,\mu_J) \nn
    & M_N\left[\mathcal{M}\left(\mu_J\right)\right]\  e^{\Gamma_{qq,N}(\mu_J,\mu_0)} \,  \widetilde \Gamma^{(1B)} (v,w | \mu_J, \mu_0)  h_N(w,\mu_0) \,,
    \label{4.71}
\end{align}
where $\Gamma_J$ and $\Gamma_{qq,N}$ are  given in Eq.~\eqref{3.22} and $\mu_J=Q/\sqrt{\Nb}$. $\widetilde \Gamma^{(1B)}(u,v | \mu_h, \mu_l)$ is the solution of the differential equation
\begin{align}
\mu \frac{\rd O(u,\mu)}{\rd \mu} &= \int_0^1 \rd v\  \widetilde \gamma^{(1B)} (u,v) O(v,\mu) \,,
\end{align}
on integrating from $\mu_l$ to $\mu_h$,
\begin{align}
O(u,\mu_h) &= \int_0^1 \rd v\  \widetilde \Gamma^{(1B)} (u,v | \mu_h , \mu_l)\, O(v,\mu_l) \,.
\label{4.70}
\end{align}
$h_N(w)$ is the moment of the PDF,
\begin{align}
h_N(w) &= \int_0^1 \rd x\ x^{N-1}\, \left[ h_q(x,w) + h_{\overline q}(x,w) \right] \,.
\end{align}
Eq.~\eqref{4.71} is obtained by matching the current at the scale $Q$, evolving the SCET operators to the scale $\mu_J$, performing the time-ordered product at $\mu_J$, and then evolving the PDF operator to $\mu_0$.

The multiplicative nature of the RGE evolution Eq.~\eqref{4.70} simplifies Eq.~\eqref{4.71} to
\begin{align}
   M_N[ x\, (g_1(x)+{g}_2(x)) ]  &=C_J(Q) \, e^{2\Gamma_J(Q,\mu_J)}\,  M_N\left[\mathcal{M}\left(\mu_J\right)\right]  e^{\Gamma_{qq,N}(\mu_J,\mu_0)}\ \times \nn
    &\int_0^1 du \int_0^1 \rd w \  C^{(1B)}(u,Q) \,   \widetilde \Gamma^{(1B)} (u,w|Q,\mu_0) \,  h_N(w,\mu_0) \,,
    \label{4.72}
\end{align}
since the matching $\mathcal{M}$ is independent of the SCET labels. As in Eq.~\eqref{3.22}, the anomalous dimension $\gamma_J$ contributes between $Q$ and $\mu_J$, and $P_{qq}(z)$ between $\mu_J$ and $\mu_0$. In addition $\widetilde \Gamma^{(1B)}(u,v) $ contributes between $Q$ and $\mu_0$. These anomalous dimensions do not mix the $z$ and $u$ evolution.

There are also twist-three gluon operators which contribute to $g_2$~\cite{Ji:1992eu}.  Their coefficients start at order $\alpha_s$, and mixing of the quark-gluon operators discussed above into these gluon operators is order $\lambda$, so the gluon operators do not contribute to the leading order resummation.

The even moments of the twist-three distribution are related to local operators,
\begin{align} \label{4.75}
& 2i (P^+)^N \epsilon_{\perp}^{\mu \sigma}S_{\sigma}\   M_N[h_q(x,u)+h_{\overline q}(x,u)] \nn
& =   \braket{P,S|[\overline{\chi}_{\nb}(0)]_{(1-u)Q}[ \mathcal{B}_{\nb}^{\alpha}(0)]_{uQ} \gamma_\perp^\alpha \frac{\slashed{n}}{2} \gamma^\perp_\mu \, \left( i n \cdot D \right)^{N-1}  \chi_\nb (0) |P,S} \qquad (N \ \text{even})\,,
\end{align}
which can be derived using the $PT$ symmetry relation
\begin{align} \label{4.76}
& \braket{P,S|  [\overline{\chi}_{\nb}(0)]_{(1-u)Q}[ \mathcal{B}_{\nb}^{\alpha}(0)]_{uQ} \gamma_\perp^\alpha \frac{\slashed{n}}{2} \gamma^\perp_\mu \, \chi_\nb (n \xi) |P,S} \nn
&= \braket{P,S| \overline{\chi}_{\nb} (n\xi) \gamma_\perp^\mu \frac{\slashed{n}}{2} \gamma^\perp_\alpha [ \mathcal{B}_{\nb}^{\alpha}(0)]_{uQ} [\chi_\nb (0)]_{(1-u)Q} |P,S}^* \nn
&= \braket{P,-S| \overline{\chi}_{\nb} (-n\xi) \gamma_\perp^\mu \frac{\slashed{n}}{2} \gamma^\perp_\alpha [ \mathcal{B}_{\nb}^{\alpha}(0)]_{uQ} [\chi_\nb (0)]_{(1-u)Q} |P,-S} \,,
\end{align}
and the same method as used for the polarized quark distribution in~\cite{Manohar:1990kr} to turn the $x$ integral into a $\delta$-function of $\xi$.
The OPE analysis determines the odd moments of $g_1$ and $g_2$, so $N$ in Eq.~\eqref{4.72} is even, because of the additional factor of $x$ on the l.h.s. Thus, we need the even moments $h_N$, which are given by the matrix elements Eq.~\eqref{4.75}. The Burkhardt-Cottingham sum rule is given by $N=0$.  There is no operator on the r.h.s.\ of Eq.~\eqref{4.75} since the power $N-1$ of $(in\cdot D)$ must be non-negative, so the sum rule is satisfied.

\section{Matching Consistency Conditions} \label{sec:consistency}

The matching conditions generate consistency conditions that relate anomalous dimensions to matching coefficients. We get five consistency conditions that provide a check of our results.

\paragraph{(a)} The QCD current in Eq.~\eqref{4.3} is $\mu$-independent, so the r.h.s.\  must also be $\mu$-independent. The order $\lambda^0$ piece gives
\begin{align} 
0 &= \mu \frac{\rd C_J(\mu)}{\rd \mu} + \gamma_J \,,
\label{9.1}
\end{align}
which is satisfied by Eq.~\eqref{3.3} and Eq.~\eqref{3.2}.

\paragraph{(b)} The order $\lambda^1$ piece gives 
\begin{align} 
0 &= \mu \frac{\rd C^{(1B)} (v,\mu)}{\rd \mu} + \int \rd u \ C^{(1B)} (u,\mu) \, \gamma^{(1B)}(u,v) \,.
\label{9.2}
\end{align}
At order $\alpha_s$, we can use the tree-level result $C^{(1B)} (u,\mu) =1$ in the second term,
\begin{align} 
0 &= \mu \frac{\rd C^{(1B)} (v,\mu)}{\rd \mu} + \int \rd u \  \gamma^{(1B)}(u,v) \,,
\label{9.3}
\end{align}
which is satisfied by Eq.~\eqref{4.18} and Eq.~\eqref{4.21}.

\paragraph{(c)} The consistency condition of the matching Eq.~\eqref{3.5} gives
\begin{align} 
2 \gamma_J \, \delta(1-z) &= \mu \frac{\rd \mathcal{M}(z)}{\rd \mu} + P_{qq}(z) + \mathcal{O}(\lambda) \,.
\label{9.4}
\end{align}
The $\mathcal{O}(\lambda)$ on the r.h.s.\ arises since the matching is performed in SCET neglecting subleading terms. Using Eq.~\eqref{3.2}, Eq.~\eqref{3.6} and Eq.~\eqref{3.8},
\begin{align} 
2\gamma_J\, \delta(1-z) &= \mu \frac{\rd \mathcal{M}(z)}{\rd \mu} + P_{qq}(z) + \frac{\alpha_s}{\pi} C_F (1+z) \,,
\label{9.5}
\end{align}
so the consistency condition Eq.~\eqref{9.4} is satisfied, since the moments of $(1+z)$ vanish as $N \to \infty$.

\paragraph{(d)} The consistency condition of the matching Eq.~\eqref{4.55} gives
\begin{align} 
\gamma_J \, \delta(1-z) \, \delta(u-v) + \gamma^{(1B)}(u,v)\,  \delta(1-z)  &= \mu \frac{\rd \mathcal{M}(z)}{\rd \mu} \delta(u-v) + \gamma(z;u,v)  + \mathcal{O}(\lambda) \,.
\label{9.6}
\end{align}
Combining with Eq.~\eqref{9.5} and using the definition Eq.~\eqref{4.21a} gives
\begin{align} 
P_{qq}(z)\, \delta(u-v) + \widetilde \gamma^{(1B)}(u,v)\, \delta(1-z)  &=  \gamma(z;u,v)  + \mathcal{O}(\lambda) \,,
\label{9.7}
\end{align}
which was verified in Eq.~\eqref{4.25}. This relation provides a non-trivial check that the matching coefficient $\mathcal{M}$ for the quark-gluon PDF $H^\mu_q$ is the same as for the twist-two quark PDFs.

\paragraph{(e)} We have verified that the $1D$ coefficients in Eq.~\eqref{4.18} satisfy the consistency condition
\begin{align} 
0 &= \mu \frac{\rd C^{(1D)} (v,\mu)}{\rd \mu} + \int \rd u \  \gamma^{(1D)}(u,v) \,,
\label{9.8}
\end{align}
using $ \gamma^{(1D)}(u,v) = -v/u\, \gamma_{(1c)}(u,v) $ with $\gamma_{(1c)}(u,v)$ given in~\cite[(3.13)]{Goerke:2017lei} (see footnote~\ref{foot:11}), and the tree-level matching $C^{(1D)}(u)=1$.
 
\section{DIS Coefficients as \texorpdfstring{$N \to \infty$}{N to Infinity}} \label{sec:coeff}

The form of the SCET operators and the $W^{\mu \nu}$ tensor structure Eq.~\eqref{2.20} give non-trivial information on the QCD coefficient functions as $x \to 1$. We compare the SCET results with the known expressions~\cite{DeRujula:1976baf,Bardeen:1978yd,Ahmed:1976ee,Kodaira:1978sh,Kodaira:1979ib,Kodaira:1979pa,Blumlein:2012bf} as $N \to \infty$.

The QCD current matches onto (schematically)
\begin{align}
j^\mu & \sim j^{\mu_\perp}_\text{SCET} + g \lambda \left[ O^{(1A)\mu_\parallel}  +   O^{(1B)\mu_\parallel} + O^{(1C)\mu_\perp}  + O^{(1D)\mu_\perp} \right]  + \text{h.c.}  + \mathcal{O}(\lambda^2) \,,
\label{11.4}
\end{align}
where we have made explicit the powers of $g$ and $\lambda$, and whether the Lorentz indices are $\parallel$ or $\perp$.  
The $\perp$ and $\parallel$ indices alternate with the order in $\lambda$, with the $C,\, D$ topology indices offset by one power of $\lambda$ from those of the $A,\, B$ topology. We have checked this to order $\lambda^3$ and order $g^2$.

The quark and gluon OPE coefficients are given from the terms in the time-ordered product of $j^\mu$ and $j^\nu$ with external incoming and outgoing $\nb$-collinear single quark or gluon lines, respectively. The terms which contribute to the quark coefficient to order $\lambda^2$ are
\begin{align}
j^\mu \ j^\nu &\sim  j^{\dagger \mu_\perp}_\text{SCET}\  j^{ \nu_\perp}_\text{SCET} + g \lambda
\left[ j^{\dagger \mu_\perp}_\text{SCET}\ O^{(1A)\nu_\parallel}  + O^{\dagger (1A)\mu_\parallel}\ j^{ \nu_\perp}_\text{SCET} \right]  + g^2 \lambda^2\, O^{\dagger(1A)\mu_\parallel}   O^{ (1A)\nu_\parallel} \nn
& + j^{\dagger \mu_\perp}_\text{SCET} \ \mathcal{O}(\lambda^2) 
\label{11.5}
\end{align}
and to the gluon coefficient to order $\lambda^2$ are
\begin{align}
j^\mu \ j^\nu &\sim g^2 \lambda^2  O^{\dagger(1D)\mu_\perp}  \ O^{(1D) \nu_\perp}  \,.
\label{11.6}
\end{align}

At order $\lambda^0$, the only term which contributes to the quark coefficient is $j^{\dagger\mu_\perp}_\text{SCET}\  j^{\nu_\perp}_\text{SCET} $, and  both indices are $\perp$ indices. The OPE produces the tensor structure
$\gamma_\perp^\mu \slashed{n} \gamma_\perp^\nu = -g_\perp^{\mu \nu} \slashed{n} - i \epsilon_\perp^{\mu \nu}
\slashed{n} \gamma_5$ from the $\gamma$-matrix identity Eq.~\eqref{3.21} and Eq.~\eqref{1.4}. This combination of $g_\perp^{\mu \nu}$ and $\epsilon_\perp^{\mu \nu}$ implies that the $F_1$ and $g_1$ coefficients are equal.
Thus at leading order in $\lambda$, $C_{1q,N}=C_{\Delta q,N}$ and $C_{L,q,N}=C_{1g,N}=C_{Lg,N}=C_{\Delta g,N}=0$. 

At order $\lambda$, the quark coefficient has contributions from the interference of $j^{\dagger\mu_\perp}_\text{SCET}$ and $O^{(1A)\nu_\parallel} $, but these terms vanish as explained at the end of Section~\ref{sec:matchingJ}. There are no contributions of order $\lambda$ to the twist-two OPE coefficients. In general, the only terms that survive are even powers of $\lambda$.

At order $\lambda^2$, the quark coefficients get contributions from the two $\lambda^2$ terms in Eq.~\eqref{11.5}.  
$ O^{\dagger(1A)\mu_\parallel}   O^{ (1A)\nu_\parallel} $ has both $\parallel$ indices, and contributes to $F_L$. The product
$ j^{\dagger \mu_\perp}_\text{SCET} \ \mathcal{O}(\lambda^2) $ has a $\perp$ index, and contributes to $F_1$ or $g_1$. The $A,
B$-topology order $\lambda^2$ operators are given in~\cite{Luke:2022ops}. The operators that contribute to $F_1$ or $g_1$ have a $\gamma_\perp^\nu$ index. The only Lorentz structure possible after doing the loop graph is $\gamma_\perp^\mu \slashed{n} \gamma_\perp^\nu$, and gives the same contribution to $F_1$ and $g_1$. The difference between $F_1$ and $g_1$ starts at order $\lambda^4 \sim 1/N^2$. The product of two order $\lambda^2$ operators has enough $\gamma$-matrix structure to generate $\gamma_\perp^\alpha \gamma_\perp^\mu \slashed{n} \gamma_\perp^\nu \gamma_{\perp \alpha}  $, which gives $F_1= - g_1$, and so contributes to $C_{\Delta q,N}-C_{1q,N}$.

\begin{figure}
\begin{center}
\includegraphics[width=4cm]{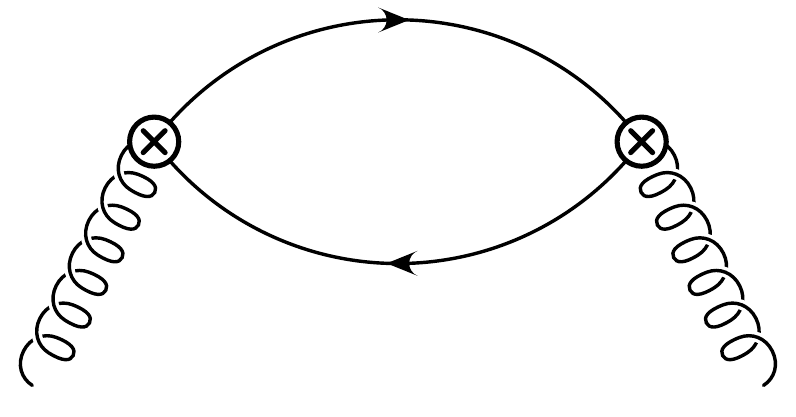}
\end{center}
\caption{\label{fig:gluon} One loop contribution of order $\alpha_s$ to the gluon coefficient functions from the order $\lambda$ operators $O^{(1D)\mu}$.}
\end{figure}

The analysis of gluon coefficients is much simpler, since there is no order $\lambda^0$ piece in the current.
At order $\lambda^2$, the gluon coefficients get contributions of order $\alpha_s$ from  $O^{\dagger(1D)\mu_\perp}  \ O^{(1D) \nu_\perp}$, the graph shown in Fig.~\ref{fig:gluon}.
The index on $O^{(1D)\mu}$ is a $\perp$ index, so Fig.~\ref{fig:gluon} can only produce the tensor structures which multiply $F_1$ and $g_1$, and does not contribute to $F_L$. $O^{(1D)\mu}$ has the $\gamma$-matrix structures
\begin{align}
\gamma_\perp^\mu \frac{\slashed{\nb}}{2} \slashed{\varepsilon}_\perp  &= - \frac{\slashed{\nb}}{2}
\left( \varepsilon_\perp^\mu - i \sigma^{\mu_\perp \alpha} \varepsilon^\perp_\alpha \right)\,,  &
\slashed{\varepsilon}_\perp  \frac{\slashed{\nb}}{2} \gamma_\perp^\mu
&= - \frac{\slashed{\nb}}{2}
\left( \varepsilon_\perp^\mu + i \sigma^{\mu_\perp \alpha} \varepsilon^\perp_\alpha \right) \,.
\label{13.4}
\end{align}
After evaluating the graph, there are three possible combinations of the external gluon polarization tensors
\begin{align}
(\varepsilon_1 \cdot \varepsilon_2^*) g^{\mu \nu} - \varepsilon_1^\mu \varepsilon_2^\nu, \quad
\varepsilon_1^\nu \varepsilon_2^{*\mu},\quad \varepsilon_1^\mu \varepsilon_2^{*\nu} \,,
\label{13.5}
\end{align}
where $\varepsilon_1$ is the incoming gluon polarization, and $\varepsilon_2^*$ is the outgoing gluon polarization.
The first two lead to $F_1=g_1$, and the third to $F_1=-g_1$. Since the graph Fig.~\ref{fig:gluon} is for the product 
 $j^{\dagger\mu}\  j^{\nu}$, the $j^\nu$ piece leads to the structures $\slashed{\nb}\varepsilon_1^\nu  $ and
$\sigma^{\nu_\perp \alpha} \slashed{\nb}\varepsilon_{1\alpha}  $ and $j^{\dagger\mu}$ to
$\slashed{\nb}\varepsilon_2^{*\mu}  $ and
$\sigma^{\mu_\perp \alpha} \slashed{\nb}\varepsilon_{2\alpha}^*  $. The incoming and outgoing polarization vectors
are correlated with $\mu$ and $\nu$. The only possible $\gamma$-matrix structure generated by the fermion propagators after doing the loop integral is $\slashed{n}$. The possible non-zero structures for the graph are
\begin{align}
\varepsilon_1^\nu \varepsilon_2^{*\mu} \,  \text{Tr} ( \slashed{\nb} \slashed{n} \slashed{\nb} \slashed{n} ) &=
32 \varepsilon_1^\nu \varepsilon_2^{*\mu} \nn
\varepsilon_{1\alpha} \varepsilon_{2\beta}^*   \text{Tr}(  \sigma^{\nu_\perp \alpha} \slashed{\nb}  \slashed{n} 
\sigma^{\mu_\perp \beta} \slashed{\nb}  \slashed{n} ) &= 32 \left[(\varepsilon_1 \cdot \varepsilon_2^*) g^{\mu \nu} - \varepsilon_1^\mu \varepsilon_2^\nu\right] \,.
\label{13.6}
\end{align}
The graph can only produce the first two linear combinations in Eq.~\eqref{13.5} and so gives $F_1=g_1$.
Evaluating the graph gives
\begin{align}
C_{1g} &=C_{\Delta g} =  \frac{\alpha_s}{4\pi} 4 T_F \left[ \ln \left( \frac{Q^2}{\mu^2} \frac{1-x}{x} \right) - 1 \right] \,,
\label{11.1}
\end{align}
with moments
\begin{align}
C_{1g,N} &=C_{\Delta g,N} =  \frac{\alpha_s}{4\pi} 4 T_F \frac{1}{N} \left[\lQ - H_n -1 + \frac{1}{N} \right] \,.
\label{11.2}
\end{align}
For $N \to \infty$, at $Q=\mu$
\begin{align}
C_{1g,N} &=C_{\Delta g,N} \to  \frac{\alpha_s}{4\pi} 4 T_F \frac{1}{ N} \left[ -\ln \Nb  -1 \right] \,,
\label{11.3}
\end{align}
which agrees with the known results. There is no gluon contribution to $F_L$ at this order. For $F_1$ and $g_1$ at order $\lambda^4\sim 1/N^2 $, we need the interference term between order $\lambda$ and $\lambda^3$ operators since the order $\lambda^2$ operators only have a $\parallel$ index. The $D$-topology amplitude that contributes to Fig.~\ref{fig:gluon} to order $\lambda^3$ is given in Appendix~\ref{app:D}. It involves $p_\perp$ momenta of the fermions. There is no $\perp$ direction in the momentum integral, so $p_\perp^\alpha p_\perp^\beta \to g_\perp^{\alpha \beta}$ on angular averaging. As a result, the $\lambda^3$ operators have the same structure as Eq.~\eqref{13.4}, so $F_1=g_1$ also at order $\lambda^4$.

Summarizing the above results, we find at one-loop order as $N \to \infty$ (up to $\ln \Nb$ powers in the higher order terms):
\begin{align}
C_{1q,N} &\to 1 + \frac{\alpha_s}{4 \pi}  C_F \left[2 \ln^2\Nb + 3  \ln \Nb
-\frac{\pi^2}{3} - 9 \right] \,, \nn
C_{\Delta q, N} - C_{1q,N} & \sim \mathcal{O}\left(\frac{1}{N^2} \right) \,, \nn
C_{Lq,N} &\to \frac{\alpha_s}{4 \pi}  \frac{4 C_F}{N} \,, \nn
C_{1g,N} &\to  \frac{\alpha_s}{4 \pi} 4 T_F \frac{1}{N} \left[- \ln \Nb -1 \right] \,, \nn
C_{\Delta g, N} - C_{1g,N} & \sim \mathcal{O}\left(\frac{1}{N^3} \right) \,, \nn
C_{Lg,N} & \sim \mathcal{O}\left(\frac{1}{N^2} \right) \,.
\label{11.7}
\end{align}
These agree with the known results~\cite{DeRujula:1976baf,Bardeen:1978yd,Ahmed:1976ee,Kodaira:1978sh,Kodaira:1979ib,Kodaira:1979pa,Blumlein:2012bf} as $N \to \infty$.

\begin{figure}
\begin{center}
\includegraphics[width=5cm]{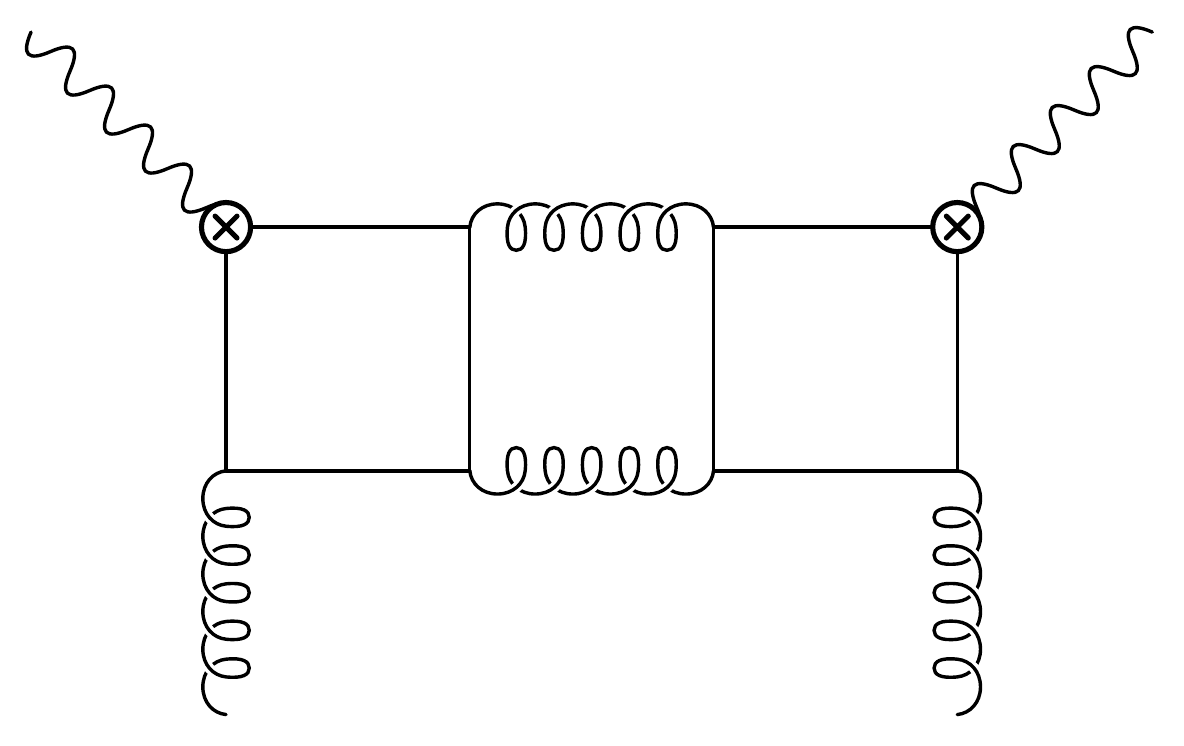}
\end{center}
\caption{\label{fig:12} Three-loop contribution to the gluon coefficient. The color factor is proportional to $d_{abc}d_{abc} n_F^2/N_A$, where $N_A$ is the dimension of the adjoint. }
\end{figure}

\begin{figure}
\begin{center}
\includegraphics[width=4cm]{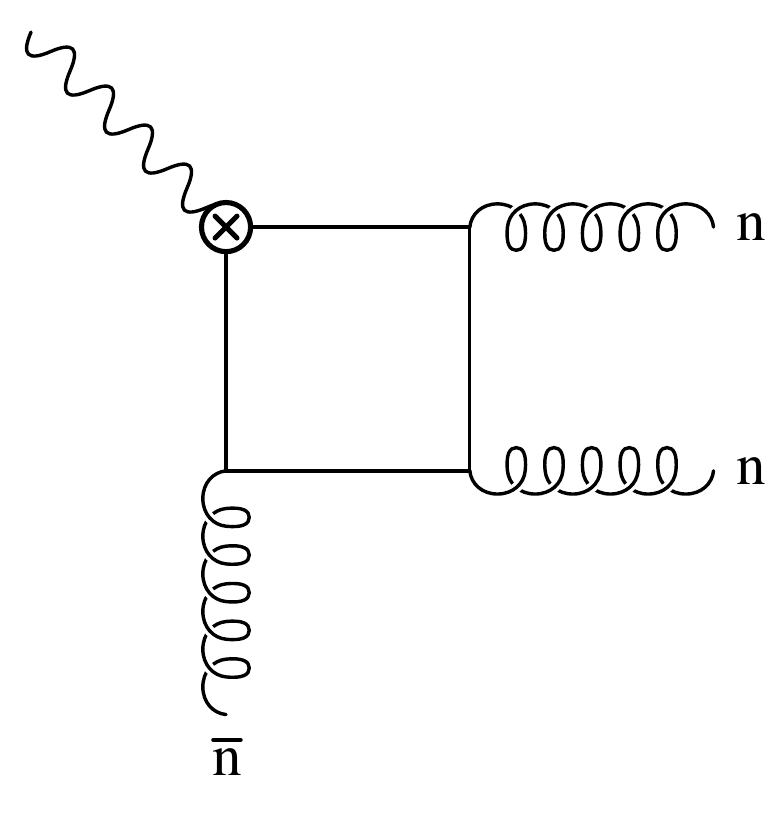}
\end{center}
\caption{\label{fig:13} One-loop contribution to the order $\lambda$ expansion of the current $j^\mu$. The color factor is proportional to $n_F d_{abc}$.}
\end{figure}

The above arguments can be extended to higher orders in perturbation theory.\footnote{We thank J.~Bl\"umlein for suggesting we should look at the higher order corrections, for providing {\tt Mathematica} notebooks with the QCD coefficient functions and anomalous dimensions to three-loop order, and for answering our questions. The QCD results are taken from~\cite{Larin:1991tj,Mertig:1995ny,Vogelsang:1995vh,Larin:1996wd,Moch:2004pa,Vogt:2004mw,Moch:2014sna,Blumlein:2021ryt,Blumlein:2021enk,Blumlein:2022gpp}. The $N \to \infty$ expansions were done with the {\tt HarmonicSums} package~\cite{Ablinger:2009ovq,Ablinger:2012ufz,Ablinger:2017tan}.}
\begin{figure}
\begin{center}
\includegraphics[width=4cm]{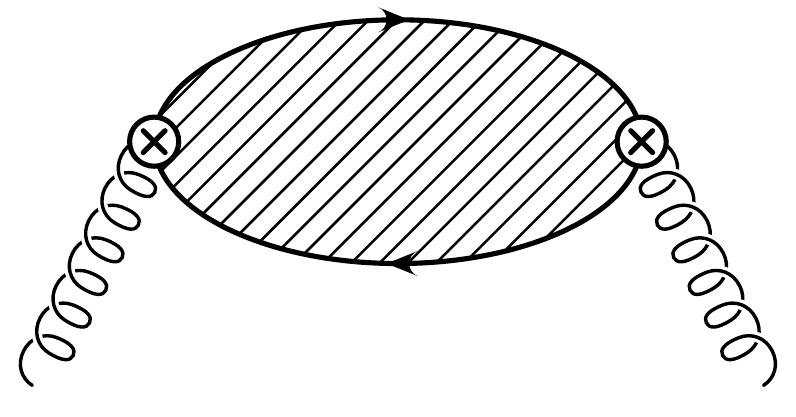}
\end{center}
\caption{\label{fig:11} Gluon coefficient functions from the order $\lambda$ operators $O^{(1D)\mu}$. The shaded blob represents arbitrary QCD corrections.}
\end{figure}
Consider radiative corrections to the gluon coefficient from $O^{\dagger(1D)\mu_\perp}  \ O^{(1D) \nu_\perp}$ at higher orders, as shown in Fig.~\ref{fig:11}. The final structure after evaluating the loop integrals must still be of the form Eq.~\eqref{13.6}, since the graph can only depend on $n$ and $\nb$. Possible $\gamma$-matrix contractions from internal gluons, such as
\begin{align}
\gamma^\tau \gamma_\perp^\mu \frac{\slashed{\nb}}{2} \slashed{\varepsilon}_\perp \gamma_\tau  &= - \frac{\slashed{\nb}}{2}
\left( (2-d) \varepsilon_\perp^\mu - i (d-6)\sigma^{\mu_\perp \alpha} \varepsilon^\perp_\alpha \right)\,,
\label{13.7}
\end{align}
do not change the correlation between $\varepsilon$, $\varepsilon^*$, $\mu$ and $\nu$, and the graph still gives $F_1=g_1$. The first place where $F_1=g_1$ might be violated is the graph Fig.~\ref{fig:12}.
The fermion loop can be hard, so that Fig.~\ref{fig:13}
generates a one-loop contribution of order $\lambda$ to the current $j^\mu$,
\begin{align}
j^\mu \propto \alpha_s\ n_F\, d_{abc}\, \mathcal{B}^{a \alpha_\perp}_\nb \,  \mathcal{B}^b_{n \alpha_\perp} \, \mathcal{B}^{c \mu_\perp}_n \,.
\label{13.8}
\end{align}
Fig.~\ref{fig:12} then has a contribution from a loop graph with two insertions of Eq.~\eqref{13.8} where the intermediate two-gluon state has invariant mass $M_J$, which can break the correlation between gluon polarizations and $\mu,\nu$. Thus, the first violation of $F_1=g_1$ in the gluon coefficients is at order $\alpha_s^3 n_F^2 d_{abc}d_{abc}/N_A  \times 1/N$, where $N_A$ is the dimension of the adjoint.

\begin{figure}
\begin{center}
\includegraphics[width=5cm]{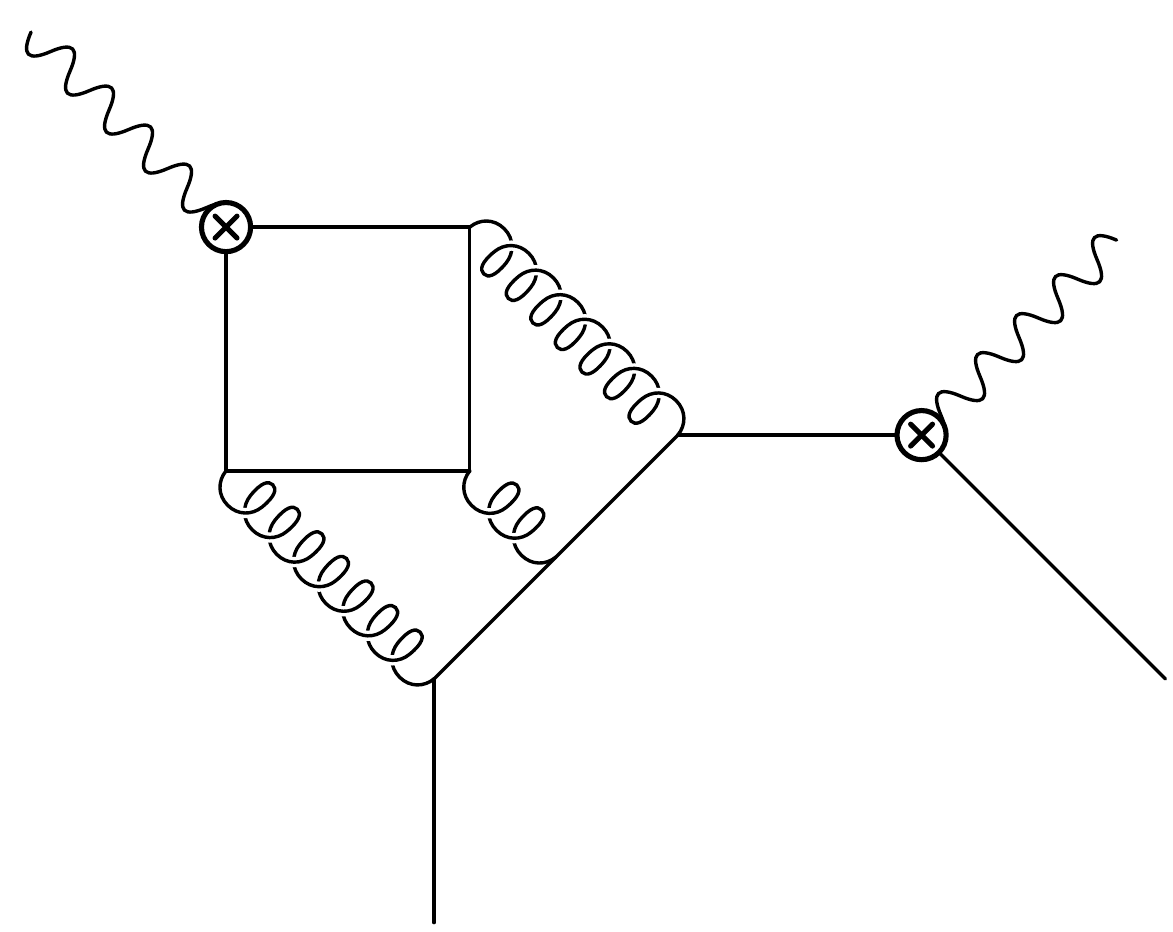}
\end{center}
\caption{\label{fig:14} Three-loop contribution to the quark coefficient. The color factor is proportional to $n_F d_{abc}d_{abc}/N_c$.}
\end{figure}

\begin{figure}
\begin{center}
\includegraphics[width=4cm]{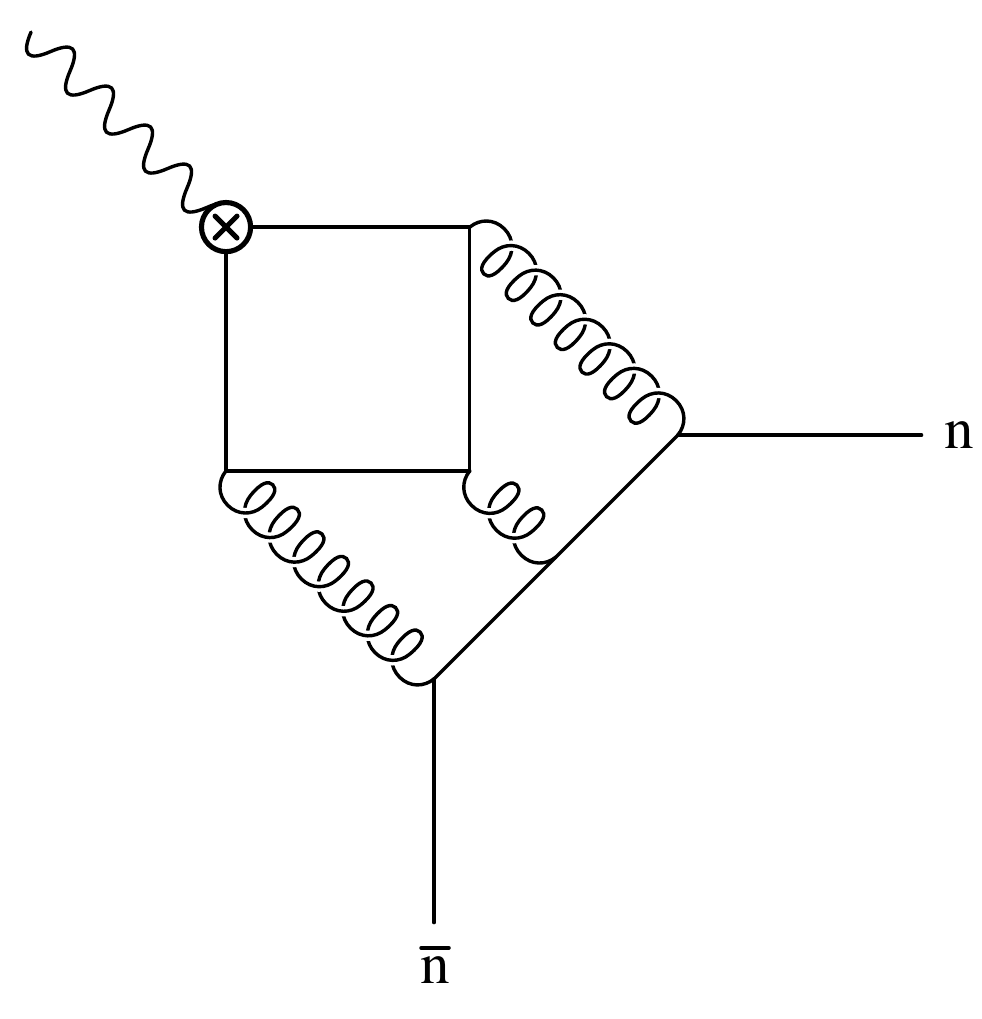}
\end{center}
\caption{\label{fig:15}  Three-loop contribution to the order $\lambda$ expansion of the current $j^\mu$. The color factor is proportional to $n_F d_{abc}d_{abc}/N_c$.}
\end{figure}

There are also $d_{abc}d_{abc}$ contributions to the quark coefficient at order $\alpha_s^3$ from the graph in Fig.~\ref{fig:14}.
In this case, the hard three-loop graph shown in Fig.~\ref{fig:15} generates the operator
\begin{align}
j^\mu \propto \alpha_s^3\ n_F \frac{d_{abc}d_{abc} }{N_c} \, \overline \chi_n \slashed{n} \chi_{\nb}\,.
\label{13.9}
\end{align}
where $N_c$ is the number of colors.
This is an $\alpha_s^3$ correction to the leading order $\lambda^0$ operator, and does not change the $\gamma$-matrix structure, so $F_1=g_1$.

The PDF power counting discussed in Appendix~\ref{app:PDF} shows that there are no additional operators other than Eq.~\eqref{13.9} at the powers of $\lambda$  that we need for our analysis. The only effect of higher order corrections in $\alpha_s$ is to generate radiative corrections to coefficients of  operators we have already considered, as from Fig.~\ref{fig:15}.

In summary, to order $\alpha_s^3$, we find (up to $\ln \Nb$ terms)
\begin{align}
C_{1q,N} &\sim \mathcal{O}\left(1 \right) \nn
C_{\Delta q, N} - C_{1q,N} & \sim \mathcal{O}\left(\frac{1}{N^2} \right) \,, \nn
C_{Lq,N} &\sim\mathcal{O}\left(\frac{1}{N} \right) \,, \nn
C_{1g,N} &\sim \mathcal{O}\left(\frac{1}{N} \right) \,, \nn
C_{\Delta g, N} - C_{1g,N} & \sim \mathcal{O}\left(\frac{1}{N^3} \right) +\frac{\alpha_s^3 d_{abc}d_{abc}}{N_A}  \mathcal{O}\left(\frac{1}{N} \right)\,, \nn
C_{Lg,N} & \sim \mathcal{O}\left(\frac{1}{N^2} \right) \,.
\label{13.10}
\end{align}
We have checked explicitly that these relations are satisfied, using the known three-loop QCD coefficients~\cite{Larin:1991tj,Larin:1996wd,Moch:2004pa,Vogt:2004mw,Moch:2014sna,Blumlein:2021ryt,Blumlein:2021enk,Blumlein:2022gpp}. These relations are expected to hold to all orders (where the $d_{abc}d_{abc}$ term can have contributions beyond $\alpha_s^3$), since the quark coefficients $C_{1q}$ and $C_{\Delta q}$ at order $\lambda^2$ are from the interference between the order $\lambda^0$ and $\lambda^2$ currents, and so maintain the relation $F_1=g_1$.
For $SU(N)$ gauge theories, there are symmetric invariants with more indices, $d_{a_1 \cdots a_r}$ with $r \le N$, which give operators with $r$ $\mathcal{B}_{\nb}^\alpha$ fields, and contribute to $C_{\Delta g, N} - C_{1g,N} $ at order $\alpha_s^r d_{a_1 \cdots a_r}d_{a_1 \cdots a_r}/N^{r-2}$.

The relations Eq.~\eqref{13.10} have to be consistent with the RGE equations for PDF evolution. In the singlet channel,
the RGE for $F_1$ is (with $-\gamma$ the moments of the evolution kernel $P$, as is standard notation in the literature)
\begin{align}
\mu \frac{\rd}{\rd \mu} \begin{bmatrix} C_{1Q} & C_{1G} \end{bmatrix} &= \begin{bmatrix} C_{1Q} & C_{1G} \end{bmatrix}   \begin{bmatrix} \gamma_{QQ} & \gamma_{QG} \\
\gamma_{GQ} & \gamma_{GG} \end{bmatrix} \,,
\label{13.11}
\end{align}
with a similar equation for $C_{Lq}$ and $C_{Lg}$ for $F_L$. The RGE in the polarized case can be written as
\begin{align}
 \mu \frac{\rd}{\rd \mu} \begin{bmatrix} C_{\Delta Q} -C_{1Q}  & C_{\Delta G} -C_{1G} \end{bmatrix} &= \begin{bmatrix} C_{\Delta Q} -C_{1Q}  & C_{\Delta G} -C_{1G} \end{bmatrix} \begin{bmatrix} \Delta \gamma_{ Q Q} & \Delta \gamma_{Q G} \\
\Delta\gamma_{GQ} &\Delta \gamma_{GG} \end{bmatrix} \nn
& +  \begin{bmatrix} C_{1Q} & C_{1G} \end{bmatrix}   \begin{bmatrix} \Delta \gamma_{ Q Q} -  \gamma_{ Q Q}  & \Delta \gamma_{Q G} - \gamma_{Q G}  \\
\Delta\gamma_{GQ} - \gamma_{GQ}  &\Delta \gamma_{GG}-\gamma_{GG} \end{bmatrix} \,,
\label{13.12}
\end{align}
where $\Delta \gamma$ are the polarized anomalous dimensions.
Requiring that Eq.~\eqref{13.10} is consistent with Eq.~\eqref{13.11}, Eq.~\eqref{13.12} gives
\begin{align}
\gamma_{Q G} &   \sim \mathcal{O}\left(\frac{1}{N} \right) \,, \nn
\Delta \gamma_{ Q Q} -  \gamma_{ Q Q} & \sim \mathcal{O}\left(\frac{1}{N^2} \right) \,, \nn
\Delta \gamma_{Q G} - \gamma_{Q G} & \sim \mathcal{O}\left(\frac{1}{N^3} \right) \,, \nn
\Delta\gamma_{GQ} - \gamma_{GQ} & \sim \mathcal{O}\left(\frac{1}{N} \right) \,,\nn
\Delta \gamma_{GG}-\gamma_{GG} &  \sim \mathcal{O}\left(\frac{1}{N^2} \right) \,.
\label{13.13}
\end{align}
The $\gamma_{QQ}$ relations also hold in the non-singlet sector, since the RGE is given by deleting the gluon contribution in the above equations.
We have checked explicitly that these relations are satisfied, using the known QCD anomalous dimensions to three-loop order~\cite{Larin:1991tj,Mertig:1995ny,Vogelsang:1995vh,Larin:1996wd,Moch:2004pa,Vogt:2004mw,Moch:2014sna,Blumlein:2021ryt,Blumlein:2021enk,Blumlein:2022gpp}. They are expected to hold to all orders to maintain the structure of Eq.~\eqref{13.10}. The relations Eq.~\eqref{13.13} are the $N$-space version of relations already noted in~\cite{Moch:2014sna}.

\section{ \texorpdfstring{$g_2$: QCD vs SCET}{g2: QCD vs SCET}} \label{sec:qcd_vs_scet}

The QCD analysis of $g_2$ uses the trilocal quark-gluon operators
\begin{align}
\overline \psi (n \xi_2) \left[ \pm i g G_{n \sigma} (0) \slashed{n}\gamma_5+ g \widetilde G_{n \sigma}(0) 
\slashed{n} \right] \psi(n \xi_1) \,,
\label{8.1}
\end{align}
with proton matrix elements giving the PDF distribution $f(x_1,x_2)$ as defined in  Eq.~\eqref{2.35}, Eq.~\eqref{2.36}. The series expansion of the operators in $\xi_1$ and $\xi_2$ using
\begin{align}\label{8.1a}
    \psi(n\xi) = \sum_{m=0}^\infty \frac{\xi^m}{m!}(n \cdot D)^m\psi(0) \,,
\end{align}
generates the operators in Eq.~\eqref{2.30} in the expression for the moments of $g_2$.

It was shown previously that the QCD anomalous dimension of $g_2$ generated by tree-level matching simplified as $x \to 1$ into a single variable anomalous dimension with kernel~\cite{Ali:1991em}
 \begin{align} \label{4.32}
    P_{g_2}(z) = \frac{\alpha_s}{4\pi} \left\{ 4C_F \left[ \frac{2}{(1-z)_+} + \frac{3}{2}\delta(1-z) \right] - 2C_A \delta(1-z) \right\} \,.
\end{align}
In our SCET analysis in Section~\ref{sec:matchingJ}, we found that the OPE generated a PDF where $\xi_1 \sim 1/Q$ and $\xi_2 \sim 1/M_J$ or $\xi_2 \sim 1/Q$ and $\xi_1 \sim 1/M_J$. The PDF evolution kernel in SCET involves scales $1/M_J$, so the operator Eq.~\eqref{8.1} either has $\xi_1 \to 0$ or $\xi_2 \to 0$, and turns into a single variable object with a single variable kernel, naturally explaining the simplification found in~\cite{Ali:1991em,Geyer:1996isa}.

In SCET, the subleading matching coefficient at tree-level is $C^{(1B)}(u)=1$, and the anomalous dimension integrated with the tree-level coefficient is
\begin{align} \label{4.33}
    \int_0^1 {\rm{d}}u \, \gamma(z;u,v) = P_{g_2}(z) - \frac{\alpha_s}{4\pi}\left(4C_F - 2C_A\right) \,  \delta(1-z) \left[ 1 + \frac{\log(1-v)}{v} \right] \,.
\end{align}
This differs from Eq.~\eqref{4.32} by the second term, which only vanishes as $v \to 0$.\footnote{
Ref.~\cite{Ali:1991em} determines the $Q^2$ evolution of $g_2$ in Eq.~\eqref{4.32} for the lowest eigenvalue of the anomalous dimension matrix. The corresponding eigenvector is proportional to the tree-level matching Eq.~\eqref{6.6}, and corresponds to assuming the PDF $h_q(x,v)$ is proportional to $\delta(v)$. This assumption is only valid for asymptotically large $Q^2$, which is an idealization and not relevant for experiments. In this limit $g_1(x)$ would be proportional to $\delta(x)$, since only its first moment would be non-zero.}
 We have checked in Section~\ref{sec:consistency} that our anomalous dimensions are consistent with the matching coefficients, which require the second term in Eq.~\eqref{4.33}.
We can see the above simplification more directly. The $\nb$ collinear sector below the jet scale is the same as full QCD, since there is only a single collinear sector. The twist-three quark-gluon operators in Eq.~\eqref{2.30} can be written as $\nb$-sector operators
\begin{align}
V_{r,s}^\sigma &= i g  \overline \chi_\nb   (i n \cdot \partial)^{r-1} [W_\nb^\dagger \, G^{\sigma \alpha}\, n_\alpha\, W_\nb]  (i n \cdot \partial)^{s-1} \slashed{n} \gamma_5   \chi_\nb \,, \nn
U_{r,s}^\sigma &= - \epsilon_\perp^{\sigma \beta} \, g  \overline \chi_\nb   (i n \cdot \partial)^{r-1} [W_\nb^\dagger \,    G_{\beta \alpha}\,  n^\alpha\, W_\nb]  (i n \cdot \partial)^{s-1}    \slashed{n}  \chi_\nb \,,
\label{6.1}
\end{align}
where we have used $\widetilde G^{\sigma \alpha} n_\alpha = - \epsilon_\perp^{\sigma \beta} G_{\beta \alpha}\,  n^\alpha$. The collinear Wilson lines $W_{\bar n}$ (which are in the $n$ direction) convert the covariant derivatives into ordinary derivatives. We introduce analogous operators involving the $\mathcal{B}_{\bar n}$ field (note the sign convention):
\begin{align}
V_{r,s}^{(\mathcal{B})\sigma} &=    \overline \chi_\nb   (i n \cdot \partial)^{r-1} \mathcal{B}_\nb^\sigma  (i n \cdot \partial)^{s-1} \slashed{n} \gamma_5   \chi_\nb \,, \nn
U_{r,s}^{(\mathcal{B})\sigma} &= i \epsilon_\perp^{\sigma \beta} \,   \overline \chi_\nb   (i n \cdot \partial)^{r-1} \mathcal{B}_\nb^\beta  (i n \cdot \partial)^{s-1}    \slashed{n}  \chi_\nb \,.
\label{6.2}
\end{align}
The identity
\begin{align}
[n \cdot \partial, \mathcal{B}_\nb^\sigma ] =W_\nb^\dagger\,   g  G^{\sigma \alpha} n_\alpha  W_\nb
\label{6.3}
\end{align}
gives
\begin{align}
V_{r,s}^\sigma &= V_{r+1,s}^{(\mathcal{B})\sigma} - V_{r,s+1}^{(\mathcal{B})\sigma}\,, &
U_{r,s}^\sigma &= U_{r+1,s}^{(\mathcal{B})\sigma} - U_{r,s+1}^{(\mathcal{B})\sigma}\,.
\label{6.4}
\end{align}
The tree-level twist-three contribution to $g_2$ using the equations of motion is given in Eq.~\eqref{a2.34}. Using Eq.~\eqref{6.4} and dropping the quark mass term, Eq.~\eqref{a2.34} simplifies to
\begin{align}
\frac{N-1}{N} O_{3,q,N}^\sigma &=  -  \frac{1}{2N }  \left[ (N-1) \left( U_{N-1,1}^{(\mathcal{B})\sigma} - U_{1,N-1}^{(\mathcal{B})\sigma} \right)
- (N-3) \left( V_{N-1,1}^{(\mathcal{B})\sigma} + V_{1,N-1}^{(\mathcal{B})\sigma} \right) \right] \nn
& -  \frac{1}{N }  \sum_{r=2}^{N-2}  V_{r,N-r}^{(\mathcal{B})\sigma} \,.
\label{6.7}
\end{align}
As $N \to \infty$, this expression further simplifies to
\begin{align}
\frac{N-1}{N} O_{3,q,N}^\sigma &=  -  \frac{1}{2 }  \left[ \left( U_{N-1,1}^{(\mathcal{B})\sigma} - U_{1,N-1}^{(\mathcal{B})\sigma} \right)
- \left( V_{N-1,1}^{(\mathcal{B})\sigma} + V_{1,N-1}^{(\mathcal{B})\sigma} \right) \right] \,.
\label{6.6}
\end{align}
The only operators that survive are those with derivatives on \emph{one} side of $\mathcal{B}$ in Eq.~\eqref{6.2}, which is the moment space version of $\xi_1=0$ or $\xi_2=0$, as shown schematically in Fig.~\ref{fig:qcd_vs_scet}. The simplification occurs because terms cancel when using Eq.~\eqref{6.4}, and only the endpoint terms where $r=1$ or $s=1$ survive. This cancellation uses operators with $\mathcal{B}_\nb^\sigma$, which acts like a gauge field, and whose derivative gives the field-strength tensor $G^{\sigma \alpha}n_\alpha$, Eq.~\eqref{6.3}. $\mathcal{B}_\nb^\sigma$ is a local operator in SCET below the scale $Q$, but is non-local in QCD. The SCET expansion generates $O^{(1B)\mu}$ operators where $\chi_{\overline n}$ and $\mathcal{B}^\sigma_{\overline n}$ are at the same point, so it automatically produces operators of the form Eq.~\eqref{6.2} with $r=1$ or $s=1$, which are the operators in Eq.~\eqref{6.6}.

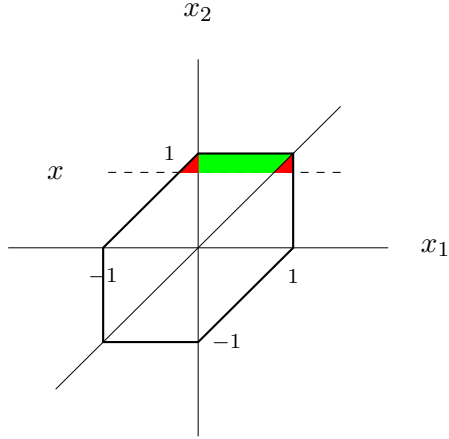
\begin{figure}
\begin{center}
\begin{tikzpicture}[scale=1.25]
\draw[very thin] (-2,0) -- (2,0);
\draw[very thin] (0,-2) -- (0,2);
\draw (2.5,0) node {$x_1$};
\draw (0,2.5,0) node {$x_2$};
\draw[dashed] (1.5,0.8) -- (-1,0.8);
\filldraw[green] (0.8,0.8) -- (1,1) -- (0,1) -- (0,0.8)  -- cycle;
\filldraw[red] (0,0.8) -- (-0.2,0.8) -- (0,1) -- cycle;
\filldraw[red] (0.8,0.8) -- (1,1) -- (1,0.8) -- cycle;
\draw[thick] (-1,0) -- (0,1) -- (1,1) -- (1,0) -- (0,-1) -- (-1,-1) -- cycle;
\draw[very thin] (-1.5,-1.5) -- (1.5,1.5);

\draw (1,-0.3) node {$\scriptstyle 1$};
\draw (-1,-0.3) node {$\scriptstyle -1$};
\draw (-0.3,1) node {$\scriptstyle 1$};
\draw (0.3,-1) node {$\scriptstyle -1$};
\draw(-1.5,0.8) node {$x$};
\end{tikzpicture}
\end{center}
\caption{\label{fig:region} The region (thick black outline) in the $x_1-x_2$ plane where the trilocal PDFs are non-vanishing. The green region is the contribution to $g_2(x)$ due to the PDF with an incoming quark and gluon, and an outgoing quark. }
\end{figure}

We can see the emergence of bilocal operators from the trilocal lightcone PDF distributions in Eq.~\eqref{8.1}.
\begin{figure}
\centering
\includegraphics[width=6cm]{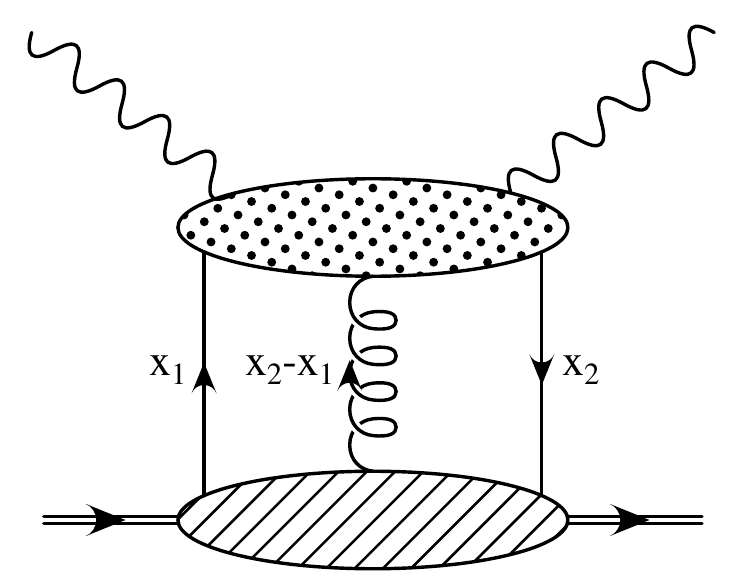}
\caption{\label{fig:twist3} Twist three quark-gluon operator contribution to DIS. The upper blob is the hard interaction, and the lower blob is the proton matrix element. $x_1$ is the incoming momentum fraction of the quark and $x_2$ is the outgoing momentum fraction of the quark, both w.r.t. the upper blob. The gluon is incoming if $x_2 - x_1 >0$ and outgoing if $x_2-x_1 < 0$.}
\end{figure}
The PDF distribution shown in Fig.~\ref{fig:twist3} has momentum fractions $-1 \le x_1 \le 1$, $-1 \le x_2 \le 1$ and $-1 \le x_2-x_1 \le 1$. The sign of $x_2-x_1$ determines whether the gluon is incoming or outgoing. The sign of $x_1$ determines whether we have an incoming quark or outgoing antiquark, and similarly for $x_2$. Here, incoming or outgoing is defined w.r.t.\ the hard interaction blob.  A detailed discussion of the interpretation of PDF distributions is given in~\cite{Jaffe:1983hp}. The allowed region in the $x_1-x_2$ plane is shown in Fig.~\ref{fig:region}.
If the matrix element has $x_1>0$, $x_2-x_1>0$ and $x_2>0$, then the allowed integration region for $g_2(x)$ is  $x \le x_2 \le 1$, $x_1 >0$ and $x_2-x_1>0$, shown in green in Fig.~\ref{fig:region}. We can add in the triangular areas shown in red in Fig.~\ref{fig:region}. These regions have an area of order $(1-x)^2$, and vanish relative to the green region with area of order $1-x$. The $x_1$ integral can be extended to $[-\infty,\infty]$ since the PDF vanishes outside the black outlined region shown in the figure. This sets the lightcone separation $\xi_1$ to zero, and the operator becomes bilocal. Similarly, one can work through the other possible signs for $x_1$, $x_2$, and $x_2-x_1$. In each case, one gets a region like the green region in the figure, or its counterparts along the edges $x_1=\pm 1$ or $x_2=-1$, and the operator becomes bilocal because either $\xi_1=0$ or $\xi_2=0$. Each bilocal operator can be identified with one of $H_q(x,u), \, H_q^\dagger(x,u), \, H_{\overline{q}}(x,u)$ or $H_{\overline{q}}^\dagger(x,u)$ based on the incoming and outgoing assignments of the particles.

For completeness, we have verified the anomalous dimensions of the twist-three QCD operators. The calculation of the RGE for the QCD operators simplifies if one uses the $\gamma$-matrix structure
\begin{align}
Y^\sigma_{r,s} &=  i g\, \overline \psi (i n \cdot D)^{r-1}  \gamma_\perp^\sigma \frac{\slashed{n}}{2} \gamma_\perp^\tau G_{\tau  \alpha } n^\alpha (in \cdot D)^{s-1}\,
\gamma_5\, \psi \,,
\label{6.11}
\end{align}
with $N=r+s+1$
instead of the QCD operators Eq.~\eqref{2.30}. These operators have the same $\gamma$-matrix structure as the $H^\mu_q$ operators Eq.~\eqref{4.50} in the OPE for $g_2$.
Using Eq.~\eqref{4.11} gives
\begin{align}
Y^\sigma_{r,s} &= - \frac12 \left( V^\sigma_{r,s} + U^\sigma_{r,s} \right)\,,
\label{6.12}
\end{align}
with hermitian (and charge) conjugate
\begin{align}
Y^{\sigma \dagger}_{r,s} &= (-1)^N \frac12 \left(- V^\sigma_{s,r} + U^\sigma_{s,r} \right)
=\mathcal{C} Y^\sigma_{r,s} \mathcal{C}^{-1} \,,
\label{6.13}
\end{align}
so $\{ V^\sigma_{r,s},\ U^\sigma_{r,s}\}$ can be obtained from linear combinations of $\{ Y^\sigma_{r,s},\ Y^{\sigma \dagger}_{r,s}\} $. DIS involves the charge-conjugate even combination
\begin{align}
Y^\sigma_{r,s}  + \mathcal{C} Y^\sigma_{r,s} \mathcal{C}^{-1}  &= 
Y^\sigma_{r,s}  + Y^{\sigma \dagger}_{r,s} = - \frac12 \left( V^\sigma_{r,s} + U^\sigma_{r,s} \right) + (-1)^N \frac12 \left(- V^\sigma_{s,r} + U^\sigma_{s,r} \right) \,,
\end{align}
which for odd $N$ is the linear combination that occurs in the OPE for $g_2$. The RGE for $Y^\sigma_{r,s} $ is the same as
for the charge-conjugate operators $\mathcal{C} Y^\sigma_{r,s} \mathcal{C}^{-1}$, so the $\{V^\sigma_{r,s}\}$ and $\{U^\sigma_{r,s}\}$
RGE can be obtained from that for $\{Y^\sigma_{r,s} \}$ using Eq.~\eqref{6.12} and Eq.~\eqref{6.13}.

The RGE for $Y^\sigma_{r,s} $ is
\begin{align}
\mu \frac{\rd}{\rd \mu} Y_{r,s}^\sigma &= \frac{\alpha_s}{2\pi} \Biggl[ \frac{2 C_F}{s (s+1)(s+2)} M^\sigma_{n+1} \nn
&+ Y_{r,s}^\sigma  \biggl\{  C_F \left(3 - 2 H_r -2  H_s \right) +  \left(2 C_F - C_A \right) \left( \frac{2 (-1)^s}{s (s+1) (s+2)} - \frac{(-1)^{r}}{r+1} +  \frac1n  \right) \nn
&+ C_A \left( \frac{1}{s}-\frac{1}{s+1}-\frac{1}{s+2} -\frac{1}{r+1}  - H_r -  H_s \right) \biggr\} \nn
&+ \sum_{m=1}^{r-1} Y_{r-m,s+m}^\sigma \biggl\{  \left(2 C_F - C_A \right) (-1)^m \left( \frac{ \binom{ n-1 }{s+m} } {\binom{n-1}{s}}\frac{n+m}{nm}
- \underbrace{ (-1)^r }  \frac{ \binom{ r-1 }{ m } } {r+1}\right) \nn 
&\hspace{8.5cm} + C_A \frac{(r-m)(r-m+1)}{ r(r+1) m}  \biggr\} \nn
&+ \sum_{m=1}^{s-1} Y_{r+m,s-m}^\sigma \biggl\{   \left(2 C_F  - C_A \right) (-1)^m \left( \frac{ \binom{ n-1 }{r+m} } {\binom{n-1}{r}}\frac{n+m}{nm} 
 + 2 \underbrace{  (-1)^s  }  \frac{ \binom{s}{ m} } { s (s+1)(s+2)}\right) \nn
& + C_A \frac{(s-m+1)(s-m+2)}{(s+1) (s+2)m} \biggr\} \Biggr] \,,
\label{6.14}
\end{align}
where $n=r+s$, $\binom{a}{b}$ are binomial coefficients and the mass operator is
\begin{align}
M^\sigma_{n+1} &=  m \overline \psi   (in \cdot D)^{n-1} (i \sigma^{\sigma n} \gamma_5)  \psi  \,.
\label{6.15}
\end{align}
These operators contribute to the $(n+1)^\text{th}$ moment.

The RGE Eq.~\eqref{6.14} has been computed previously in~\cite{Bukhvostov:1983eob,Ratcliffe:1985mp,Ji:1990br}. We agree with~\cite[(27)]{Ji:1990br} up to the exchange $(-1)^r \leftrightarrow (-1)^s$ of the two terms denoted by an underbrace. Since the results for $g_2$ have $r+s=n=N-1$ even, $(-1)^r = (-1)^s$ and the RGE for $g_2$ is the same as~\cite[(27)]{Ji:1990br}.
The first term in the RGE involving the mass operator differs from~\cite{Ji:1990br} by a factor of two because of a corresponding difference in the normalization of the mass operator. There are  small differences with the results of~\cite{Bukhvostov:1983eob,Ratcliffe:1985mp} which are explained in~\cite{Ji:1990br}.

\section{Conclusion} \label{sec:conclusions}

We have resummed the Sudakov double logarithms $[\alpha_s \ln(1-x)]^n$ in the structure functions $g_1$ and $g_2$ of  polarized DIS using SCET. $g_2$ is given by a quark-gluon PDF distribution $h_q(x,u)$ where $x$ is a parton momentum fraction, and $u$ is a SCET momentum fraction label. We computed the anomalous dimension of the quark-gluon PDF operator. These results give the leading order resummation for $g_2$ including the full $\alpha_s$ correction as $x \to 1$.

We computed the one-loop matching coefficients for the order $\lambda$ operators in the matching of the current onto SCET. These results can also be used for one-loop and next-to-leading log analyses of other hard QCD scattering processes such as Drell-Yan or dijet production.

The extension of our results to higher order in $\alpha_s$ has no conceptual difficulties, though the calculation will be significantly more complicated than the results presented here. Power corrections lead to terms suppressed by $1-x$ or $M_p^2/Q^2$.
Power corrections to the structure functions beyond the leading contribution computed here, which have additional factors of $1-x$, can be computed using results derived for $F_1$ in~\cite{Luke:2022ops} and the arguments in Section~\ref{sec:coeff}. Power corrections which lead to higher twist PDFs with additional partons have $1/u$ endpoint singularities in their coefficient function, which are not present in the leading term. The treatment of these endpoint divergences is non-trivial, and is an active area of investigation, e.g.~Ref.~\cite{Beneke:2020ibj}.


\acknowledgments

We thank J.~Bl\"umlein, Matthew Inglis-Whalen, Julie Pag\`es and Iain Stewart for helpful discussions,
and J.~Bl\"umlein for {\tt Mathematica} files with the QCD formul\ae. We also thank Alamjeet Singh for help in comparing our results to the 3-loop QCD results in the literature. JSG acknowledges the support of the Natural Sciences and Engineering Research Council of Canada (NSERC), [PGS-D 587640]. JSG also thanks the Galileo Galilei Institute for Theoretical Physics, Florence and the Institut d'Etudes Scientifiques de Carg\`ese for their hospitality during the completion of this work. This work was partially supported by
the U.S. Department of Energy (DOE) award number DE-SC0009919. JR is supported by the U.S. Department of Energy under grant contract number DE-FG02-05ER41367 and by the DOE Quark-Gluon Tomography Topical Collaboration under award number DE-SC0023646.

\appendix

\section{Properties of Twist-3 Operators}\label{app:ops}

The operators in Eq.~\eqref{2.30} have the hermitian conjugates ($N=r+s+1$)
\begin{align}
(V_{N,r,s}^\sigma) ^\dagger &= (-1)^{N} V_{N,s,r}^\sigma \,, &
(U_{N,r,s}^\sigma) ^\dagger  &=  (-1)(-1)^{N}    U_{N,s,r}^\sigma \,,
\label{a1.18}
\end{align}
and transform under charge conjugation as
\begin{align}
V_{N,r,s}^\sigma & \to (-1)^{N}  V_{N,s,r}^\sigma  \,, &
U_{N,r,s}^\sigma & \to (-1)(-1)^{N}  U_{N,s,r}^\sigma \,.
\label{a2.31}
\end{align}
Under $PT$,
\begin{align}
V_{N,r,s}^\sigma & \to V_{N,r,s}^\sigma \,, &
U_{N,r,s}^\sigma  & \to   U_{N,r,s}^\sigma \,.
\label{a1.17}
\end{align}

The moment formul\ae\ for $g_2$ involve charge-conjugation even operators with odd $N$, so the hermitian and charge conjugation even combinations which enter DIS are
\begin{align}
V_{N,r,s}^\sigma -V_{N,s,r}^\sigma \qquad U_{N,r,s}^\sigma+U_{N,s,r}^\sigma &
\label{a2.32}
\end{align}
If $N=2k+1$ so that $r+s=2k$, the charge-conjugation even combinations are $V_{N,r,s}^\sigma -V_{N,s,r}^\sigma$ for $1 \le r < k$ and $U_{N,r,s}^\sigma+U_{N,s,r}^\sigma$ for $1 \le r \le k$ for a total of $2k-1$ independent operators. 

\section{Equation of Motion}\label{app:eom}

The equation of motion relation between the twist-three quark operators and the quark-gluon operators is
\begin{align}
O_{3,q,N}^\sigma &=M^\sigma_N  -  \frac{1}{2(N-1)}  \sum_{r=1}^{N-2} (N-r-1) \left( U^\sigma_{r,N-r-1} + U^\sigma_{N-r-1,r}+  V^\sigma_{r,N-r-1} - V^\sigma_{N-r-1,r} \right) \,,
\label{a2.34}
\end{align}
where
\begin{align}
M^\sigma_N &=  m \overline \psi   (in \cdot D)^{N-2} (i \sigma^{\sigma n} \gamma_5)  \psi  \,,
\label{a2.35}
\end{align}
and $m$ is the quark mass. This identity can be obtained using $(i \slashed{D}-m) \psi=0$. Eq.~\eqref{a2.34} agrees with~\cite[(5)]{Kodaira:1994ge} up to the sign of the second term, and keeping only the hermitian part of the mass term in~\cite[(5)]{Kodaira:1994ge}.

The anomalous dimension of the mass operator is~\cite{Bukhvostov:1983eob,Ratcliffe:1985mp,Ji:1990br,Kodaira:1994ge}
\begin{align}
 \mu \frac{\rd M^\sigma_N}{\rd \mu} &=  \frac{\alpha_s}{4\pi}C_F\, (- 8\, H_{N-1} )\,  M^\sigma_N \,,
\end{align}
including the anomalous dimension of $m$,
\begin{align}
 \mu \frac{\rd m}{\rd \mu} &=  \frac{\alpha_s}{4\pi} (-6 \, C_F)\, m \,.
\end{align}
We will neglect quark masses in this paper.

\section{Axial Operators in the BMHV Scheme} \label{app:BMHV}

The QCD analysis of $g_1$ requires operators with $\gamma_5$, Eq.~\eqref{2.10}, which arise from the $\gamma$-matrix identity Eq.~\eqref{3.21}. In dimensional regularization, $\gamma_5$ in the BMHV scheme~\cite{tHooft:1972tcz,Breitenlohner:1977hr} gives a consistent definition of $\gamma_5$. Often, the NDR scheme in which $\gamma_5$ anticommutes with all the $\gamma$ matrices is used, but this scheme is known to have inconsistencies.

Renormalization of the axial vector current requires care and was studied in~\cite{TRUEMAN1979331}. The proper definition in dimensional regularization to three-loop order was given in~\cite{Larin:1991tj,Larin:1993tq}. It was shown that to maintain current conservation and the axial Ward identities for the non-singlet axial current, the axial current needs to be defined as
\begin{align}
j^\mu_A  &= Z_5\, \overline \psi t^a \gamma^\mu \gamma_5 \psi \,,
\label{c.1}
\end{align}
where $t^a$ is a flavor matrix, and
\begin{align}
Z_5 &=  1 + \frac{\alpha_s}{4\pi} (-4C_F) \,,
\label{c.2}
\end{align}
is an additional \emph{finite} renormalization. $Z_5$ was computed to two-loop order in~\cite{Larin:1991tj}. A similar analysis for the singlet axial current leads to a $Z_5$ factor which differs from the non-singlet factor starting at two loops~\cite{Larin:1991tj}. In DIS, we need a definition of the axial operators
\begin{align}
O_{q,A,N}  &= Z_{5,N} \, \frac12 \overline \psi \slashed{n}  \gamma_5 (i n \cdot D)^{N-1} \psi \,, \qquad \text{(BMHV)} \,.
\label{c.3}
\end{align}
The matrix elements of $O_{q,A,N}$ give the moments of the polarized quark PDF $f_{\Delta q}(x)$. A natural way to fix $Z_{5,N}$ is to require that the matrix elements of the axial operators in an off-shell quark state are the helicity $h$ of the quark times the matrix element of the corresponding vector operators
\begin{align}
O_{q,V,N}  &=  \frac12 \overline \psi \slashed{n}   (i n \cdot D)^{N-1} \psi \, .
\label{c.4}
\end{align}
This determines
\begin{align}
Z_{5,N} &=  1 + \frac{\alpha_s}{4\pi}  \frac{ -8 C_F}{N(N+1)} \,,
\label{c.5}
\end{align}
which agrees with Eq.~\eqref{c.2} for the axial current, which has $N=1$. The singlet and non-singlet operators have the same $Z_{5,N}$ factor at one loop.  In $x$-space,
\begin{align}
Z_5(x) &=  1 - \frac{\alpha_s}{4\pi}  8 C_F\,(1-x) \,, & Z_{5,N} &= M_N[Z_5(x) ]\,.
\label{c.6}
\end{align}
For large moments, $Z_{5,N} \to 1$ as $N \to \infty$. $Z_{5,N}$ has been computed to order $\alpha_s^3$ in~\cite{Matiounine:1998re,Blumlein:2022gpp}.

If the coefficient functions $C_{\Delta q,N}$ are computed by using on-shell matching, the on-shell matrix element of the operators Eq.~\eqref{c.3} is given by their tree-level value, since loop graphs are scaleless and vanish. This means that the matrix elements have an $\alpha_s$ piece from $Z_{5,N}$, which affects the order $\alpha_s$ contribution to $C_{\Delta q,N}$, and is needed to get the known matching coefficients. One gets the same result using on-shell or off-shell matching in the BMHV scheme, including $Z_{5,N}$.\footnote{$Z_{5,1}$ is the difference between the two results for the first moment of $g_T(x)$ in~\cite{Ji:2000ny}. See (24) therein and the comment below it.}

The matching computation simplifies if one uses NDR with an anticommuting $\gamma_5$, since one does not have to break up the $\gamma$-matrices into four-dimensional and fractional-dimensional pieces.  In the NDR scheme, there is no $Z_{5,N}$ factor in the axial operators. The off-shell quark matrix element of the axial operators Eq.~\eqref{c.3} in NDR is the helicity $h$ times the corresponding vector operators, since $\gamma_5$ can be pulled out of the loop graphs to act on the incoming quark line. The vector operator matrix element does not depend on the $\gamma_5$ scheme, and is the same in BMHV and NDR.  Since in the BMHV scheme, $Z_{5,N}$ is chosen so that the axial operator matrix element is $h$ times the vector operator, the axial operator matrix element is the same in both schemes. The QCD computation only involves vector currents and does not depend on the $\gamma_5$ scheme. As a result, the one-loop off-shell matching
is the same in both schemes. For NDR with on-shell matching, the axial matrix element is unity, since loop graphs don't contribute and there is no $Z_{5,N}$ factor. For BMHV with on-shell matching, the axial matrix element is $Z_{5,N}$. Again, the QCD on-shell calculation is the same in both schemes, since it only involves vector currents. Thus, NDR with on-shell matching gives an incorrect matching coefficient $C_{\Delta q,N}$, since it is missing $Z_{5,N}$.

We have avoided the complications of $\gamma_5$ for $g_2$ by using PDF operators with no $\gamma_5$. The spin $S^\sigma$ arises from the hadronic matrix element of a vector operator, which is proportional to $\epsilon_\perp^{\mu \sigma} S_\sigma$.

\section{Form of PDF Operators} \label{app:PDF}

In this appendix, we derive the general form of PDF operators using SCET. Since PDF operators involve only a single SCET sector, the results also hold for QCD.
The analysis is closely related to that using ``good'' and ``bad'' components of fields in lightcone quantization in lightcone gauge~\cite{Kogut:1969xa,Jaffe:1991ra,Ji:2001bm}.

In SCET in the $\nb$ sector, the momentum $i (n \cdot D)=p^+$ is a \large label momentum of order $\lambda^0$, so that $i (n \cdot D)$  can be inverted. Writing the Dirac equation, Maxwell's equation, and the Bianchi identity
$[D_\alpha ,G_{\beta\gamma} ]+[ D_\beta , G_{\gamma \alpha} ]+ [D_\gamma, G_{\alpha \beta} ]= 0$ in lightcone components, it is easy to see that
\begin{itemize}
\item The fermion field can be decomposed as $\psi = \psi_\nb + \psi_n$, $\psi_\nb = P_\nb \psi$, $\psi_n = P_n \psi$ where $P_\nb$ and $P_n$ are the projectors in Eq.~\eqref{3.1a}, and $\psi_n$ is eliminated using the equation of motion.
$\psi_\nb$ is the ``good'' fermion component and $\psi_n$ is the ``bad'' component, in the terminology of~\cite{Jaffe:1991ra,Ji:2001bm}.

\item The $G^{\alpha \beta} n_\alpha \nb_\beta$, $G^{\alpha_\perp \beta} \nb_\beta$ and $G^{\alpha_\perp \beta_\perp}$ components of the gauge field can be eliminated, leaving only $G^{\alpha_\perp \beta}n_\beta$.

\item $\nb \cdot D$ derivatives acting on $\psi_\nb$ and $G^{\alpha_\perp \beta}n_\beta$ can be eliminated.

\item $i (n \cdot D)$ acting on the fields is a label momentum, and so does not generate a new object. Equivalently, using Eq.~\eqref{8.1a} it can be converted into a lightcone derivative which, on integration by parts, produces a factor of $x P^+$ in PDFs (see e.g.\ Eq.~\eqref{3.20} and Eq.~\eqref{3.13}).

\end{itemize}
The elimination of ``bad'' components in SCET is automatic, since these components of the fields have been integrated out, and do not appear in the EFT Lagrangian.

Since the PDFs are gauge invariant, they are Fourier transforms of lightcone correlators of $\chi_\nb$, $\overline \chi_\nb$, and $W_\nb^\dagger G^{\alpha \beta_\perp} n_\alpha W_\nb$ with possible $D_\perp$ derivatives. Using Eq.~\eqref{6.3},
$W_\nb^\dagger G^{\alpha \beta_\perp} n_\alpha W_\nb$ can be replaced by $\mathcal{B}_\nb^{\beta_\perp}$.
 The identities
 \begin{align}
i \partial_\perp^\alpha\, \chi_\nb &= \partial_\perp^\alpha  \left( W_\nb^\dagger \psi_\nb \right)
  = - \mathcal{B}_\nb^{\alpha_\perp} \chi_\nb + W_\nb^\dagger ( i D_\perp^\alpha \psi_\nb ) \nn
i \partial_\perp^\alpha \, \mathcal{B}_\nb^{\beta_\perp} &= \mathcal{B}_\nb^{\alpha_\perp \beta_\perp} +
[\mathcal{B}_\nb^{\beta_\perp}, \mathcal{B}_\nb^{\alpha_\perp} ]
 \label{13.1}
 \end{align}
 eliminate $W_\nb^\dagger ( i D_\perp^\alpha \psi_\nb)$ and $ \mathcal{B}_\nb^{\alpha_\perp \beta_\perp}$ 
 in terms of $i \partial_\perp^\alpha\, \chi_\nb $ and $i \partial_\perp^\alpha \, \mathcal{B}_\nb^{\beta_\perp} $, so we only need $\chi_\nb$, not $\psi_\nb$, and $\mathcal{B}$ with a single $\perp$ index.
 
 The general PDF is then
\begin{align}
f(x_1,\cdots,x_r) &= \int \rd \xi_1 \cdots \rd \xi_r e^{-i \xi_1 x_1 P^+} \cdots e^{-i \xi_r x_r P^+} \times \nn
& \braket{P,S | A_1(n \xi_1) \ A_2(n \xi_2) \ \cdots A_r(n \xi_r) \ A_{r+1} (0) | P,S}
\label{13.2}
\end{align}
where $A_r$ is one of $\chi_\nb$, $\overline \chi_\nb$ or $\mathcal{B}_\perp^\alpha$ with possible $\partial_\perp$
derivatives, and $\gamma$-matrices have been omitted. Note that not all $\gamma$ matrices are allowed. The allowed non-zero combinations are $\overline \chi_\nb \slashed{n} \chi_\nb $, $\overline \chi_\nb \slashed{n} \gamma_5 \chi_\nb$, $\overline \chi_\nb  \gamma_\perp^\alpha  \slashed{n}  \chi_\nb$, and $\overline \chi_\nb  \gamma_\perp^\alpha  \slashed{n} \gamma_5 \chi_\nb$.

$\chi_\nb$, $\overline \chi_\nb$ or $\mathcal{B}_\perp^\alpha$ and $ \partial_\perp$ are each order $\lambda$ in the power counting, so the PDF is order $\lambda^{t-2}$ where $t$ is the total number of order $\lambda$ objects, and $-2$ comes from the states. Since the PDF Eq.~\eqref{13.2} is in the $\nb$ SCET sector, it is also valid in QCD where $\mathcal{B}$ is replaced by the field-strength tensor using Eq.~\eqref{6.3}. For example, the QCD gluon PDF is
\begin{align}
 g(x)  &= -\frac{1}{x P^+} \int \frac{{\rd} \xi}{2\pi}  e^{-i \xi x P^+}  \braket{P,S |\ [W_{\nb}^\dagger n_\alpha G^{\lambda_\perp \alpha }W_\nb] (n \xi) \ \
[W_\nb^\dagger n_\beta G_{ \lambda_\perp}{}^{\beta} \, W_\nb] (0)\ | P,S} \,,\nn
&= -\frac{x P^+}{g^2 } \int \frac{{\rd} \xi}{2\pi}  e^{-i \xi x P^+}  \braket{P,S |\  \mathcal{B}_\nb^{\lambda_\perp} (n \xi) \
\mathcal{B}_{{\nb}{\lambda_\perp}} (0) \ | P,S}\,,
\label{13.3}
\end{align}
and is order $\lambda^0$. In general, $[W_{\nb}^\dagger n_\alpha G^{\lambda_\perp \alpha }W_\nb] (n \xi) \to 
\mathcal{B}_\nb^{\lambda_\perp} (n \xi)/(i x P^+)$ where $xP^+$ is the momentum flowing out of $\mathcal{B}$ and into the hadronic matrix element.

\section{ \texorpdfstring{$D$-topology $q \overline{q} g$ Amplitude to Order $\lambda^4$}{D-topology qqg Amplitude to Order Lambda4}}\label{app:D}

The QCD 1-gluon tree-level amplitude Fig.~\ref{fig:subleading}, expanded to order $\lambda^4$ in the $D$-topology in the Breit-frame gives 
\begin{align}
	i\mathcal{M}^\mu_D = gT^a \overline{u}(p_1&) \bigg[  \frac{1}{Q}\left( \frac{\slashed{\varepsilon}_\perp^a}{u} \frac{\slashed{\nb}}{2}\gamma_\perp^\mu - \gamma_\perp^\mu \frac{\slashed{\nb}}{2} \frac{\slashed{\varepsilon}_\perp^a}{\overline{u}} \right) - \frac{p_{1\perp}.\varepsilon^a_\perp}{u\overline{u}Q^2}  (n^\mu + \nb^\mu) \slashed{\nb} \nn
	&+ \frac{p_{1\perp}.\varepsilon^a_\perp}{u\overline{u}Q^3} \left(  \frac{\slashed{p}_{1\perp}}{u} \slashed{\nb} \gamma_\perp^\mu -   \gamma_\perp^\mu \slashed{\nb} \frac{\slashed{p}_{1\perp}}{\overline{u}}   \right)  - \frac{p_{1\perp}.\varepsilon^a_\perp \, p_{1\perp}^2}{u^2(\overline{u})^2 Q^4} \nb^\mu \slashed{\nb} \bigg] v(p_2) 
\end{align}
where $u$ and $\overline{u}$ are the momentum fractions of the outgoing quark and antiquark, respectively. The above expansion contains all the order $g$ operators in the SCET $D$-topology. There are other $q\overline{q}g^r$ operators generated at order $\lambda^{r}$ where $r$ denotes the number of gluons in the amplitude. 

\section{\texorpdfstring{Endpoint Behavior of the $h_q(x,u)$ PDF}{Endpoint Behavior of the hq(x,u) PDF}}\label{app:F}

%
\begin{figure}
\begin{center}
\begin{align*}
\begin{array}{ccc}
\includegraphics[width=4cm]{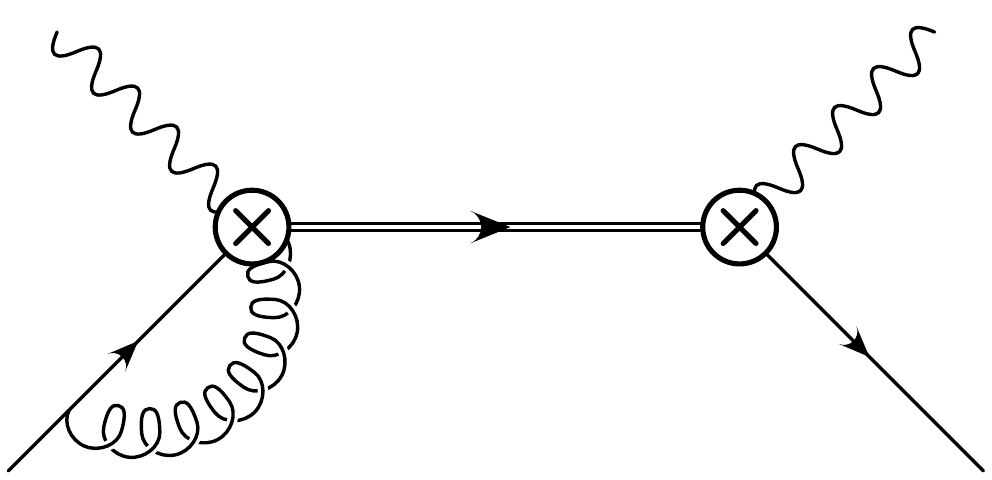} & \hspace{2cm} &
\includegraphics[width=4cm]{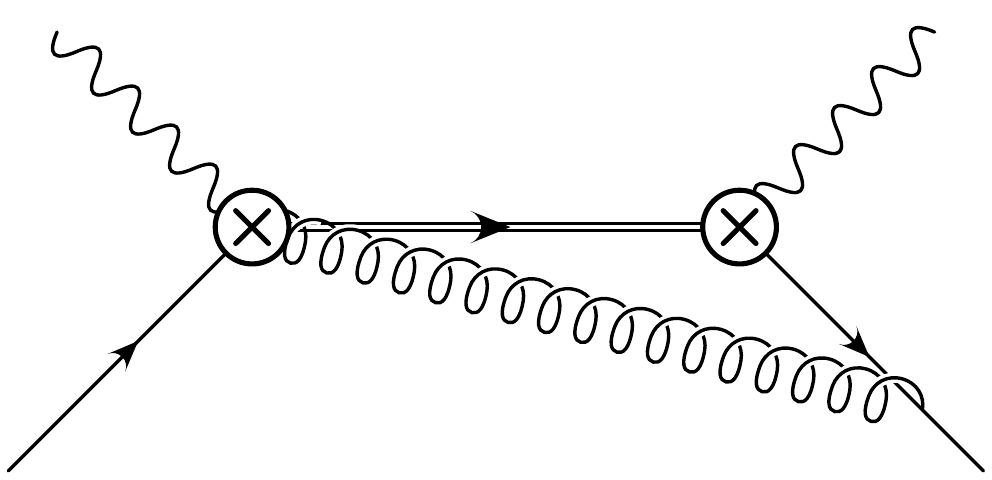} \\
\text{(a)} & & \text{(b)} 
\end{array}
\end{align*}
\end{center}
\caption{\label{fig:16} Graphs contributing to the matrix element of the PDF operator $H_q^\mu(x P^+,u)$ in a quark state. There are also graphs where the gluon couples to Wilson lines, and from the two-gluon contribution of  $[ \mathcal{B}_{\nb}^{\alpha}(0)]_{uQ}$, which vanish because of the $\perp$-projectors in the definition of the operator.}
\end{figure}
%
%
The behavior of the PDF $h_q(x,u)$ near the endpoint can be studied using the method of~\cite{vanBijleveld:2025ekz}, by taking the matrix element of the PDF operator $H_q^\mu(x p^+,u)$, Eq.~\eqref{4.50}, in a massive quark state with momentum $p$ and spin $s$. The mass $m$ of the quark introduces a scale, so that the graphs are no longer scaleless and zero, and also allows for the quark to be transversely polarized. The matrix element $\braket{p,s | H_q^\mu(x p^+,u) | p,s}$ starts at one-loop order, and the relevant graphs are shown in Fig.~\ref{fig:16}. Graphs where the gluon from $[ \mathcal{B}_{\nb}^{\alpha}(0)]_{uQ}$  couples to the Wilson lines in $ \mathcal{B}_{\nb}$, $\chi_{\nb}$ or $\overline \chi_{\nb}$ vanish because of the transverse projector on $\gamma_\perp^\alpha$ in the definition of $H_q^\mu(x p^+,u)$. The graph from the two-gluon contribution of  $ \mathcal{B}_{\nb}$ vanishes for the same reason. These vanishing graphs have not been shown in the figure.

Fig.~\ref{fig:16}(b) vanishes because the constraint that the incoming momentum of the gluon is $uQ$ cannot be satisfied. 
The $\gamma$-matrix structure of the graph in Fig.~\ref{fig:16}(a) can be simplified using the $\gamma$-matrix identity Eq.~\eqref{3.21}
and the Dirac equation, $\slashed{p} u(p,s) = m u(p,s)$, to give
\begin{align}
\overline u(p,s)  \gamma_\perp^\mu \frac{\slashed{n}}{2}   \frac{ \slashed{\overline n}}{2} u(p,s) 
&=  \frac{i}{2}  \epsilon_{ \perp \mu  \lambda}   \overline u(p,s)    \gamma_\lambda \gamma_5  u(p,s) =  i \epsilon_{ \perp \mu  \lambda}  s^\lambda\,, \nn
m\, \overline u(p,s)  \gamma_\perp^\mu \frac{\slashed{n}}{2} u(p,s) &=i\, p^+ \, \epsilon_{ \perp \mu  \lambda}  s^\lambda \,.
\label{f.2}
\end{align}
These have the structure of the matrix element Eq.~\eqref{4.53a}, and so give $h_q(x,u)$ for a quark state,
\begin{align}
h_q (x,u) &= -\frac{\alpha_s C_F}{4 \pi} \Gamma(\epsilon) e^{\epsilon \gamma_E} \delta(1-x) 
 (1- \epsilon)  u \left(\frac{\mu^2}{m^2 u^2} \right)^{\epsilon} \,,
 \label{f.3}
\end{align}
in $4-2\epsilon$ dimensions. Expanding in $\epsilon$,
\begin{align}
h_q (x,u) &=   -\frac{\alpha_s C_F}{4 \pi} \delta(1-x)  u  \left[ \frac{1}{\epsilon} -1 - \log \frac{m^2 u^2}{\mu^2} \right] \,,
\label{f.4}
\end{align}
where the $1/\epsilon$ UV divergence is cancelled by the PDF renormalization counterterm. We see that $h_q(x,u)$ behaves in $d$ dimensions as $u^{1-2\epsilon}$ and $(1-u)^0$ near the $u \to 0$ and $u\to 1$ endpoints,  or as $u^{1}$ and $(1-u)^0$ in four dimensions. These satisfy the $u$ integrability condition for the PDF that $h_q(x,u)$ should behave as $1/u^a$ or $1/(1-u)^b$ with $a,b < 1$.

\bibliographystyle{JHEP}
\bibliography{refs}

\providecommand{\href}[2]{#2}\begingroup\raggedright\begin{thebibliography}{10}

\bibitem{H1:2015ubc}
{\scshape H1, ZEUS} collaboration, H.~Abramowicz et~al., \emph{{Combination of
  measurements of inclusive deep inelastic ${e^{\pm }p}$ scattering cross
  sections and QCD analysis of HERA data}},
  \href{https://doi.org/10.1140/epjc/s10052-015-3710-4}{\emph{Eur. Phys. J. C}
  {\bfseries 75} (2015) 580},
  [\href{https://arxiv.org/abs/1506.06042}{{\ttfamily 1506.06042}}].

\bibitem{COMPASS:2016jwv}
{\scshape COMPASS} collaboration, C.~Adolph et~al., \emph{{Final COMPASS
  results on the deuteron spin-dependent structure function $g_1^{\rm d}$ and
  the Bjorken sum rule}},
  \href{https://doi.org/10.1016/j.physletb.2017.03.018}{\emph{Phys. Lett. B}
  {\bfseries 769} (2017) 34--41},
  [\href{https://arxiv.org/abs/1612.00620}{{\ttfamily 1612.00620}}].

\bibitem{JeffersonLabHallAg2p:2022qap}
{\scshape Jefferson Lab Hall A g2p} collaboration, D.~Ruth et~al.,
  \emph{{Proton spin structure and generalized polarizabilities in the strong
  quantum chromodynamics regime}},
  \href{https://doi.org/10.1038/s41567-022-01781-y}{\emph{Nature Phys.}
  {\bfseries 18} (2022) 1441--1446},
  [\href{https://arxiv.org/abs/2204.10224}{{\ttfamily 2204.10224}}].

\bibitem{AbdulKhalek:2021gbh}
R.~Abdul~Khalek et~al., \emph{{Science Requirements and Detector Concepts for
  the Electron-Ion Collider}: {EIC Yellow Report}},
  \href{https://doi.org/10.1016/j.nuclphysa.2022.122447}{\emph{Nucl. Phys. A}
  {\bfseries 1026} (2022) 122447},
  [\href{https://arxiv.org/abs/2103.05419}{{\ttfamily 2103.05419}}].

\bibitem{Bauer:2000ew}
C.~W. Bauer, S.~Fleming and M.~E. Luke, \emph{{Summing Sudakov logarithms in $B
  \rightarrow X_s \gamma$ in effective field theory}},
  \href{https://doi.org/10.1103/PhysRevD.63.014006}{\emph{Phys. Rev.}
  {\bfseries D63} (2000) 014006},
  [\href{https://arxiv.org/abs/hep-ph/0005275}{{\ttfamily hep-ph/0005275}}].

\bibitem{Bauer:2001yt}
C.~W. Bauer, D.~Pirjol and I.~W. Stewart, \emph{{Soft collinear factorization
  in effective field theory}},
  \href{https://doi.org/10.1103/PhysRevD.65.054022}{\emph{Phys. Rev.}
  {\bfseries D65} (2002) 054022},
  [\href{https://arxiv.org/abs/hep-ph/0109045}{{\ttfamily hep-ph/0109045}}].

\bibitem{Bauer:2001ct}
C.~W. Bauer and I.~W. Stewart, \emph{{Invariant operators in collinear
  effective theory}},
  \href{https://doi.org/10.1016/S0370-2693(01)00902-9}{\emph{Phys. Lett.}
  {\bfseries B516} (2001) 134--142},
  [\href{https://arxiv.org/abs/hep-ph/0107001}{{\ttfamily hep-ph/0107001}}].

\bibitem{Bauer:2002nz}
C.~W. Bauer, S.~Fleming, D.~Pirjol, I.~Z. Rothstein and I.~W. Stewart,
  \emph{{Hard scattering factorization from effective field theory}},
  \href{https://doi.org/10.1103/PhysRevD.66.014017}{\emph{Phys. Rev. D}
  {\bfseries 66} (2002) 014017},
  [\href{https://arxiv.org/abs/hep-ph/0202088}{{\ttfamily hep-ph/0202088}}].

\bibitem{Beneke:2002ph}
M.~Beneke, A.~P. Chapovsky, M.~Diehl and T.~Feldmann, \emph{{Soft collinear
  effective theory and heavy to light currents beyond leading power}},
  \href{https://doi.org/10.1016/S0550-3213(02)00687-9}{\emph{Nucl. Phys. B}
  {\bfseries 643} (2002) 431--476},
  [\href{https://arxiv.org/abs/hep-ph/0206152}{{\ttfamily hep-ph/0206152}}].

\bibitem{Manohar:2003vb}
A.~V. Manohar, \emph{{Deep inelastic scattering as $x \to 1$ using
  soft-collinear effective theory}},
  \href{https://doi.org/10.1103/PhysRevD.68.114019}{\emph{Phys. Rev.}
  {\bfseries D68} (2003) 114019},
  [\href{https://arxiv.org/abs/hep-ph/0309176}{{\ttfamily hep-ph/0309176}}].

\bibitem{Inglis-Whalen:2021bea}
M.~Inglis-Whalen, M.~Luke, J.~Roy and A.~Spourdalakis, \emph{{Factorization of
  power corrections in the Drell-Yan process in EFT}},
  \href{https://doi.org/10.1103/PhysRevD.104.076018}{\emph{Phys. Rev. D}
  {\bfseries 104} (2021) 076018},
  [\href{https://arxiv.org/abs/2105.09277}{{\ttfamily 2105.09277}}].

\bibitem{Luke:2022ops}
M.~Luke, J.~Roy and A.~Spourdalakis, \emph{{Factorization at subleading power
  in deep inelastic scattering in the $x\to 1$ limit}},
  \href{https://doi.org/10.1103/PhysRevD.107.074023}{\emph{Phys. Rev. D}
  {\bfseries 107} (2023) 074023},
  [\href{https://arxiv.org/abs/2210.02529}{{\ttfamily 2210.02529}}].

\bibitem{Goerke:2017lei}
R.~Goerke and M.~Inglis-Whalen, \emph{{Renormalization of dijet operators at
  order $1/Q^{2}$ in soft-collinear effective theory}},
  \href{https://doi.org/10.1007/JHEP05(2018)023}{\emph{JHEP} {\bfseries 05}
  (2018) 023}, [\href{https://arxiv.org/abs/1711.09147}{{\ttfamily
  1711.09147}}].

\bibitem{Ali:1991em}
A.~Ali, V.~M. Braun and G.~Hiller, \emph{{Asymptotic solutions of the evolution
  equation for the polarized nucleon structure function $g_2 (x, Q^2)$}},
  \href{https://doi.org/10.1016/0370-2693(91)90753-D}{\emph{Phys.Lett.}
  {\bfseries B266} (1991) 117--125}.

\bibitem{Geyer:1996isa}
B.~Geyer, D.~Mueller and D.~Robaschik, \emph{{The evolution of the nonsinglet
  twist - three parton distribution function}},  in \emph{{3rd Meeting on the
  Prospects of Nucleon-Nucleon Spin Physics at HERA}}, 11, 1996,
  \href{https://arxiv.org/abs/hep-ph/9611452}{{\ttfamily hep-ph/9611452}}.

\bibitem{Larin:1991tj}
S.~A. Larin and J.~A.~M. Vermaseren, \emph{{The $\alpha_s^3$ corrections to the
  Bjorken sum rule for polarized electroproduction and to the Gross-Llewellyn
  Smith sum rule}},
  \href{https://doi.org/10.1016/0370-2693(91)90839-I}{\emph{Phys. Lett.}
  {\bfseries B259} (1991) 345--352}.

\bibitem{Larin:1993tq}
S.~A. Larin, \emph{{The Renormalization of the axial anomaly in dimensional
  regularization}},
  \href{https://doi.org/10.1016/0370-2693(93)90053-K}{\emph{Phys. Lett.}
  {\bfseries B303} (1993) 113--118},
  [\href{https://arxiv.org/abs/hep-ph/9302240}{{\ttfamily hep-ph/9302240}}].

\bibitem{Jaffe:1988up}
R.~L. Jaffe and A.~Manohar, \emph{{Deep Inelastic Scattering from Arbitrary
  Spin Targets}},
  \href{https://doi.org/10.1016/0550-3213(89)90347-7}{\emph{Nucl. Phys. B}
  {\bfseries 321} (1989) 343}.

\bibitem{Hoodbhoy:1988am}
P.~Hoodbhoy, R.~L. Jaffe and A.~Manohar, \emph{{Novel Effects in Deep Inelastic
  Scattering from Spin 1 Hadrons}},
  \href{https://doi.org/10.1016/0550-3213(89)90572-5}{\emph{Nucl. Phys. B}
  {\bfseries 312} (1989) 571--588}.

\bibitem{Becher:2006mr}
T.~Becher, M.~Neubert and B.~D. Pecjak, \emph{{Factorization and Momentum-Space
  Resummation in Deep-Inelastic Scattering}},
  \href{https://doi.org/10.1088/1126-6708/2007/01/076}{\emph{JHEP} {\bfseries
  01} (2007) 076}, [\href{https://arxiv.org/abs/hep-ph/0607228}{{\ttfamily
  hep-ph/0607228}}].

\bibitem{Hoang:2015iva}
A.~H. Hoang, P.~Pietrulewicz and D.~Samitz, \emph{{Variable Flavor Number
  Scheme for Final State Jets in DIS}},
  \href{https://doi.org/10.1103/PhysRevD.93.034034}{\emph{Phys. Rev. D}
  {\bfseries 93} (2016) 034034},
  [\href{https://arxiv.org/abs/1508.04323}{{\ttfamily 1508.04323}}].

\bibitem{Goerke:2017ioi}
R.~Goerke and M.~Luke, \emph{{Power Counting and Modes in SCET}},
  \href{https://doi.org/10.1007/JHEP02(2018)147}{\emph{JHEP} {\bfseries 02}
  (2018) 147}, [\href{https://arxiv.org/abs/1711.09136}{{\ttfamily
  1711.09136}}].

\bibitem{Manohar:1992tz}
A.~V. Manohar, \emph{{An introduction to spin dependent deep inelastic
  scattering}},  \href{https://arxiv.org/abs/hep-ph/9204208}{{\ttfamily
  hep-ph/9204208}}.

\bibitem{Manohar:1990jx}
A.~V. Manohar, \emph{{Polarized parton distribution functions}},
  \href{https://doi.org/10.1103/PhysRevLett.66.289}{\emph{Phys. Rev. Lett.}
  {\bfseries 66} (1991) 289--292}.

\bibitem{Manohar:1990kr}
A.~V. Manohar, \emph{{Parton distributions from an operator viewpoint}},
  \href{https://doi.org/10.1103/PhysRevLett.65.2511}{\emph{Phys. Rev. Lett.}
  {\bfseries 65} (1990) 2511--2514}.

\bibitem{Burkhardt:1970ti}
H.~Burkhardt and W.~N. Cottingham, \emph{{Sum rules for forward virtual Compton
  scattering}},
  \href{https://doi.org/10.1016/0003-4916(70)90025-4}{\emph{Annals Phys.}
  {\bfseries 56} (1970) 453--463}.

\bibitem{Shuryak:1981pi}
E.~V. Shuryak and A.~I. Vainshtein, \emph{{Theory of Power Corrections to Deep
  Inelastic Scattering in Quantum Chromodynamics. 2. $Q^{-4}$ Effects:
  Polarized Target}},
  \href{https://doi.org/10.1016/0550-3213(82)90377-7}{\emph{Nucl. Phys. B}
  {\bfseries 201} (1982) 141}.

\bibitem{Kodaira:1978sh}
J.~Kodaira, S.~Matsuda, T.~Muta, K.~Sasaki and T.~Uematsu, \emph{{QCD Effects
  in Polarized Electroproduction}},
  \href{https://doi.org/10.1103/PhysRevD.20.627}{\emph{Phys. Rev.} {\bfseries
  D20} (1979) 627}.

\bibitem{Kodaira:1979ib}
J.~Kodaira, S.~Matsuda, K.~Sasaki and T.~Uematsu, \emph{{QCD Higher Order
  Effects in Spin Dependent Deep Inelastic Electroproduction}},
  \href{https://doi.org/10.1016/0550-3213(79)90329-8}{\emph{Nucl. Phys.}
  {\bfseries B159} (1979) 99--124}.

\bibitem{Kodaira:1979pa}
J.~Kodaira, \emph{{QCD Higher Order Effects in Polarized Electroproduction:
  Flavor Singlet Coefficient Functions}},
  \href{https://doi.org/10.1016/0550-3213(80)90310-7}{\emph{Nucl. Phys.}
  {\bfseries B165} (1980) 129--140}.

\bibitem{Kodaira:1994ge}
J.~Kodaira, Y.~Yasui and T.~Uematsu, \emph{{Spin structure function $g_2 (x,
  Q^2)$ and twist - three operators in QCD}},
  \href{https://doi.org/10.1016/0370-2693(94)01550-V}{\emph{Phys. Lett. B}
  {\bfseries 344} (1995) 348--354},
  [\href{https://arxiv.org/abs/hep-ph/9408354}{{\ttfamily hep-ph/9408354}}].

\bibitem{Bukhvostov:1983eob}
A.~P. Bukhvostov, E.~A. Kuraev and L.~N. Lipatov, \emph{{Deep inelastic
  electron scattering by a polarized target in quantum chromodynamics}},
  {\emph{JETP Lett.} {\bfseries 37} (1983) 482--486}.

\bibitem{Ratcliffe:1985mp}
P.~G. Ratcliffe, \emph{{Transverse Spin and Higher Twist in {QCD}}},
  \href{https://doi.org/10.1016/0550-3213(86)90495-5}{\emph{Nucl. Phys. B}
  {\bfseries 264} (1986) 493--512}.

\bibitem{Ji:1990br}
X.-D. Ji and C.-h. Chou, \emph{{QCD radiative corrections to the transverse
  spin structure function $g_2 (x, Q^2)$: 1. Nonsinglet operators}},
  \href{https://doi.org/10.1103/PhysRevD.42.3637}{\emph{Phys. Rev. D}
  {\bfseries 42} (1990) 3637--3644}.

\bibitem{Balitsky:1987bk}
I.~I. Balitsky and V.~M. Braun, \emph{{Evolution Equations for QCD String
  Operators}}, \href{https://doi.org/10.1016/0550-3213(89)90168-5}{\emph{Nucl.
  Phys. B} {\bfseries 311} (1989) 541--584}.

\bibitem{Jaffe:1983hp}
R.~Jaffe, \emph{{Parton Distribution Functions for Twist Four}},
  \href{https://doi.org/10.1016/0550-3213(83)90361-9}{\emph{Nucl.Phys.}
  {\bfseries B229} (1983) 205}.

\bibitem{Manohar:2006nz}
A.~V. Manohar and I.~W. Stewart, \emph{{The Zero-Bin and Mode Factorization in
  Quantum Field Theory}},
  \href{https://doi.org/10.1103/PhysRevD.76.074002}{\emph{Phys. Rev.}
  {\bfseries D76} (2007) 074002},
  [\href{https://arxiv.org/abs/hep-ph/0605001}{{\ttfamily hep-ph/0605001}}].

\bibitem{Freedman:2011kj}
S.~M. Freedman and M.~Luke, \emph{{SCET, QCD and Wilson Lines}},
  \href{https://doi.org/10.1103/PhysRevD.85.014003}{\emph{Phys. Rev. D}
  {\bfseries 85} (2012) 014003},
  [\href{https://arxiv.org/abs/1107.5823}{{\ttfamily 1107.5823}}].

\bibitem{Manohar:2002fd}
A.~V. Manohar, T.~Mehen, D.~Pirjol and I.~W. Stewart, \emph{{Reparameterization
  invariance for collinear operators}},
  \href{https://doi.org/10.1016/S0370-2693(02)02029-4}{\emph{Phys. Lett. B}
  {\bfseries 539} (2002) 59--66},
  [\href{https://arxiv.org/abs/hep-ph/0204229}{{\ttfamily hep-ph/0204229}}].

\bibitem{Ji:2000ny}
X.-D. Ji, W.~Lu, J.~Osborne and X.-T. Song, \emph{{One loop factorization of
  the nucleon $g_2$ structure function in the nonsinglet case}},
  \href{https://doi.org/10.1103/PhysRevD.62.094016}{\emph{Phys. Rev. D}
  {\bfseries 62} (2000) 094016},
  [\href{https://arxiv.org/abs/hep-ph/0006121}{{\ttfamily hep-ph/0006121}}].

\bibitem{vanBijleveld:2025ekz}
R.~van Bijleveld, E.~Laenen, C.~Marinissen, L.~Vernazza and G.~Wang,
  \emph{{Next-to-leading power jet functions in the small-mass limit in QED}},
  \href{https://doi.org/10.1007/JHEP07(2025)257}{\emph{JHEP} {\bfseries 07}
  (2025) 257}, [\href{https://arxiv.org/abs/2503.10810}{{\ttfamily
  2503.10810}}].

\bibitem{Beneke:2020ibj}
M.~Beneke, M.~Garny, S.~Jaskiewicz, R.~Szafron, L.~Vernazza and J.~Wang,
  \emph{{Large-x resummation of off-diagonal deep-inelastic parton scattering
  from d-dimensional refactorization}},
  \href{https://doi.org/10.1007/JHEP10(2020)196}{\emph{JHEP} {\bfseries 10}
  (2020) 196}, [\href{https://arxiv.org/abs/2008.04943}{{\ttfamily
  2008.04943}}].

\bibitem{Freedman:2014uta}
S.~M. Freedman and R.~Goerke, \emph{{Renormalization of Subleading Dijet
  Operators in Soft-Collinear Effective Theory}},
  \href{https://doi.org/10.1103/PhysRevD.90.114010}{\emph{Phys. Rev. D}
  {\bfseries 90} (2014) 114010},
  [\href{https://arxiv.org/abs/1408.6240}{{\ttfamily 1408.6240}}].

\bibitem{Goerke:2018pns}
R.~Goerke, \emph{{A New Formalism for Soft Collinear Effective Theory with
  Applications}}, Ph.D. thesis, Toronto U., 2018.

\bibitem{Ji:1992eu}
X.-D. Ji, \emph{{Gluon correlations in the transversely polarized nucleon}},
  \href{https://doi.org/10.1016/0370-2693(92)91375-J}{\emph{Phys. Lett. B}
  {\bfseries 289} (1992) 137--142}.

\bibitem{DeRujula:1976baf}
A.~De~Rujula, H.~Georgi and H.~D. Politzer, \emph{{Demythification of
  Electroproduction, Local Duality and Precocious Scaling}},
  \href{https://doi.org/10.1016/S0003-4916(97)90003-8}{\emph{Annals Phys.}
  {\bfseries 103} (1977) 315}.

\bibitem{Bardeen:1978yd}
W.~A. Bardeen, A.~J. Buras, D.~W. Duke and T.~Muta, \emph{{Deep Inelastic
  Scattering Beyond the Leading Order in Asymptotically Free Gauge Theories}},
  \href{https://doi.org/10.1103/PhysRevD.18.3998}{\emph{Phys. Rev.} {\bfseries
  D18} (1978) 3998}.

\bibitem{Ahmed:1976ee}
M.~Ahmed and G.~G. Ross, \emph{{Polarized Lepton - Hadron Scattering in
  Asymptotically Free Gauge Theories}},
  \href{https://doi.org/10.1016/0550-3213(76)90328-X}{\emph{Nucl.Phys.}
  {\bfseries B111} (1976) 441}.

\bibitem{Blumlein:2012bf}
J.~Bl{\"u}mlein, \emph{The theory of deeply inelastic scattering},
  \href{https://doi.org/10.1016/j.ppnp.2012.09.006}{\emph{Progress in Particle
  and Nuclear Physics} {\bfseries 69} (Mar., 2013) 28--84}.

\bibitem{Mertig:1995ny}
R.~Mertig and W.~L. van Neerven, \emph{{The Calculation of the two loop spin
  splitting functions $P_{ij}^{(1)}(x)$}},
  \href{https://doi.org/10.1007/s002880050138}{\emph{Z. Phys.} {\bfseries C70}
  (1996) 637--654}, [\href{https://arxiv.org/abs/hep-ph/9506451}{{\ttfamily
  hep-ph/9506451}}].

\bibitem{Vogelsang:1995vh}
W.~Vogelsang, \emph{{A Rederivation of the spin dependent next-to-leading order
  splitting functions}},
  \href{https://doi.org/10.1103/PhysRevD.54.2023}{\emph{Phys. Rev. D}
  {\bfseries 54} (1996) 2023--2029},
  [\href{https://arxiv.org/abs/hep-ph/9512218}{{\ttfamily hep-ph/9512218}}].

\bibitem{Larin:1996wd}
S.~A. Larin, P.~Nogueira, T.~van Ritbergen and J.~A.~M. Vermaseren, \emph{{The
  Three loop QCD calculation of the moments of deep inelastic structure
  functions}}, \href{https://doi.org/10.1016/S0550-3213(97)80038-7}{\emph{Nucl.
  Phys. B} {\bfseries 492} (1997) 338--378},
  [\href{https://arxiv.org/abs/hep-ph/9605317}{{\ttfamily hep-ph/9605317}}].

\bibitem{Moch:2004pa}
S.~Moch, J.~A.~M. Vermaseren and A.~Vogt, \emph{{The three-loop splitting
  functions in QCD: The non-singlet case}},
  \href{https://doi.org/10.1016/j.nuclphysb.2004.03.030}{\emph{Nucl. Phys.}
  {\bfseries B688} (2004) 101--134},
  [\href{https://arxiv.org/abs/hep-ph/0403192}{{\ttfamily hep-ph/0403192}}].

\bibitem{Vogt:2004mw}
A.~Vogt, S.~Moch and J.~A.~M. Vermaseren, \emph{{The three-loop splitting
  functions in QCD: The singlet case}},
  \href{https://doi.org/10.1016/j.nuclphysb.2004.04.024}{\emph{Nucl. Phys.}
  {\bfseries B691} (2004) 129--181},
  [\href{https://arxiv.org/abs/hep-ph/0404111}{{\ttfamily hep-ph/0404111}}].

\bibitem{Moch:2014sna}
S.~Moch, J.~A.~M. Vermaseren and A.~Vogt, \emph{{The Three-Loop Splitting
  Functions in QCD: The Helicity-Dependent Case}},
  \href{https://doi.org/10.1016/j.nuclphysb.2014.10.016}{\emph{Nucl. Phys. B}
  {\bfseries 889} (2014) 351--400},
  [\href{https://arxiv.org/abs/1409.5131}{{\ttfamily 1409.5131}}].

\bibitem{Blumlein:2021ryt}
J.~Bl{\"u}mlein, P.~Marquard, C.~Schneider and K.~Sch{\"o}nwald, \emph{{The
  three-loop polarized singlet anomalous dimensions from off-shell operator
  matrix elements}}, \href{https://doi.org/10.1007/JHEP01(2022)193}{\emph{JHEP}
  {\bfseries 01} (2022) 193},
  [\href{https://arxiv.org/abs/2111.12401}{{\ttfamily 2111.12401}}].

\bibitem{Blumlein:2021enk}
J.~Bl{\"u}mlein, P.~Marquard, C.~Schneider and K.~Sch{\"o}nwald, \emph{{The
  three-loop unpolarized and polarized non-singlet anomalous dimensions from
  off shell operator matrix elements}},
  \href{https://doi.org/10.1016/j.nuclphysb.2021.115542}{\emph{Nucl. Phys. B}
  {\bfseries 971} (2021) 115542},
  [\href{https://arxiv.org/abs/2107.06267}{{\ttfamily 2107.06267}}].

\bibitem{Blumlein:2022gpp}
J.~Bl{\"u}mlein, P.~Marquard, C.~Schneider and K.~Sch{\"o}nwald, \emph{{The
  massless three-loop Wilson coefficients for the deep-inelastic structure
  functions F$_{2}$, F$_{L}$, xF$_{3}$ and g$_{1}$}},
  \href{https://doi.org/10.1007/JHEP11(2022)156}{\emph{JHEP} {\bfseries 11}
  (2022) 156}, [\href{https://arxiv.org/abs/2208.14325}{{\ttfamily
  2208.14325}}].

\bibitem{Ablinger:2009ovq}
J.~Ablinger, \emph{{A Computer Algebra Toolbox for Harmonic Sums Related to
  Particle Physics}},  Master's thesis, Linz U., 2009.

\bibitem{Ablinger:2012ufz}
J.~Ablinger, \emph{{Computer Algebra Algorithms for Special Functions in
  Particle Physics}}, Ph.D. thesis, Linz U., 4, 2012.
\newblock \href{https://arxiv.org/abs/1305.0687}{{\ttfamily 1305.0687}}.

\bibitem{Ablinger:2017tan}
J.~Ablinger, A.~Behring, J.~Bl{\"u}mlein, A.~De~Freitas, A.~von Manteuffel and
  C.~Schneider, \emph{{The three-loop splitting functions $P_{qg}^{(2)}$ and
  $P_{gg}^{(2, N_F)}$}},
  \href{https://doi.org/10.1016/j.nuclphysb.2017.06.004}{\emph{Nucl. Phys. B}
  {\bfseries 922} (2017) 1--40},
  [\href{https://arxiv.org/abs/1705.01508}{{\ttfamily 1705.01508}}].

\bibitem{tHooft:1972tcz}
G.~'t~Hooft and M.~J.~G. Veltman, \emph{{Regularization and Renormalization of
  Gauge Fields}},
  \href{https://doi.org/10.1016/0550-3213(72)90279-9}{\emph{Nucl. Phys. B}
  {\bfseries 44} (1972) 189--213}.

\bibitem{Breitenlohner:1977hr}
P.~Breitenlohner and D.~Maison, \emph{{Dimensional Renormalization and the
  Action Principle}}, \href{https://doi.org/10.1007/BF01609069}{\emph{Commun.
  Math. Phys.} {\bfseries 52} (1977) 11--38}.

\bibitem{TRUEMAN1979331}
T.~Trueman, \emph{{Chiral symmetry in perturbative QCD}},
  \href{https://doi.org/https://doi.org/10.1016/0370-2693(79)90480-5}{\emph{Physics
  Letters B} {\bfseries 88} (1979) 331--334}.

\bibitem{Matiounine:1998re}
Y.~Matiounine, J.~Smith and W.~L. van Neerven, \emph{{Two loop operator matrix
  elements calculated up to finite terms for polarized deep inelastic lepton -
  hadron scattering}},
  \href{https://doi.org/10.1103/PhysRevD.58.076002}{\emph{Phys. Rev. D}
  {\bfseries 58} (1998) 076002},
  [\href{https://arxiv.org/abs/hep-ph/9803439}{{\ttfamily hep-ph/9803439}}].

\bibitem{Kogut:1969xa}
J.~B. Kogut and D.~E. Soper, \emph{{Quantum Electrodynamics in the Infinite
  Momentum Frame}}, \href{https://doi.org/10.1103/PhysRevD.1.2901}{\emph{Phys.
  Rev. D} {\bfseries 1} (1970) 2901--2913}.

\bibitem{Jaffe:1991ra}
R.~Jaffe and X.-D. Ji, \emph{{Chiral odd parton distributions and Drell-Yan
  processes}},
  \href{https://doi.org/10.1016/0550-3213(92)90110-W}{\emph{Nucl.Phys.}
  {\bfseries B375} (1992) 527--560}.

\bibitem{Ji:2001bm}
X.-D. Ji and J.~Osborne, \emph{{An Analysis of the next-to-leading order
  corrections to the $g_T( = g_1 + g_2)$ scaling function}},
  \href{https://doi.org/10.1016/S0550-3213(01)00249-8}{\emph{Nucl. Phys. B}
  {\bfseries 608} (2001) 235--278},
  [\href{https://arxiv.org/abs/hep-ph/0102026}{{\ttfamily hep-ph/0102026}}].

\end{thebibliography}\endgroup

\end{document}